\documentclass[prb,aps,reprint,a4paper,superscriptaddress,citeautoscript,floatfix]{revtex4-2}

\usepackage[charter]{mathdesign}
\DeclareSymbolFont{usualmathcal}{OMS}{cmsy}{m}{n}
\DeclareSymbolFontAlphabet{\mathcal}{usualmathcal}
\usepackage{amsmath}
\usepackage[pdftex]{graphicx}
\usepackage{microtype}
\usepackage[svgnames]{xcolor}
\usepackage[colorlinks=True,linkcolor=DarkRed,citecolor=ForestGreen,urlcolor=MediumBlue,
	pdfstartview=FitH,bookmarks=False,pdfpagemode=UseNone]{hyperref}

\let\vec\mathbf
\usepackage{bm}

\def\up{\uparrow}
\def\down{\downarrow}
\def\pdagger{^{\phantom\dagger}}

\begin{document}

\title{Angular momentum of vortex-core Majorana zero modes}

\author{Giulia Venditti}
\email{giulia.venditti@unige.ch}
\affiliation{Department of Quantum Matter Physics, University of Geneva, 24 quai Ernest-Ansermet, 1211 Geneva, Switzerland}
\author{Christophe Berthod}
\affiliation{Department of Quantum Matter Physics, University of Geneva, 24 quai Ernest-Ansermet, 1211 Geneva, Switzerland}
\author{Louk Rademaker}
\affiliation{Department of Quantum Matter Physics, University of Geneva, 24 quai Ernest-Ansermet, 1211 Geneva, Switzerland}
\affiliation{Institute-Lorentz for Theoretical Physics, Leiden University, PO Box 9506, 2300 Leiden, The Netherlands}

\date{January 5, 2026}

\begin{abstract}

Majorana zero modes (MZMs) are highly sought-after states with a possible application in quantum computation. Here, we show that vortex-core MZMs can carry a nontrivial angular momentum. This establishes distinct `flavors' of Majorana modes, independent of the Chern classification of topological superconductors. The MZM angular momentum is explicitly calculated for a microscopic model of a $d+id$ superconductor placed on a three-dimensional topological insulator ($d+id+\phantom{}$Dirac model) using both exact diagonalization and the Chebyshev expansion. We classify all possible quantum numbers of MZMs depending on the windings of the order parameter and underlying normal state. The topological protection of the MZM is set by the bulk gap, quasiparticle poisoning by trivial in-gap states, and its localization length. All these severely limit the stability of MZMs in the $d+id+\phantom{}$Dirac model, in contrast to earlier claims. Nevertheless, the possibility of having different flavors of MZM---in the form of angular momentum or something else---can provide a unique path forward for the study of MZMs.

\end{abstract}

\maketitle

\section{Introduction}

The search for Majorana particles \cite{Yazdani.2023} is important both for fundamental understanding and possible topological quantum computation \cite{Aasen.2025}. Since Majorana particles are an equal superposition of particle and antiparticle, they are naturally sought in superconductors, in the form of Majorana zero modes (MZMs). From Kitaev's seminal paper \cite{Kitaev.2001}, the quest to engineer devices capable of hosting MZMs is still ongoing.

Fu and Kane proposed to find MZMs in a vortex core at the interface between an $s$-wave superconductor and a strong topological insulator (STI) \cite{Fu.2008}. Evidence for such MZMs in vortex cores has been found in the iron-based FeTe$_{0.55}$Se$_{0.45}$ \cite{Pathak.2021, PengZhang.2018, DongfeiWang.2018, Kong.2019}, providing a natural platform for the Fu--Kane model, and in heterostructures of Bi$_2$Te$_3$/NbSe$_2$ \cite{Xu.2015, Sun.2016}, and SnTe/Pb \cite{Liu.2024}. The big challenge, however, is to separate the MZMs from other in-gap states, a phenomenon that has been colorfully called `quasiparticle poisoning' of the Majorana. Naturally, a larger critical $T_c$ of the superconductor and thus a larger gap can provide a more stable platform to engineer MZMs. As such, it is natural to look at the systems with a higher $T_c$, such as the cuprates. This deviates from the original Fu--Kane proposal, however, as the cuprates' $d$-wave pairing should be taken into account \cite{Tsuei.2000, Zhongbo.2018, TaoLiu.2018, Wang.2013, Su-YangXu.2014, Zi-XiangLi.2015}. Experimental studies have been performed with Bi$_2$Se$_3$ or Bi$_2$Te$_3$ combined with the cuprate Bi$_2$Sr$_2$CaCu$_2$O$_{8+\delta}$ \cite{Zareapour.2012, Wang.2013,Su-YangXu.2014}. However, quasi-particle poisoning is an even bigger problem in $d$-wave cuprates, since they are nodal (gapless)  superconductors.

A recent proposal for ``high-temperature'' MZMs sought to alleviate this problem by putting a fully gapped $d+id$ superconductor on the surface of a STI \cite{Mercado.2022}. This superconducting order parameter might be realized in twisted cuprates \cite{Can.2021, Volkov.2025, Zhao.2023, Zhu.2021, Volkov.2023, Martini.2023}, but also in elemental tin \cite{Wu.2025, Ming.2017, Ming.2023, Wu.2025}. A $d+id$ order parameter was also predicted for graphene heterostructures \cite{Kennes.2018, Törmä.2022, Pantaleón.2023}, as well as twisted bilayer WSe$_2$ \cite{Akbar.2024}, where superconductivity was recently discovered \cite{Xia.2025, Guo.2025}. In Sr$_2$RuO$_4$, the possibility of time reversal symmetry breaking and spin-singlet superconductivity could also provide a $d+id$ platform, although the symmetry of the order parameter is still debated \cite{Roising.2019, Grinenko.2021}.

From the theoretical point of view, a close inspection of the vortex-core MZM in such a $d+id+\phantom{}$Dirac system reveals novel properties of the zero mode itself. In this paper we show that, depending on microscopic parameters, the MZM carries a nontrivial \emph{angular momentum}. Specifically, this angular momentum will be a consequence of the three chiralities involved in the system: the Dirac cone winding number, the $d+id$ order parameter winding, and the winding of the vortex. The Bogoliubov--de Gennes equation now puts constraints on the relation between these chiralities and the real-space phase winding of the electron and hole components of the MZM. Surprisingly, the angular momentum of the MZM is independent of the Chern number of the topological superconducting phase, which in this setup can be $C = \pm1$ ($p\pm ip$ symmetry) or $C=\pm 3$ ($f\pm if$ symmetry). The theory of MZM angular momentum is developed in Sec.~\ref{sec:theory}.

Having established the existence of the MZM and their angular momentum, we revisit the question of their topological protection. In other words, what are the limits of the MZM stability? Given the unusual combination of $d+id$ with a Dirac cone, it turns out that not $T_c$, but the much lower gap at the Fermi level, governs the stability and localization length of the MZM. Moreover, the stability of MZMs is further limited by additional in-gap states within vortices, the Caroli--de Gennes--Matricon (CdGM) states \cite{Caroli.1964}. Both effects are discussed in Sec.~\ref{sec:protection}. The resulting enormous size of the MZM wavefunction provides another challenge for the accurate numerical solution of a lattice model, which we address in two ways discussed in Sec.~\ref{Sec:ChallengesNumerical}: by redefining appropriate values of the parameters involved in the pairing, and by using an expansion on Chebyshev polynomials.

We end this paper with a brief outlook on how realistic the detection and use of vortex-core MZM can be, including the role of the previously unrecognized angular momentum (Sec.~\ref{Sec:Outlook}).

\section{Theory of zero-mode angular momentum}
\label{sec:theory}

The core result of this paper is that vortex MZMs in topological superconductors carry an angular momentum quantum number $\ell$. This number is in principle independent of the Chern number $C$ of the superconductor. While the existence of the MZM requires an odd Chern number, the value of the angular momentum $\ell$ depends on a nontrivial combination of microscopic parameters and the winding of the vortex.

In this section, we derive the theory of MZM angular momentum, starting with a continuum model of MZM in a $d+id$ superconductor on top of the surface of a three-dimensional topological insulator, inspired by Ref.~\cite{Mercado.2022}. The general derivation of the angular-momentum quantum number $\ell$ is supplemented by a numerical calculation on a lattice model, which includes a generalization of the notion of angular momentum of vortex states to the case of a square lattice with discrete rotational symmetry.

\subsection{\boldmath Continuum model of $d+id+\phantom{}$Dirac}

Our starting point is a continuum description of a specific model with a spin-Dirac cone and a $d+id$ pairing interaction \cite{Ohashi.2021}, summarized as ``$d+id+\phantom{}$Dirac''. Both the Dirac cone and the pairing term have a \emph{winding number}, which we will use to derive the Chern number of the corresponding topological superconducting phase. Since the total Chern number is odd, there will be MZMs in vortex cores. We finally analyze the nontrivial angular momentum $\ell$ of those modes.

\subsubsection{Winding numbers of a Dirac cone with $d\pm id$ pairing}

Let us consider a general model with a dispersion $h_\vec{k}$ and a singlet superconducting order parameter $\Delta_\vec{k}$ induced by proximity effect, which is a generalization of the Fu--Kane model \cite{Fu.2008}. The Bogoliubov--de Gennes (BdG) Hamiltonian is explicitly written as
\begin{equation}
	H_\mathrm{BdG}({\vec{k}}) = \Psi^\dagger_\vec{k}
		\begin{pmatrix}
			h_\vec{k} - \mu \sigma^0 & \Delta_\vec{k} \sigma^0 \\
			\Delta^*_\vec{k} \sigma^0 & -h_\vec{k} + \mu \sigma^0
		\end{pmatrix}
		\Psi_\vec{k},
	\label{Eq:HamiltonianContinuum_k}
\end{equation}
where we have introduced a chemical potential $\mu$, while the Nambu spinor 
\begin{equation}
	\Psi_\vec{k} = \big(c\pdagger_{\vec{k}\up}, c\pdagger_{\vec{k}\down},
	c^\dagger_{-{\vec{k}}\down}, -c^\dagger_{-{\vec{k}}\up}\big)^\intercal
\end{equation}
follows the convention adopted in Ref.~\cite{Fu.2008}. With this convention, the superconducting terms are diagonal (proportional to $\sigma^0$) in Nambu space for spin-singlet superconductivity. 

The dispersion $h_\vec{k}$ represents the surface of a three-dimensional topological insulator, which can be described by a single Dirac cone,
\begin{align}
	\nonumber
	h_\vec{k} &= v \left(\sigma^y k_x + n_X \sigma^x k_y\right) \\
    &= \begin{pmatrix}
        0 & -i v|\vec{k}| e^{i n_X \theta_\vec{k}} \\
        i v|\vec{k}| e^{- i n_X \theta_\vec{k}} &  0
    \end{pmatrix},
    \label{Eq:NormalH_withNx}
\end{align}
where $v$ is the Dirac velocity (setting $\hbar=1$), $\sigma$ are Pauli matrices referring to the electron spin \cite{Hasan.2010}, $\tan \theta_\vec{k} = k_y / k_x$, and $n_X=\pm 1$. With $n_X=-1$, this is equivalent to the often-used notation
\begin{equation}
    h'_\vec{k} = v\, \bm{\sigma} \cdot {\vec{k}}.
    \label{Eq:NormalH_sigma-dot-k}
\end{equation}
A $\pi/2$ rotation $R_{\pi/2} = e^{i(\pi/4)\sigma^z}$ of the spin along the $z$-axis changes one dispersion relation into the other, according to $h_\vec{k} = R^\dagger_{\pi/2} h'_\vec{k} R_{\pi/2}$. Both expressions are therefore interchangeable. Refs.~\cite{Fu.2008} and \cite{Hasan.2010} use Eq.~\eqref{Eq:NormalH_sigma-dot-k}, while we will use Eq.~\eqref{Eq:NormalH_withNx} like in Ref.~\cite{Mercado.2022}. Note that the normal-state Hamiltonian $h_\vec{k}$ carries a winding number $n_X$, as manifested in the off-diagonal matrix elements that are proportional to $e^{\pm i n_X \theta_\vec{k}}$. We shall refer to $n_X$ as \emph{the winding number of the Dirac cone}.

The superconducting order parameter is momentum-independent for $s$-wave pairing, whereas in the case of $d \pm id$ pairing it reads
\begin{equation}
	\Delta_\vec{k} = \frac{\Delta}{4}\left(k_x^2 - k_y^2\right) + i \frac{n_\Delta}{2} \Delta' k_x k_y,
	\label{Eq:OrderParameterContinuum}
\end{equation}
where $k_x$ and $k_y$ are dimensionless coordinates in reciprocal space and $n_\Delta=\pm2$. The factor $1/4$ is chosen for consistency with the lattice model (Sec.~\ref{sec:lattice-model}), which gives a peak-to-peak spectral gap of $2 \Delta$ in accordance with the general cuprate literature. Similar to the Dirac cone, the $d\pm id$ pairing carries a winding number. This is most clearly seen in the chiral limit $\Delta' = \Delta/2$, where Eq.~\eqref{Eq:OrderParameterContinuum} can be rewritten as
\begin{equation}
	\Delta_\vec{k} = \frac{\Delta}{4}|\vec{k}|^2 e^{i n_\Delta \theta_\vec{k}}
	\label{Eq:nDelta}
\end{equation}
with winding $n_\Delta = \pm 2$.

The Hamiltonian defined by Eqs.~\eqref{Eq:HamiltonianContinuum_k} to \eqref{Eq:OrderParameterContinuum} has an anti-unitary anti-commuting particle-hole symmetry $\{\mathcal{P}, H\}=0$, where the particle-hole symmetry operator is defined as $\mathcal{P}=\sigma^y\tau^y\mathcal{K}$ with $\mathcal{P}^2=\openone$, $\mathcal{K}$ the complex conjugation, and $\tau$ are the Pauli matrices in the particle-hole sector. Time-reversal symmetry is broken in the case of $d\pm id$ pairing, which makes the system fall into symmetry class D according to Ref.~\cite{Teo.Kane.2008}. 

Finally, the spectrum of the Hamiltonian is
\begin{equation}
    E_\vec{k}=\pm\sqrt{(\pm v|{\vec{k}}|-\mu)^2+|\Delta_\vec{k}|^2}.
\end{equation}

\subsubsection{Effective order-parameter winding and Chern number}

\begin{table}[tb]
	\setlength{\tabcolsep}{8pt}\def\arraystretch{1.3}
	\begin{tabular*}{0.85\columnwidth}{rr|@{\extracolsep{\fill}}rc|@{\extracolsep{\fill}}rc}
    \hline\hline
    \multicolumn{1}{c}{$n_X$} &
    \multicolumn{1}{c|}{$n_\Delta$} &
    \multicolumn{2}{l|}{$C~~(\mu<0$)} &
    \multicolumn{2}{l}{$C~~(\mu>0$)} \\
    \hline
    $+1$ & $+2$ & $+1$  & $p+ip$ & $-3$ & $f-if$ \\
    $  $ & $-2$ & $-3$  & $f-if$ & $+1$ & $p+ip$ \\
    $-1$ & $+2$ & $+3$  & $f+if$ & $-1$ & $p-ip$ \\
    $  $ & $-2$ & $-1$  & $p-ip$ & $+3$ & $f+if$ \\
    \hline\hline
    \end{tabular*}
	\caption{Chern number ($C$) as given by Eq.~\eqref{Eq:ChernFormula} for the topological superconducting state arising from a hole-doped ($\mu<0$) or electron-doped ($\mu>0$) Dirac cone with winding number $n_X$ and a $d\pm id$ pairing with winding $n_\Delta$.}
        \label{tab:windings&OP}
\end{table}

The combination of the Dirac cone dispersion with the singlet $d\pm id$ pairing gives rise on the Fermi surface to an effective triplet pairing. Indeed, upon diagonalizing the normal-state Hamiltonian, the winding of the Dirac cone is transferred to the pairing term, changing its winding from $\pm2$ to $\pm1$ or $\pm3$. To show this, we diagonalize the normal-state part of the Hamiltonian with the transformation \footnote{Note that this transformation should have a smooth gauge; otherwise new sources of Berry curvature are introduced in the normal part of the BdG Hamiltonian. In that case, the winding of the order parameter would no longer be equal to the Chern number.} 
\begin{subequations}\label{Eq:UmatrixBdG}\begin{align}
	\mathcal{U}_\mathrm{BdG} &=
		\begin{pmatrix}
			\mathcal{U}_\vec{k} & 0 \\
			0 & \sigma^y\,\mathcal{U}^*_{-\vec{k}} 
		\end{pmatrix} \\
	\mathcal{U}_\vec{k} &= \frac{1}{\sqrt{2}}
		\begin{pmatrix}
			1 & i e^{ i n_X \theta_\vec{k}} \\
			i e^{- i n_X \theta_\vec{k}} & 1
		\end{pmatrix},
\end{align}\end{subequations}
which is such that $\mathcal{U}^\dagger_\vec{k} h\pdagger_\vec{k} \mathcal{U}\pdagger_\vec{k} = \begin{pmatrix}  v |{\vec{k}}| & 0 \\ 0 & - v |{\vec{k}}| \end{pmatrix}$. The transformed Hamiltonian is
\begin{multline*}
	H_\mathrm{BdG}' = \mathcal{U}^\dagger_\mathrm{BdG} H\pdagger_\mathrm{BdG} \mathcal{U}\pdagger_\mathrm{BdG} \\
    = \begin{pmatrix}
    	v|\vec{k}|-\mu & 0 & \Delta_\vec{k}e^{i n_X \theta_\vec{k}} & 0 \\
		0 & -v|\vec{k}|-\mu  & 0 & -\Delta_\vec{k}e^{-i n_X \theta_\vec{k}} \\
		\Delta^*_\vec{k}e^{-i n_X \theta_\vec{k}} & 0 & -v|\vec{k}|+\mu & 0 \\
		0 & -\Delta^*_\vec{k}e^{i n_X \theta_\vec{k}} & 0 & v|\vec{k}|+\mu
	\end{pmatrix},
\end{multline*}
where we have used the relation $e^{i \theta_{-\vec{k}}} = - e^{i \theta_\vec{k}}$. Upon reordering, we obtain two $2\times2$ blocks, one corresponding to the electron band with order parameter $\Delta_\vec{k}e^{i n_X \theta_\vec{k}}$, the other corresponding to the hole band with order parameter $-\Delta_\vec{k}e^{-i n_X \theta_\vec{k}}$. If $\mu<0$, the pairing opens a gap in the hole band and the effective winding of the order parameter on the Fermi surface is therefore $n_\Delta-n_X$. If $\mu>0$, the electron band is gapped with effective winding number $n_\Delta+n_X$. The two cases can be summarized as
\begin{equation}
	n'_{\Delta} = n_\Delta + \mathrm{sgn}(\mu) n_X.
\end{equation}
The Chern number $C$ of the topological superconducting phase corresponds to the order-parameter winding contained \emph{within} the Fermi surface. For the hole band, $C$ is equal to the order-parameter winding, while for the electron band $C$ is opposite to the winding. Collecting both cases, we get
\begin{equation}
	C = - n_X - \mathrm{sgn}(\mu) n_\Delta.
	\label{Eq:ChernFormula}
\end{equation}
The same argument applies to the original Fu--Kane model \cite{Fu.2008} where, starting from $n_\Delta=0$ ($s$-wave), one can recover an effective $p+ip$ ($p-ip$) superconductor with linear electron (hole) dispersion. In the case of a $d\pm id$ order parameter discussed here, the transformation can lead to either $p\pm ip$ or $f\pm if$.

The Chern numbers of the $d+id+\phantom{}$Dirac model are summarized in Table~\ref{tab:windings&OP}. We have verified these Chern numbers using the lattice model introduced in Sec.~\ref{sec:lattice-model}, using the gauge-independent numerical method of Ref.~\cite{Fukui.2005}.

\subsubsection{Existence of MZMs in vortex cores}

\begin{figure}[tb]
    \includegraphics[width=\columnwidth]{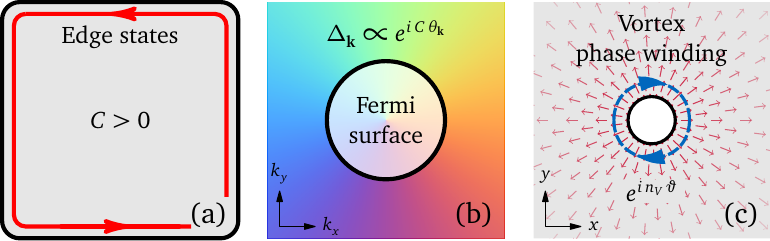}
    \caption{Origin of vortex Majorana zero modes (MZMs) in a chiral superconductor. (a) A superconductor with a nonzero Chern number has chiral edge states. (b) The Chern number derives from a winding of the complex order-parameter phase around the Fermi surface in momentum space. (c) In real space, a vortex adds an additional winding to the order parameter. Behaving as a void in the superconductor, the vortex core must have edge modes around the void (blue arrows).}
    \label{fig:IntuitiveReason}
\end{figure}

We just found that the combination of a $d\pm id$ superconductor and a normal-state Dirac cone yields a topological superconductor characterized by an odd Chern number $C$. It has been known for some time \cite{Read.2000, Fu.2008, Fukui.2010, Chamon.2010, Bernevig-Hughes.2013} that $C=1$ $p+ip$ superconductors host a Majorana zero mode in the core of a vortex. In fact, this is true for all topological superconductors with an odd Chern number \cite{Jackiw.1976, Jackiw.1981, Volovik.1999, Zlotnikov.2023, Chiu.2016}. Intuitively, the idea is that since all Chern superconductors have topologically protected edge states [see Fig.~\ref{fig:IntuitiveReason}(a)], we can treat the vortex core as an ``edge'' which, therefore, should host gapless modes. Let us now show the mathematical reasoning behind this intuitive picture. 

A nonzero Chern number arises from a winding of the order-parameter phase around the Fermi surface \cite{Pavarini:884084}. A representative topological superconductor can thus be described by a BdG Hamiltonian in momentum space, where the Chern number $C$ is captured in a winding of the order parameter, 
\begin{equation}
    \Delta_\vec{k} = |\Delta_\vec{k}| e^{ i \, C \, \theta_\vec{k}},
\end{equation}
similar to Eq.~\eqref{Eq:nDelta} [Fig.~\ref{fig:IntuitiveReason}(b)]. For example, a $p+ip$ superconductor has $ \Delta_\vec{k} \propto k_x + i k_y = |\vec{k}| e^{ i \theta_\vec{k}}$.

To add a \emph{vortex} structure, we need to express the corresponding BdG Hamiltonian in real space, which amounts to transforming $e^{i C \theta_\vec{k}}$ in momentum space to $e^{i C \vartheta}$ in real space. In addition, a vortex configuration has a phase winding of the order parameter, $\Delta (r,\vartheta) = \Delta(r) e^{i n_V \vartheta}$ with $n_V = \pm1$, and a vortex core where the order-parameter amplitude $\Delta(r)$ vanishes, see Fig.~\ref{fig:IntuitiveReason}(c). Note that $n_V$ sets how many magnetic flux quanta are trapped and is also called \emph{vorticity}. Combining these two sources of winding, the order parameter term in the real-space BdG Hamiltonian has an angular dependence of the form $e^{i (C + n_V) \vartheta}$. The vortex core, at the same time, acts as a ``puncture'' in the superconductor, around which we expect edge states. This is the same angular momentum along the edge of a nano flake \cite{Ptok.2020}.

The problem of finding vortex-core bound states is thus the same as finding edge modes of the chiral superconductor around such a puncture, with the constraint that the electron $u(r,\vartheta)$ and hole $v(r,\vartheta)$ components' angular phase winding differs by $e^{i (C + n_V) \vartheta}$. Chiral edge modes have a well-defined energy dispersion $E \sim k$ with $k$ the momentum along the edge; this translates for the circular puncture into $E \sim L_z$, where $L_z$ is the angular momentum around the puncture. These puncture edge states are thus described by the wavefunction Ansatz
\begin{equation}
    \begin{bmatrix}
        u(r,\vartheta) \\ v(r,\vartheta)
    \end{bmatrix}
    = g(r) \; e^{i L_z \vartheta} \; \begin{bmatrix}
        - e^{i (C+n_V) \vartheta /2} \\e^{-i (C+n_V) \vartheta /2}
    \end{bmatrix}
    \label{Eq:WaveFunctionsPuncture}
\end{equation}
with $g(r)$ containing the radial properties of the edge mode.

\begin{figure}[tb]
    \includegraphics[width=0.85\columnwidth]{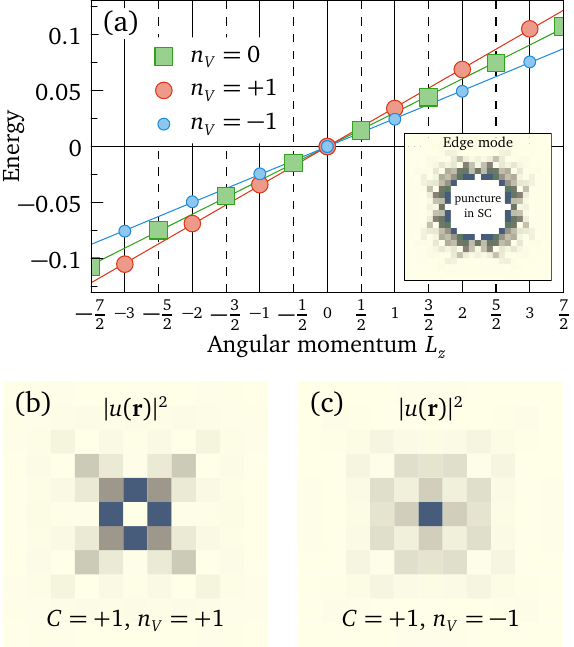}
    \caption{(a) Energy of the edge modes around a puncture in a topological $p+ip$ superconductor on a square lattice, vs angular momentum $L_z$. For odd Chern number $C$, $L_z$ is half-integer in the absence of vortex ($n_V = 0$) and integer in the presence of a vortex of vorticity $n_V = \pm 1$, $L_z=0$ corresponding to a MZM. (Inset) Example of edge mode. (b) Vortex MZM carrying an angular winding and without amplitude at the center of the vortex when $n_V$ has the same sign as $C$. (c) MZM carrying no winding and with amplitude at the center of the vortex when $n_V$ and $C$ have opposite signs.
    }
    \label{fig:IntuitiveAngular}
\end{figure}

In Fig.~\ref{fig:IntuitiveAngular}, we show how this works for a $p+ip$ superconductor on a square lattice with nearest-neighbor hopping $t$ and with model parameters $\mu = -0.5t$, $\Delta_p = 0.2t$. By cutting out a circular region as a ``puncture'', we gain edge states [inset of Fig.~\ref{fig:IntuitiveAngular}(a)] with a well-defined linear dependence on $L_z$ [Fig.~\ref{fig:IntuitiveAngular}(a)]. These results display a general property of the wavefunctions in Eq.~\eqref{Eq:WaveFunctionsPuncture}: When there is no vortex ($n_V = 0$), $L_z\pm C/2$ must be an integer for the wavefunctions to be single valued and it follows that the angular momentum must be \emph{half-integer}. Therefore, there is no state with zero-energy in the absence of a vortex. On the other hand, with an odd vorticity the same condition on single-valuedness implies that the angular momentum $L_z$ must be integer. This includes the $L_z=0$ state, which we now identify as the Majorana zero mode (MZM). 

When shrinking the vortex core to a point, the windings of the components $u$ and $v$ become relevant for the shape of the MZM. When $C = n_V = 1$, even though the total angular momentum is zero ($L_z=0$), the separate components $u(r,\vartheta)$ and $v(r,\vartheta)$ still have a nonzero angular winding $\propto e^{\pm i \vartheta}$. This implies that the radial component $g(r)$ must go to zero when $r \rightarrow 0$, to avoid multivaluedness of the wavefunction. By contrast, in the case where $C + n_V = 0$, there is no angular winding in $u$ and $v$ and the radial component $g(r)$ can stay nonzero at the vortex center. This difference can be seen in Figs.~\ref{fig:IntuitiveAngular}(b) and \ref{fig:IntuitiveAngular}(c), where we show the MZM wavefunctions for a $p+ip$ superconductor for a vortex and an antivortex, where the MZM does or does not exhibit spectral weight at the center of the vortex.

\subsubsection{Angular momentum of Majorana vortex states}

Having discussed the general case of MZMs appearing in the vortex cores of odd-$C$ superconductors, we now go back to the $d+id+\phantom{}$Dirac model and add the vortex structure. Compared to the simple $p+ip$ case, the resulting physics is much richer, including nontrivial relations for the MZM angular momentum.

To describe the vortex states, we use the real-space representation of the normal-state Hamiltonian Eq.~\eqref{Eq:NormalH_withNx}, $\hat{h}(\vec{r})=v\left(\sigma^y \hat{p}_x + n_X \sigma^x \hat{p}_y\right)$, with $\hat{p}_{x,y}=-i\partial_{x,y}$. By means of the operator $\partial_{n_X}=\partial_x+in_X\partial_y$, the Bogoliubov--de Gennes Hamiltonian can be recast as
\begin{equation}
	\hat{H}_\mathrm{BdG}(\vec{r}) = \begin{pmatrix}
		-\mu & -v\partial_{n_X} & \hat\Delta(\vec{r}) & 0 \\
		v\partial_{-n_X} & -\mu & 0 & \hat\Delta(\vec{r}) \\
		\hat\Delta^*(\vec{r}) & 0 & \mu & v\partial_{n_X} \\
		0 & \hat\Delta^*(\vec{r}) & -v\partial_{-n_X} & \mu
        \end{pmatrix}.
	\label{Eq:HamiltonianContinuum_r}
\end{equation}
With the components of the wavefunction written as $\Psi=\left(u_\up,u_\down, v_\down, -v_\up\right)^\intercal$, one readily verifies that for each solution $\Psi$ with energy $E$, there exists a solution with energy $-E$ and components $(v^*_\up, v^*_\down, u^*_\down, -u^*_\up)^\intercal$, which expresses the builtin particle-hole symmetry of $\hat{H}_\mathrm{BdG}$. We have seen in the previous section that such a pair of states with energy $E=0$ exists in our case. The sum of both states in this pair yields a state whose components obey the relations $v_\up=u^*_\up$ and $v_\down=u^*_\down$ and whose eigenvalue under particle-hole transformation is $+1$. This eigenstate has the property of a Majorana mode. For this zero-energy state, Eq.~\eqref{Eq:HamiltonianContinuum_r} reduces to the pair of equations
\begin{subequations}\begin{align}
	-\mu u_\up - v \partial_{n_X} u_\down + \hat{\Delta}  u^*_\down &= 0 \\
	-\mu u_\down + v \partial_{-n_X} u_\up - \hat{\Delta} u^*_\up &= 0.
\end{align}\label{Eq:BdGforMZM}\end{subequations}
We study these equations in a polar-coordinates system $(r,\vartheta)$, as appropriate in the presence of a vortex at the origin. We have $\partial_{n_X}=e^{i n_X \vartheta}(\partial_r + i n_X \partial_\vartheta / r)$. On the other hand, we show in Appendix~\ref{app:continuum limit} that the continuum limit of a vortex order parameter on the lattice with $d+i(n_\Delta/2)d$ symmetry in the chiral limit, neglecting the spatial variation of the amplitude, is $\hat{\Delta}=-|\Delta|e^{i(n_\Delta+n_V)\vartheta}\,\hat{\mathcal{D}}$, where $n_V=\pm1$ for a vortex or antivortex and $\hat{\mathcal{D}}$ is a differential operator that does not change the phase of the wavefunctions. With this, we see that the angular part of Eq.~\eqref{Eq:BdGforMZM} is solved by the Ansatz 
\begin{equation}
    u_{\up/\down}(r,\vartheta)=u_{\up/\down}(r)e^{in_{\up/\down}\vartheta},
    \label{Eq:Ansatz}
\end{equation}
provided that the angular momentum is the same in each of the terms, i.e., $n_\up=n_X+n_\down=n_\Delta+n_V-n_\down$ and $n_\down=-n_X+n_\up=n_\Delta+n_V-n_\up$. Combining these relations, we deduce
\begin{subequations}\label{Eq:nu&nd}\begin{align}
	n_\up + n_\down &= n_\Delta + n_V \\
    n_\up - n_\down &= n_X.
    \label{Eq:nu-nd}
\end{align}\end{subequations}
The possible values of $n_\up$ and $n_\down$ are collected in Table~\ref{tab:nup_ndown}.

The wavefunction components $u_{\up/\down}(r,\vartheta)$ are eigenstates of the angular momentum $\hat{L}_z = \hat{x} \hat{p}_y - \hat{y} \hat{p}_x = -i \partial_\vartheta$ with eigenvalues $n_{\up/\down}$. It follows that the Majorana mode as a whole is not an eigenstate of $\hat{L}_z$ unless $n_\up=n_\down$, which is forbidden by Eq.~\eqref{Eq:nu-nd}. This is expected, because $\hat{L}_z$ is not a symmetry of the Hamiltonian. However, it is possible to construct a true rotational symmetry of the BdG Hamiltonian, by combining the regular angular momentum operator $\hat{L}_z$ with specific phase factors for the different $u/v$ components. The lattice version of this rotational symmetry is discussed later, in Sec.~\ref{Sec:Effective angular momentum of the MZMs}.

The radial structure of the MZM wavefunction can be studied in a simplified picture, if we modify the order-parameter operator to $\hat{\Delta} = - |\Delta| e^{i (n_\Delta + n_V) \vartheta}$, which amounts to replacing $\hat{\mathcal{D}}$ by the identity. Introducing the localization length $\xi=v/|\Delta|$ and the Fermi wavelength $1/k_{\mathrm{F}}=v/\mu$, the radial part of the BdG Eq.~\eqref{Eq:BdGforMZM} simplifies to
\begin{subequations}\begin{align}
	\left(1 + \tfrac{\xi}{r} n_X n_\up + \xi\, \partial_r\right)u_\up(r) &= k_{\mathrm{F}}\xi\, u_\down(r) \\
	\left(1 - \tfrac{\xi}{r} n_X n_\down + \xi\, \partial_r\right)u_\down(r) &= -k_{\mathrm{F}}\xi\, u_\up(r).
\end{align}\end{subequations}
The solution that is regular at the origin $r=0$ is
\begin{subequations}\begin{align}
	u_\up(r) &= \mathcal{N} e^{-r/\xi} J_{n_\up}\left(k_{\mathrm{F}}r\right) \\
	u_\down(r) &= -\mathcal{N} e^{-r/\xi} J_{n_\down}\left(k_{\mathrm{F}}r\right),
\end{align}\label{Eq:pair}\end{subequations}
where $\mathcal{N}$ is a normalization, $J_n(x)=J_{-n}(x)$ is the Bessel function of the first kind, and use has been made of Eq.~\eqref{Eq:nu&nd}. Since $J_n(x)\sim x^n$ for small $x$ and the local density of states at $E=0$ is proportional to $|u_\up(r)|^2+|u_\down(r)|^2$, only $J_0$ is able to provide intensity at the center of the vortex. Therefore, in this simplified description, the MZM has its maximum probability at the center of the vortex only if $n_\up n_\down=0$, or equivalently $|n_\Delta+n_V|=1$. This is consistent with the requirement of single-valuedness imposing that, whenever $n_{\up/\down} \neq 0$, the wavefunction amplitude must vanish at $r=0$. The numerical simulations on the lattice support this qualitative picture.

\begin{table}[tb]
	\setlength{\tabcolsep}{8pt}\def\arraystretch{1.3}
	\begin{tabular*}{0.7\columnwidth}{@{\extracolsep{\fill}}rrr|rr|r}
    \hline\hline
    \multicolumn{1}{c}{$n_X$} &
    \multicolumn{1}{c}{$n_\Delta$} &
    \multicolumn{1}{c|}{$n_V$} &
    \multicolumn{1}{c}{$n_\up$} &
    \multicolumn{1}{c|}{$n_\down$} &
    \multicolumn{1}{c}{$\ell$} \\
    \hline
    $+1$ & $+2$ & $+1$ & $+2$ & $+1$ & $+2$ \\
    $  $ & $  $ & $-1$ & $+1$ & $ 0$ & $ 0$ \\
    $  $ & $-2$ & $+1$ & $ 0$ & $-1$ & $ 0$ \\
    $  $ & $  $ & $-1$ & $-1$ & $-2$ & $-2$ \\
    \hline
    $-1$ & $+2$ & $+1$ & $+1$ & $+2$ & $-2$ \\
    $  $ & $  $ & $-1$ & $ 0$ & $+1$ & $ 0$ \\
    $  $ & $-2$ & $+1$ & $-1$ & $ 0$ & $ 0$ \\
    $  $ & $  $ & $-1$ & $-2$ & $-1$ & $+2$ \\
    \hline\hline
    \end{tabular*}
	\caption{Angular quantum numbers $n_\up$ and $n_\down$ of the Majorana zero mode in a vortex ($n_V=+1$) or antivortex ($n_V=-1$) for a Dirac cone with winding number $n_X$ and $d\pm id$ pairing with winding $n_\Delta$, as given by Eq.~\eqref{Eq:nu&nd}. The last column shows the index defined in Eq.~\eqref{Eq:ell}, which is nonzero when the MZM has no intensity at the vortex/antivortex core.}
        \label{tab:nup_ndown}
\end{table}

We conclude this section with formulating a quantum number that, in the $d+id+\phantom{}$Dirac model, tells us whether the MZM has a nonzero amplitude in the vortex core. This number is
\begin{equation}
	\ell=n_X\left(\frac{n_\Delta}{2}+n_V\right),
	\label{Eq:ell}
\end{equation}
which vanishes whenever $n_\up n_\down=0$ and otherwise takes the values $\pm2$ (see Table~\ref{tab:nup_ndown}). A nonzero value of this index implies that the MZM has no intensity at the vortex/antivortex core. One sees that although both vortices and antivortices host MZMs, their physical properties are systematically different: Whenever the configuration of the windings $n_X$ and $n_\Delta$ is such that the vortex hosts a MZM with maximum at the vortex center, the antivortex doesn't, and conversely. In the next subsection, we show that the same qualitative conclusions can be drawn from the analysis and numerical simulation of a lattice model.

\subsection{Lattice model}
\label{sec:lattice-model}

Although numerical simulations of vortex-core states in the continuum are possible (see, e.g., Refs.~\cite{Gygi.1990, Hayashi.1998, Franz.1998}), lattice models provide a more practical and controllable approach. Lattice simulations are physically relevant as real superconductors are almost always crystalline solids.

When addressing the very existence and the topological protection of MZMs, finite-size effects must be dealt with. Here, we will use exact diagonalization (ED) when working with sufficiently small lattices---typically $\lesssim 200 \times 200$ sites---and the kernel polynomial method (KPM) \cite{Weisse.2006, Covaci.2010} when treating large lattices, up to $\sim 1000 \times 1000$ sites. In this subsection, we first describe the model and our numerical approaches, and then use ED to confirm the predictions based on the continuum model, which requires us to generalize the notion of angular momentum to our square-lattice system with discrete 90\textdegree{}-rotation symmetry. In the subsequent section, we will use both ED and the KPM to study the topological protection of the MZMs.

\subsubsection{Definition}
\label{sec:latticemodel}

A reliable lattice model that faithfully computes the low-energy spectrum of an anomalous Dirac cone, without fermion doubling \cite{Beenakker.2023}, can be constructed as follows on a square lattice \cite{Mercado.2022}:
\begin{align}
	\label{Eq:LatticeHBdGMomentumSpace}
	H_\mathrm{BdG} (\vec{k}) &=
		\begin{pmatrix}
			h^\mathrm{latt}_\vec{k} & \Delta_\vec{k} \sigma^0 \\
			 \Delta^*_\vec{k} \sigma^0 & -\sigma^y  (h^\mathrm{latt}_{-\vec{k}})^\intercal \sigma^y
		\end{pmatrix} \\
    \nonumber
	h^\mathrm{latt}_\vec{k} &= v (\sigma^y \sin k_x + n_X \sigma^x \sin k_y)
		+ m_\vec{k} \sigma^z- \mu \sigma^0\\
    \nonumber
	\Delta_\vec{k} &= -\frac{1}{2} \Delta_d \left( \cos k_x - \cos k_y \right)
		+ i \frac{n_\Delta}{2} \Delta_{d}' \sin k_x \sin k_y \\
    \nonumber
	m_\vec{k} &= m (2 - \cos k_x - \cos k_y ),
\end{align}
where $n_\Delta=\pm 2$ for a $d\pm id$ order parameter and we have set the lattice parameter to unity. The addition of the time-reversal symmetry breaking mass term $m_\vec{k}$ is negligible near the $\Gamma$ point and is thus not expected to affect the main physics of in-gap states \cite{Beenakker.2023}. A topological transition does occur as $m$ decreases (see Sec.~\ref{sec:large-gap}). With the minus sign of the $d_{x^2-y^2}$ term, the above lattice model returns to the continuum model defined by Eqs.~\eqref{Eq:HamiltonianContinuum_k}, \eqref{Eq:NormalH_withNx}, and \eqref{Eq:OrderParameterContinuum} in the continuum limit $\vec{k}\rightarrow \vec{0}$. Note that there is a factor 2 difference with respect to Eq.~(13) of Ref.~\cite{Mercado.2022}.

In real space, the hopping amplitudes take direction-dependent complex values. For instance, the term $t_{\vec{r}\vec{r}'}^{\up\down} c^{\dagger}_{\vec{r}\up} c\pdagger_{\vec{r}'\down}$ has $t_{\vec{r}\,\vec{r} + \hat{\vec{x}}}^{\up\down} = v/2$ and $t_{\vec{r}\,\vec{r} + \hat{\vec{y}}}^{\up\down} = i n_X v / 2$, with $\hat{\vec{x}}$ and $\hat{\vec{y}}$ unit vectors. Like the hopping terms, the order parameter now lives on the \emph{links} of the square lattice. In the homogeneous superconducting phase, the term $\Delta\pdagger_{\vec{r}\vec{r}'} c^{\dagger}_{\vec{r}\up} c^{\dagger}_{\vec{r}'\down}$ has $\Delta_{\vec{r}\,\vec{r} + \vec{x}} = -\Delta_d / 4$, $\Delta_{\vec{r}\,\vec{r} + \vec{x} + \vec{y}} = -i n_{\Delta} \Delta'_d / 8$, $\Delta_{\vec{r}\,\vec{r} + \vec{y}} = \Delta_d / 4$, and $\Delta_{\vec{r}\,\vec{r} + \vec{y} - \vec{x}} = i n_{\Delta} \Delta'_d / 8$. To define a vortex with winding number $n_V$ centered in $(0, 0)$, we multiply each $\Delta_{\vec{r}\vec{r}'}$ by the phase factor $e^{i n_V \vartheta_{\vec{r}\vec{r}'}}$, where $\tan\vartheta_{\vec{r}\vec{r}'} = (y+y') / (x+x')$. We ignore the radial variation of the order parameter, that would appear in a self-consistent solution, as this dependence does not influence the properties of the vortex states qualitatively, but only quantitatively.

\subsubsection{Numerical methods}

The Hamiltonian is defined on a square lattice of $L\times L$ sites with open boundary conditions. We take $L$ odd, such that the vortex sits at a symmetric position in the center of the system. The size of the BdG Hamiltonian is $4 L^2 \times 4L^2$, which enables ED up to $L\sim 201$ with desktop-level resources. Note that for reliable results, the system size must be much bigger than the localization length of the MZM, preventing hybridization between the zero mode localized in the vortex and its partner living at the system's boundaries. We will elaborate on the localization length further in Sec.~\ref{sec:localization}. The eigenvalues $E_n$ and eigenvectors $\Psi_n$ give access to the electron local density of states (LDOS)
\begin{multline}
	\rho(E,\vec{r})=\sum_{E_n\geqslant 0}\left\{
	\left[|u_{n\up}(\vec{r})|^2 + |u_{n\down}(\vec{r})|^2\right] \delta(E - E_n) \right.\\
	\left.+\left[|v_{n\up}(\vec{r})|^2 + |v_{n\down}(\vec{r})|^2\right] \delta(E+E_n)\right\}.
\end{multline}
The main advantage of ED is that it is free of systematic numerical errors, yielding a clear and global real-space picture of the simulated system. In favorable situations, it furthermore enables extrapolation to the thermodynamic limit through finite-size scaling.

In parameter regimes where the localization length reaches hundreds of lattice spacings---which happens in the proposal of Ref.~\cite{Mercado.2022}, see Sec.~\ref{sec:protection}---we switch from ED to the KPM, using Chebyshev polynomials $T_n(x)$ \cite{Covaci.2010, Berthod.2016}. For a lattice with two degrees of freedom per site, the LDOS is evaluated as
\begin{align}
	\nonumber
	\rho(E,\vec{r}) &\approx \frac{2}{\pi a}\mathrm{Re}\,\left\{
		\frac{1}{\sqrt{1-\widetilde{E}^2}}\left[1+\sum_{n=1}^N e^{-in\arccos(\widetilde{E})}c_n\right]\right\}\\
	c_n&=\sum_{s=\up\down}\langle \vec{r}s|T_n(\widetilde{H}_\mathrm{BdG})|\vec{r}s\rangle.
\end{align}
This representation is exact if $N=\infty$. The energy $a$ is an upper bound for the half-width of the spectrum of $H_\mathrm{BdG}$, such that the spectrum of the dimensionless matrix $\widetilde{H}_\mathrm{BdG}=H_\mathrm{BdG}/a$  falls entirely within $[-1,1]$, where the polynomials $T_n(x)$ are defined. Similarly, $\widetilde{E}=E/a$. The coefficients $c_n$ are evaluated recursively using the property $T_n(x)=2xT_{n-1}(x)-T_{n-2}(x)$, starting from a state $|\vec{r}s\rangle$ representing an electron with spin $s$ localized at site $\vec{r}$, and applying $\widetilde{H}_\mathrm{BdG}$ repeatedly. The practical use of a finite expansion order $N$ introduces a systematic error and an energy resolution $\Delta E_N\sim a/N$ \cite{Berthod.2018}, as well as Gibbs oscillations of the LDOS, that we correct using the Jackson kernel \cite{Weisse.2006}.

As it only computes the LDOS for a given site at a time, the KPM gets impractical if the full real-space information is required. It furthermore discards the phase information of the wavefunctions. We have checked that our implementations of ED and the KPM give the same results in the regimes where they can be compared.

\subsubsection{Effective angular momentum of the MZMs}
\label{Sec:Effective angular momentum of the MZMs}

\begin{figure*}[tb]
	\includegraphics[width=0.85\textwidth]{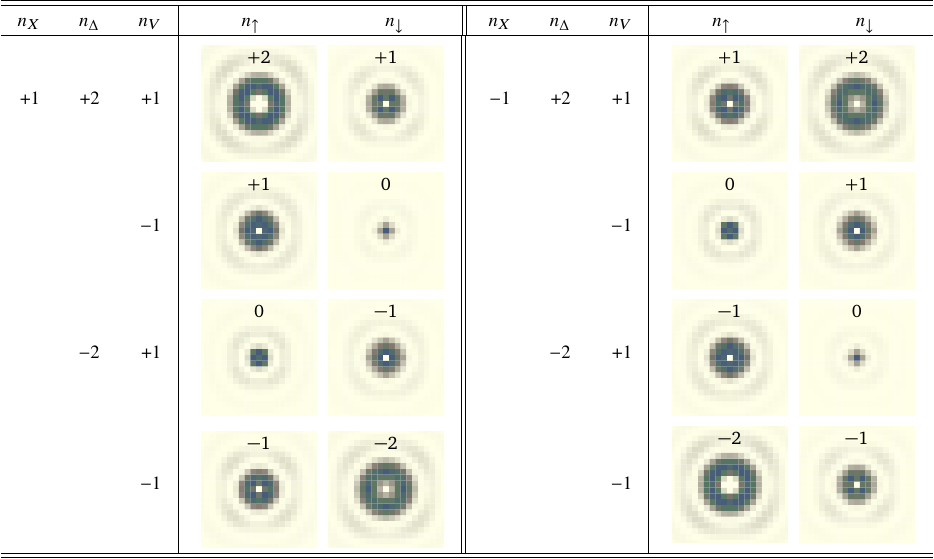}
	\caption{
Spin-resolved LDOS of the MZM near the vortex center for all values of $n_X$, $n_\Delta$, and $n_V$, calculated with $L=201$ at negative $\mu$. Similar results are obtained at positive $\mu$. The columns labeled $n_\up$ ($n_\down$) show $|u_\up(\vec{r})|^2$ [$|u_\down(\vec{r})|^2$] and the numbers on the images are the numerically determined values of $n_\up$ and $n_\down$. The color scale covers the whole data range in all images.
	}
	\label{fig:flavors}
\end{figure*}

On a square lattice, the rotation group is generated by 90\textdegree{} counterclockwise rotations $R_{\pi/2}$ acting on the unit vectors as 
\begin{equation}
	R_{\pi/2}\,\hat{\vec{x}} = \hat{\vec{y}}, \qquad R_{\pi/2}\,\hat{\vec{y}} = -\hat{\vec{x}}.     
\end{equation}
The result is a reshuffling of the lattice sites. This rotation is not a symmetry of the lattice Hamiltonian Eq.~\eqref{Eq:LatticeHBdGMomentumSpace}, since it changes the Dirac cone and the superconducting term. Under the rotation $R_{\pi/2}$, the Dirac term $\sigma^y \sin k_x + n_X \sigma^x \sin k_y$ turns into $\sigma^y \sin k_y - n_X \sigma^x \sin k_x$. To make this term invariant, we must perform a $\pi/2$ spin rotation around the spin $z$-axis, with direction depending on $n_X$. This can be realized by multiplying all spin up components with $e^{- i (\pi/4) n_X}$ and all down-components with $e^{+i (\pi/4) n_X}$. Similarly, one has to account for the sign change in the $d$-wave order parameter by multiplying the hole components with $-1$. In the case of a vortex or antivortex with winding number $n_V$, performing a 90\textdegree{} rotation adds a phase $n_V \pi/2$ to the $u$ components of the eigenstates and $- n_V \pi/2$ to the $v$ components. These must be countered by the rotation operator. As a result, the operator
\begin{equation}
	R_\mathrm{tot}=\widetilde{R}\, R_{\pi/2}
	\label{Eq:RotationSymmetry}
\end{equation}
\emph{is} a symmetry of the lattice BdG Hamiltonian, with
\begin{equation}
	\widetilde{R} = e^{-i\textstyle\frac{\pi}{4}\left[n_X\tau^0\sigma^z+(n_\Delta+n_V)\tau^z\sigma^0\right]}.
	\label{Eq:Rtilde}
\end{equation}
We verified numerically that this operator indeed commutes with the Hamiltonian. It satisfies all the properties of a 90\textdegree{} rotation, including $R_\mathrm{tot}^4 = \openone$. Consequently, the possible eigenvalues of the rotation are $1$, $i$, $-1$, and $-i$, that we can write as $e^{i(\pi/2)n_R}$ with $n_R=0$, $1$, $2$, or $3$. The winding number $n_R$ characterizes the action of $R_\mathrm{tot}$ on the non-degenerate eigenstates $\Psi$ of $H_\mathrm{BdG}$ according to $R_\mathrm{tot}\Psi=e^{i(\pi/2)n_R}\Psi$. The action of $R_{\pi/2}$ on $\Psi$ follows from Eqs.~\eqref{Eq:RotationSymmetry} and \eqref{Eq:Rtilde}:
\begin{equation}
	R_{\pi/2}\Psi = e^{i\textstyle\frac{\pi}{2}n_R}
	e^{i\textstyle\frac{\pi}{4}\left[n_X\tau^0\sigma^z+(n_\Delta+n_V)\tau^z\sigma^0\right]}\Psi.
	\label{Eq:Rotation}
\end{equation}
By analogy with the continuum description, we characterize the behavior of $\Psi=(u_\up, u_\down, v_\down, -v_\up)^\intercal$ under a 90\textdegree{} rotation by four numbers $\{n^u_\up, n^u_\down, n^v_\down, n^v_\up\}$, defined such that $u_\up(R_{\pi/2}\,\vec{r})=e^{i(\pi/2)n^u_\up}u_\up(\vec{r})$ and similarly for the other three components. Comparing with Eq.~\eqref{Eq:Rotation}, these numbers can be deduced as
\begin{subequations}\begin{align}
	2n^u_\up   &= 2n_R + n_X + (n_\Delta+n_V) \\
	2n^u_\down &= 2n_R - n_X + (n_\Delta+n_V) \\
	2n^v_\down &= 2n_R + n_X - (n_\Delta+n_V) \\
	2n^v_\up   &= 2n_R - n_X - (n_\Delta+n_V).
\end{align}\end{subequations}
Because of the particle-hole symmetry, a Majorana state has only two independent winding numbers $n_\up \equiv n^u_\up = -n^v_\up$ and $n_\down \equiv n^u_\down = -n^v_\down$. The four relations above reduce to the same two relations that were found in the continuum model, Eq.~\eqref{Eq:nu&nd}. In addition, they give the constraint that $n_R=0$ \footnote{On a square lattice, the winding number conditions are all modulo 4, so the precise condition is $2n_R~\mathrm{mod}~4 = 0$.}. In other words, the MZMs are invariant under the action of the rotational symmetry $R_\mathrm{tot}$ \cite{Fang.2014, Liu.2014, Konig.2019, Qin.2019, Kobayashi.2020, Hu.2022, Kobayashi.2023}. However, the winding numbers $n_{\uparrow/\downarrow}$ of the individual components $u_{\uparrow/\downarrow}$ can still be nontrivial, depending on the microscopic properties of the model.

We have performed ED simulations to confirm these expectations. We use the parameters defined for unit velocity $v$ as $m=0.8$, $\mu=-0.8$, $\Delta_d=0.9$, $\Delta'_d=0.6$. The gap values are in the same ratio but larger than those of Ref.~\cite{Mercado.2022}, which allows us to obtain sufficiently well-converged results on small lattices $L\gtrsim 41$. See the discussion in Sec.~\ref{sec:localization}. We compute the eigenstate with the smallest positive eigenenergy and confirm that it has unit eigenvalue ($n_R=0$) under $R_\mathrm{tot}$. We will show in the next section how this eigenenergy approaches zero with increasing system size and how this dependence varies with $\Delta_d$ and $\Delta'_d$. From the components $u_\up$ and $u_\down$, we determine $n_\up$ and $n_\down$ by applying $R_{\pi/2}$ and checking that the relation $u_{\up/\down}(R_{\pi/2}\,\vec{r})=e^{i(\pi/2)n_{\up/\down}}u_{\up/\down}(\vec{r})$ is obeyed consistently with a single value of $n_{\up/\down}$ for all lattice sites, within machine accuracy.

Our results are collected in Fig.~\ref{fig:flavors}. For each value or $n_X$, $n_\Delta$, and $n_V$, we show maps of $|u_\up(\vec{r})|^2$ and $|u_\down(\vec{r})|^2$ in the central part of the lattice, together with the calculated values of $n_\up$ and $n_\down$. We recast these values between $-2$ and $+2$, as they are defined modulo 4, which allows one to readily check that they satisfy Eq.~\eqref{Eq:nu&nd}. Hence the numerical results on the lattice are fully consistent with the conclusions drawn from the continuum model. At the qualitative level, changing the sign of $n_X$ is equivalent to exchanging the $\up$ and $\down$ components, consistently with Table~\ref{tab:nup_ndown}. At the quantitative level, however, the exact symmetry also requires flipping the sign of $v$, i.e., $\{v, n_X, \up, \down\}$ is equivalent to $\{-v, -n_X, \down, \up\}$, as can be seen e.g. from Eq.~\eqref{Eq:HamiltonianContinuum_r}. Another exact symmetry of the MZM LDOS is that $\{n_X, n_\Delta, n_V\}$ is equivalent to $\{-n_X, -n_\Delta, -n_V\}$, as seen in Fig.~\ref{fig:flavors}.

\begin{figure}[tb]
	\includegraphics[width=\columnwidth]{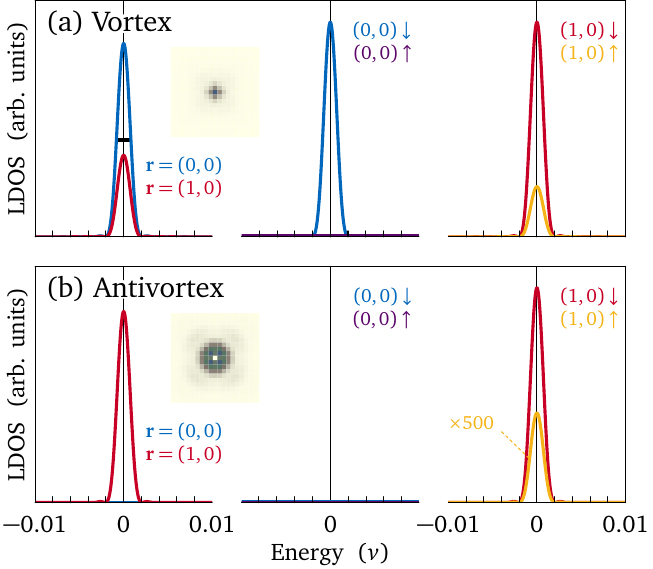}
	\caption{Total (left panels) and spin-resolved (center and right panels) LDOS at $\vec{r}=(0,0)$ and $\vec{r}=(1,0)$ for Dirac-cone winding $n_X=-1$, pairing winding $n_\Delta=-2$, and vorticity (a) $n_V=+1$ and (b) $n_V=-1$. At $\vec{r}=(0,0)$, there is no spin-$\up$ intensity in the vortex and neither spin-$\up$, nor spin-$\down$ intensity in the antivortex. The insets display the total LDOS in the vortex- and antivortex-core region. Calculations are made with $L=1001$ using the KPM at order $N=20000$. The horizontal bar in (a) indicates the energy resolution of the calculation.
	}
	\label{fig:empty}
\end{figure}

The spin imbalance visible in Fig.~\ref{fig:flavors} was already observable in the scenario with $s$-wave pairing \cite{Fu.2008}. The additional unique feature in the $d\pm id$ scenario is the possibility of having an ``empty'' vortex core in the total LDOS, in those cases where both $n_\up$ and $n_\down$ are nonzero. In this situation, searching a zero-energy peak at the vortex center is not appropriate to detect a MZM. We illustrate this in Fig.~\ref{fig:empty} for the case $(n_X, n_\Delta)=(-1, -2)$. In the vortex ($n_V=1$), the energy-dependent LDOS at $\vec{r}=(0,0)$ peaks at $E=0$, which is the behavior expected for a ``conventional'' vortex MZM. The LDOS is entirely spin-$\down$ polarized (blue curve), because $n_\up \neq 0$. At $\vec{r}=(1,0)$, the LDOS shows weak intensity at $E=0$, illustrating the fast decay of the MZM upon moving away from the vortex center, with a small admixture of spin-$\up$ polarization. In the antivortex ($n_V=-1$), the LDOS vanishes identically at $\vec{r}=(0,0)$ and $E\approx0$. A zero-energy peak is however visible at $\vec{r}=(1,0)$, manifesting an ``anomalous'' vortex MZM, which in this particular case is also dominantly spin-$\down$ polarized (red curve), since the spin-$\up$ wave function has larger angular momentum ($n_\up=-2$).

\subsubsection{Topological transition in the lattice model}
\label{sec:large-gap}

The lattice model undergoes a topological phase transition when $\Delta_d$ crosses the value $\Delta^c_d=\sqrt{(2m)^2-\mu^2}$, because of a gap closing and band inversion at the $(\pi, 0)$ and $(0,\pi)$ points of the Brillouin zone. It is important to realize that this topological transition is the result of the introduction of the lattice; it is manifestly absent in the continuum model. At this transition, the Chern numbers of the superconducting state switch from $\pm 1$ to $\pm 3$, and conversely. We also find that the flavor of the vortex MZM changes at the transition, zero-angular momentum MZMs becoming nonzero-angular momentum ones, and conversely. These behaviors may be rationalized if the band inversion is equivalent to changing $n_X$ to $-n_X$, $\mu$ to $-\mu$, and $n_\Delta$ to $-n_\Delta$. With these changes, Eqs.~\eqref{Eq:ChernFormula} and \eqref{Eq:nu&nd} predict the Chern numbers and the flavors in the new topological phase. Given the qualitative changes occurring at $\Delta^c_d$, for our numerical results we choose parameters that place the lattice model in the small-gap phase, $\Delta_d<\Delta^c_d$. This allows a direct comparison with the continuum model.

\section{Topological protection of $d+id$ vortex-core MZMs}
\label{sec:protection}

The quality of the topological protection is the main figure of merit in view of a practical application of MZMs for fault-tolerant quantum computing \cite{Yazdani.2023,Aasen.2025}. The key protective factor is the energy separation between the MZM and other electronic states. One such scale is the energy associated with the superconducting gap $\Delta$. On top of that, vortices host conventional CdGM bound states whose lowest energy varies as $\Delta^2/E_{\mathrm{F}}$ \cite{Caroli.1964}.

In addition to \emph{energy} scales, the protection of the MZMs is set by a \emph{length} scale, namely their localization length. MZMs in different vortices hybridize if their localization length exceeds the intervortex distance. As the localization length is inversely proportional to $\Delta$, a large gap is again an asset. On the other hand, a large localization length would possibly allow for the detection of the different MZMs flavors discussed above.

In this section, we discuss the topological protection owing to the actual topological gap $\Delta_{\mathrm{F}}$ and the localization length in the $d\pm id$ scenario. To enable a direct comparison with Ref.~\cite{Mercado.2022}, we use here the lattice model with the same gap parameters as there, which corresponds to $\Delta_d=0.3v$ and $\Delta'_d=0.2v$ with our convention. These parameters are suggested to be relevant for twisted cuprate bilayers deposited on top of Bi$_2$Se$_3$.

\subsection{Topological gap and localization of MZMs}
\label{sec:localization}

\begin{figure}[tb]
    \includegraphics[width=\columnwidth]{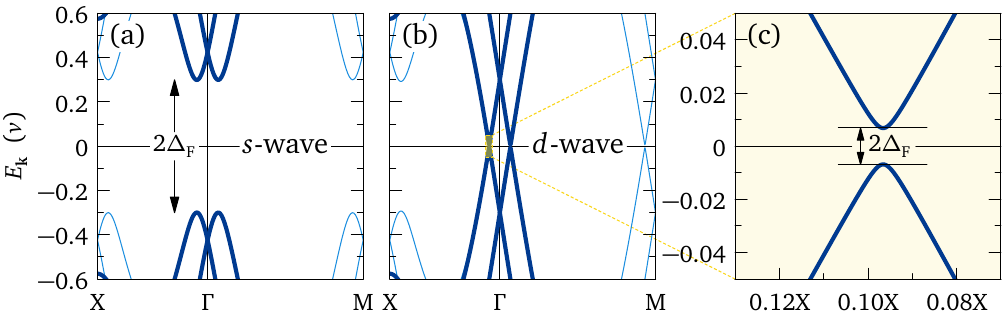}
    \caption{Low-energy spectrum of the lattice Hamiltonian, Eq.~\eqref{Eq:LatticeHBdGMomentumSpace}, for (a) $s$-wave pairing ($\Delta_\vec{k}\equiv0.3v$) and (b) $d\pm id$ pairing ($\Delta_d=0.3v$, $\Delta'_d=0.2v$). The tiny spectral gap in (b) is given approximately by Eq.~\eqref{Eq:SmallGap}, as better seen by zooming in (c). The other model parameters are $m=0.5$ and $\mu=-0.3$. The thin lines show the spectrum for $m=0$.
    }
    \label{fig:dispersion}
\end{figure}

Upon moving the chemical potential away from the Dirac point in the absence of a superconducting order parameter, a small Fermi surface forms with a Fermi wavevector given by $k_\mathrm{F} = |\mu|/v$ at small $|\mu|$. The spectral gap $2\Delta_\mathrm{F}$ of the topological superconducting state is, to lowest order in the order parameter $\Delta_d$, set by the value of $\Delta_\vec{k}$ on the Fermi surface. Because the order parameter vanishes at the zone center for a $d$-wave superconductor, the spectral gap remains small. In the chiral limit $\Delta'_d = \Delta_d/2$, it is given by
\begin{equation}
    \Delta_\mathrm{F} = \frac{\Delta_d}{4} \left( \frac{\mu}{v} \right)^2.
    \label{Eq:SmallGap}
\end{equation}
This is typically two orders of magnitude smaller than the main $d$-wave gap amplitude $\Delta_d$. Figure~\ref{fig:dispersion} illustrates the qualitative difference between $s$-wave pairing, where $\Delta_\mathrm{F}$ is equal to the magnitude of the order parameter, and $d\pm id$ pairing, where it is much smaller. For the parameters used in Ref.~\cite{Mercado.2022} and in this section, $\Delta_\mathrm{F}$ is only $\sim2$\% of $\Delta_d$.

The localization length of the vortex MZM is set by the gap to the bulk excitations, hence by $\Delta_\mathrm{F}$,
\begin{equation}
    \xi_d \approx \frac{v}{\Delta_\mathrm{F}} = \frac{4 v^3}{\mu^2\Delta_d}.
	\label{eq:loclength}
\end{equation}
Thus, the very small $\Delta_\mathrm{F}$ in the $d\pm id$ model leads to very large values of $\xi_d$, e.g., $\xi_d \sim 150$ lattice spacings with the parameters considered in this section. This is orders of magnitude longer than the Majorana localization length of the Fu--Kane model \cite{Fu.2008}, $\xi_s=v/\Delta$, e.g., $\xi_s\sim 3$ for $\Delta=0.3v$. The very large value of $\xi_d$ puts severe demands on numerical calculations, since reliable results can only be obtained with system sizes $L\gg \xi_d$.

\begin{figure}[tb]
	\includegraphics[width=0.8\columnwidth]{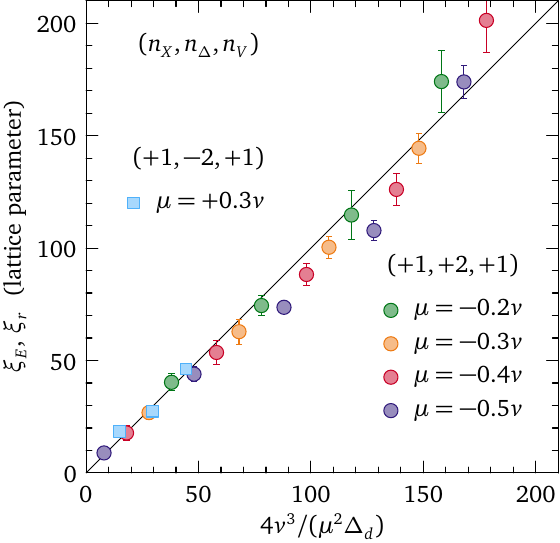}
	\caption{Numerically determined MZM localization length for various values of $\mu$ and $\Delta_d$ in the chiral limit $\Delta'_d=\Delta_d/2$. The squares show $\xi_E$ extracted from finite-size scaling of the MZM energy. The circles show $\xi_r$ extracted from the spatial decay of the zero-energy LDOS.
	}
	\label{fig:localization}
\end{figure}

We have tested Eq.~\eqref{eq:loclength} numerically using two different approaches. When using ED on finite lattices, the MZM comes out with a small finite energy, which is expected to approach zero as $\delta E\sim e^{-L/\xi_E}$ with increasing system size. The actual dependence on $L$ displays more structure (see Appendix~\ref{app:xi} for details), but nevertheless follows an overall exponential decay. The values of $\xi_E$ obtained in this way are plotted in Fig.~\ref{fig:localization} as squares. On a lattice of size $L\gg\xi_d$, the zero-energy LDOS is expected to decay as $\rho(0,\vec{r})\sim e^{-2r/\xi_r}$. Using the KPM with $L=1001$, we calculate $\rho(0,\vec{r})$ along the diagonal direction $\hat{\vec{x}}+\hat{\vec{y}}$ and deduce $\xi_r$ as shown in Fig.~\ref{fig:localization} with circles. Again, while consistent with an exponential decay, the LDOS has more structure; see Appendix~\ref{app:xi} for details. For both $\xi_E$ and $\xi_r$, we vary the data range used to fit an exponential, which leads to a distribution of values whose standard deviation is used as the error bars in Fig.~\ref{fig:localization}. Larger error bars signal stronger deviations from a pure exponential behavior. The figure confirms that Eq.~\eqref{eq:loclength} correctly captures the localization of the vortex MZMs.

\subsection{Quasiparticle poisoning}
\label{sec:CdGM}

Abrikosov vortices host subgap electronic states in their core, known as CdGM states \cite{Caroli.1964}. In a superconductor with a hard gap, like in the case of $s$ and $d\pm id$ pairing, these states are truly bound (exponentially localized) with discrete energies, while for nodal pairings like $d_{x^2-y^2}$ they are quasi-bound and form a continuum \cite{Franz.1998, Berthod.2016}. In first approximation, the discrete vortex-core states have energies $E_n=(n+1/2)\Delta^2/E_\mathrm{F}$, where $n\in\mathbb{N}$ and $\Delta$ is the spectral gap. Hence, one expects quasiparticles at energies $\sim \Delta_\mathrm{F}^2/\mu$ in the $d\pm id$ scenario. Since this energy is a small fraction of $\Delta_\mathrm{F}$, the hybridization with conventional vortex-core states becomes the most stringent factor limiting the topological protection \cite{Zlotnikov.2023}.

\begin{figure}[tb]
	\includegraphics[width=0.8\columnwidth]{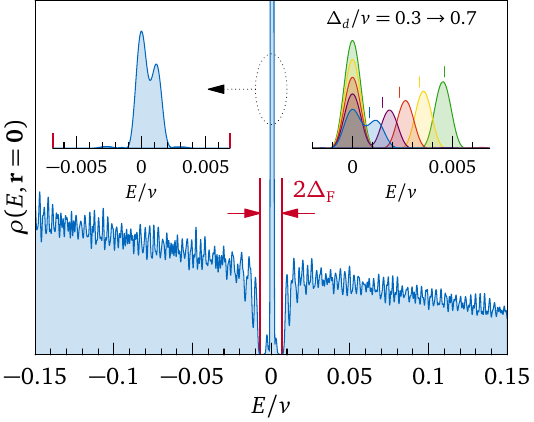}
	\caption{LDOS at the center of a vortex hosting a MZM for the parameters $\Delta_d=0.3v$, $\Delta'_d=0.2v$, and $\mu=-0.3$. The spectral gap $\Delta_\mathrm{F}\ll\Delta_d$ is marked by red bars. Left inset. Zoom-in of the gap, showing the zero-energy peak resulting from the sum of two elementary peaks. Right inset. Evolution of the in-gap spectrum with increasing $\Delta_d$ with $\Delta'_d/\Delta_d=2/3$. The vertical bars indicate the energy $\Delta_\mathrm{F}\Delta_d/(8|\mu|)$.
	}
	\label{fig:CdGM}
\end{figure}

We illustrate this in Fig.~\ref{fig:CdGM}. The main panel shows the LDOS $\rho(E,\vec{0})$ at the center of a vortex (winding numbers $n_X=+1$, $n_\Delta=-2$, and $n_V=+1$), calculated on a large lattice of size $L=1001$ using the KPM with order $N=20000$. The vortex hosts a ``conventional'' MZM with finite amplitude at the core, yielding a tall zero-energy peak in the LDOS. The width of this peak reflects the energy resolution of the calculation, which is set by the expansion order as $\Delta E_N=2\pi\sqrt{\ln 4}a/N\approx10^{-3}v$ \cite{Berthod.2018}. The resolution is sufficient to distinguish the gap edges and the in-gap states from the continuum, where the apparent noise is in fact a manifestation of the finite-lattice discrete levels, broadened to a width $\Delta E_N$. The left inset is a zoom-in of the spectral-gap range, showing the zero-energy peak. Clearly, this peak results from the superposition of two discrete states. The state at $E=0$ is the MZM, while the state at $E\approx10^{-3}v$ is a CdGM bound state.

The right inset of Fig.~\ref{fig:CdGM} shows that, indeed, the CdGM state and the MZM split further as $\Delta_d$ increases, while at the same time the MZM peak sharpens, indicating that the localization of the MZM in the core improves because of the weakened hybridization with the CdGM state. The energy of the CdGM state grows roughly as $\Delta_{\mathrm{F}}\Delta_d/(8|\mu|)$, which indicates a renormalization of the simple dependence $(1/2)\Delta_\mathrm{F}^2/|\mu|$ by a factor $(v/\mu)^2$. We conclude that the energy of the lowest poisoning quasiparticle is
\begin{equation}
	E_0=\frac{\mu\Delta_d^2}{32v^2}.
	\label{eq:E0}
\end{equation}
With the parameters adopted here, this gives a tiny energy $\sim8\times 10^{-4}v$. 

\section{ Challenges in the numerical identification of $d+id$ vortex MZMs}
\label{Sec:ChallengesNumerical}

\begin{figure*}
    \includegraphics[width=\textwidth]{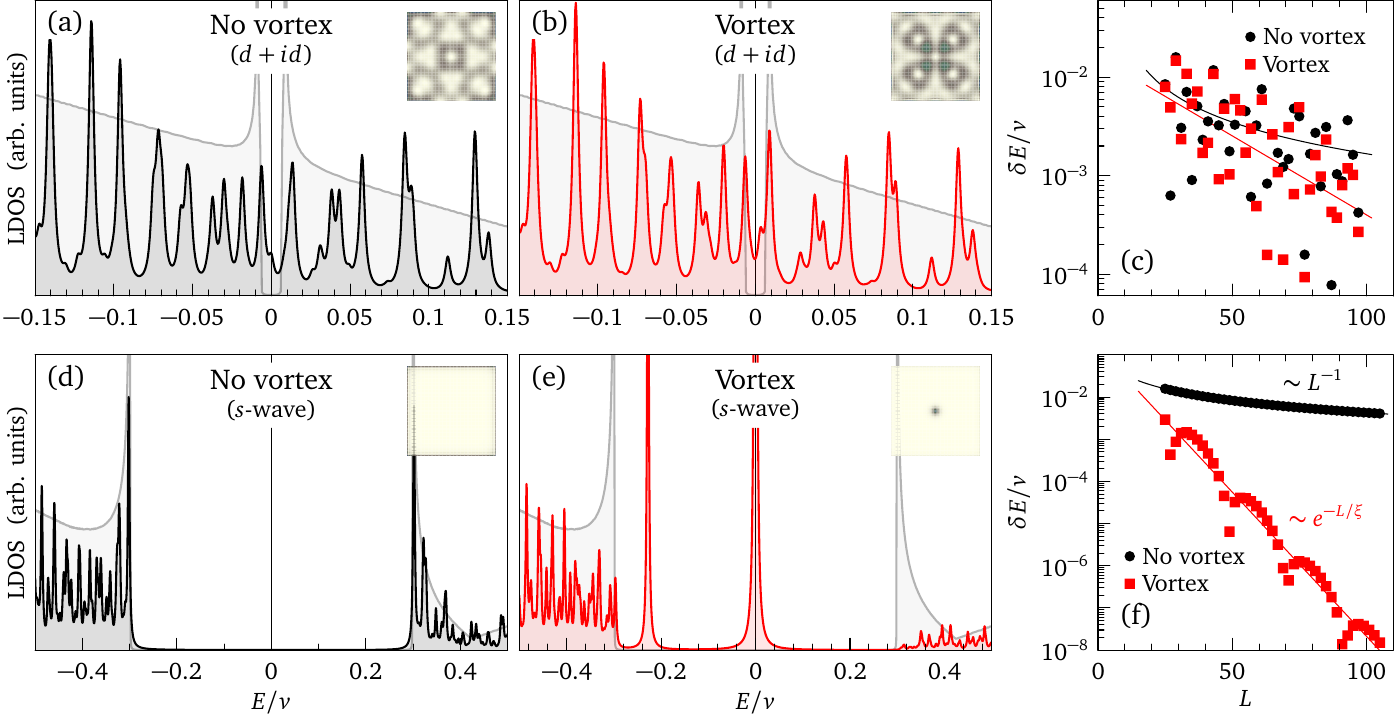}
    \caption{LDOS at $\vec{r}=\vec{0}$ (a) in the absence and (b) in the presence of a vortex with $d\pm id$ pairing (winding numbers $n_X=+1$, $n_\Delta=+2$, $n_V=-1$) for a lattice of size $L=65$. (d) and (e) show the corresponding LDOS for an $s$-wave order parameter $\Delta_\vec{k}=0.3v$. The shaded areas denote the total DOS in a translationally invariant system with $L\to\infty$. The insets show $|u_\up|^2+|u_\down|^2$ for the lowest-energy eigenstate. (c), (f) Lattice-size dependence of the lowest eigenenergy with and without the vortex. The red and black lines show fits of the form $e^{-L/\xi}$ and $1/L$, respectively. The model parameters in units of $v$ are $m = 0.5$, $\mu = -0.3$, $\Delta_d = 0.3$, and $\Delta'_d = 0.2$. A Lorentzian broadening $0.002v$ was used for the LDOS.
    }
    \label{fig:FiniteSizeED}
\end{figure*}

We have seen that the protection of $d\pm id$ vortex MZMs depends on two energy scales, the bulk gap $\Delta_\mathrm{F}$ and the quasiparticle poisoning energy $E_0$, and a localization length scale $\xi_d$. Numerical calculations performed on a finite lattice insert a new length scale $L$, a new energy scale $\Delta E_L\propto 1/L^2$ corresponding, e.g., to the average interlevel spacing, and in the case of the KPM, an energy resolution $\Delta E_N$. From a spectroscopic point of view, a MZM at $E=0$ can be distinguished from a CdGM state at $E_0$ only if $\Delta E_N<E_0$, which sets a lower bound for the expansion order $N$. This is not an issue for ED, which has perfect energy resolution. Yet, the energy of the MZM delivered by ED of a finite lattice is shifted from zero to $\delta E$, because of the hybridization of the MZM with the partner edge state. As the MZM is exponentially localized, it is expected that $\delta E$ approaches zero like $\delta E\sim\exp (-L/\xi_d)$. A problem arises when $\delta E>\Delta E_L$, as the state with the lowest energy may not be the Majorana. Another problem comes about if $\Delta E_L>\Delta_\mathrm{F}$, as the density of levels is not sufficient to resolve the topological gap.

For lattice sizes up to $L=99$, our ED simulations of the lattice model with $d+id$ pairing at $\Delta_d=0.3v$ and $\Delta'_d=0.2v$ \cite{Mercado.2022} show no clear signature of a MZM in the vortex, because $\Delta E_L>\Delta_\mathrm{F}$. As an example, we show the LDOS calculated for $L=65$ and $n_X=+1, n_\Delta=+2$, without and with a $n_V = -1$ vortex in Fig.~\ref{fig:FiniteSizeED}(a) and \ref{fig:FiniteSizeED}(b), respectively. The density of levels is not sufficient to reveal the presence of a vortex and the LDOS is qualitatively identical with it and without it. For comparison, a very clear resolution of the vortex MZM is possible with the same system size but for $s$-wave pairing [Figs.~\ref{fig:FiniteSizeED}(d) and \ref{fig:FiniteSizeED}(e)], because in that case $\Delta E_L\ll\Delta_\mathrm{F}$.

In Figs.~\ref{fig:FiniteSizeED}(c) and \ref{fig:FiniteSizeED}(f), we display the smallest eigenenergy with and without a vortex, for $d+id$ and $s$ pairing, respectively. In the $s$-wave case, the lowest energy varies as $1/L$ in the absence of vortex, while it decays exponentially with system size in the presence of a vortex, as expected for a MZM with energy $\delta E\sim\exp (-L/\xi_s)$. In contrast, no significant exponential decay is seen in the $d+id$ case up to $L=99$. More importantly, at these sizes there is no qualitative difference between having a vortex or not.

In view of this, the lattice simulations reported in Ref.~\cite{Mercado.2022} with $L<60$, for parameters leading to a tiny spectral gap $\Delta_\mathrm{F}\sim0.0068v$, comparable to the empirical peak broadening $\sim0.01v$, cannot be expected to provide convincing evidence for the existence of vortex MZMs.

\section{Discussion and outlook}
\label{Sec:Outlook}

We showed that when combining a $d+id$ superconducting order parameter with a Dirac cone on the surface of a three-dimensional topological insulator, the Majorana zero modes (MZMs) appearing inside vortices can have a nontrivial \emph{angular momentum} that is absent in simpler $s$-wave structures. The MZMs with nonzero angular momenta reside in the vicinity of the vortex core, with zero wavefunction amplitude at the phase singularity point. The MZM angular momentum results from the interplay of three distinct winding numbers in the $d+id+\phantom{}$Dirac model, one characterizing the Dirac cone in the normal state ($n_X$), one the $d\pm id$ order parameter ($n_\Delta$), and one the vorticity ($n_V$). The MZM angular momentum is often not appreciated in the existing literature: Even though all Chern-odd chiral superconductors have MZMs in vortex cores, the existence of MZM angular momentum suggests the existence of different `flavors' of MZMs. Because it depends on the specific components of the Hamiltonian, this additional angular momentum quantum number is not set solely by the Chern number but is influenced by microscopic details.

We have classified all possible cases for the $d+id+\phantom{}$Dirac model in terms of the resulting angular momentum quantum numbers $n_\up$ and $n_\down$ of the electron and hole components of the MZM. We have found, in particular, that for a given configuration of the dispersion (fixed $n_X$ and $n_\Delta$), the vortex and the antivortex systematically present distinct spectral signatures \cite{Kraus.2009, Lee.2016, Zlotnikov.2023}. This feature is peculiar, because vortices and antivortices are usually indistinguishable spectroscopically. It opens up an interesting possibility to investigate the angular momentum of MZMs in scanning tunneling experiments, by monitoring the vortex spectroscopy while reversing the magnetic field. On the one hand, the observation of a finite tunneling conductance at zero bias in the vicinity of the vortex core, well separated from the finite-energy excitations, proves topological superconductivity. A vanishing of this tunneling conductance at the core signals a finite angular momentum of the MZM, as already observed for ordinary CdGM states \cite{Hanaguri.2012, Kong.2019, Liu.2020, Chen.2021}. On the other hand, a systematic change of the conductance upon reversing the magnetic field reveals the existence of different angular momenta in the two spin sectors. Specifically, if the zero-bias conductance vanishes at the core in the vortex, it should not vanish in the antivortex, and conversely. Although challenging, the direct observation of the two spin sectors via spin-resolved STM is also possible in principle \cite{Wiesendanger.2009}.

The interplay of $n_X$ and $n_\Delta$ enables $f\pm if$ superconducting states with Chern numbers $C=\pm3$, which goes beyond the original Fu--Kane proposal limited to $p\pm ip$ with $C=\pm1$. Indeed, the windings of the Dirac cone and order parameter either cooperate or compete depending on whether the Fermi surface is electron-like or hole-like. Interestingly, the angular momentum of the MZM is unrelated to the Chern number. In particular, the topological transition occurring when the Fermi surface changes from electron-like to hole-like does not affect the vortex MZM qualitatively.

From the point of view of practical applications, we have argued that the topological protection of vortex MZMs for $d+id$ order parameters is generically weak, because the hard gap $\Delta_\mathrm{F}$ providing the actual protection is much smaller than the maximum $d$-wave gap $\Delta_d$. In Ref.~\cite{Mercado.2022}, the proposed $d+id$ vortex MZMs were dubbed ``high-temperature Majorana zero modes'', suggesting that the high critical temperature of the cuprate material with its large $d$-wave gap would provide a robust protection. This ignores the smallness of the cuprate order parameter for small momenta, which results in a value $\Delta_\mathrm{F}\ll\Delta_d$ for the twisted cuprate bilayer near the center of the Brillouin zone. As an illustration, we estimate $\Delta_\mathrm{F}$ for the proposed setup, which involves the topological insulator Bi$_2$Se$_3$ proximitized by a twisted bilayer cuprate \cite{Mercado.2022}. The lattice parameter of Bi$_2$Se$_3$ is $\sim4$~\AA{}, $v\sim3$~eV~\AA{}, and $\mu$ is limited to $\sim250$~meV \cite{Zhang.2009}. Assuming ideal proximity transfer with $\Delta_d=45$~meV, one gets the most optimistic estimate $\Delta_\mathrm{F}\sim1$~meV.

One consequence of the small $\Delta_\mathrm{F}$ is a large localization length $\propto 1/\Delta_\mathrm{F}$, leading to MZMs that are much more extended than in the $s$-wave scenario. This may be an advantage in making the MZMs more resistant to disorder and more easily observable by local probes. For example, if $d+id$ superconductivity were realized in moir\'{e} TMDs \cite{Akbar.2024, Xia.2025, Guo.2025}, the corresponding vortex-core MZMs would span several hundreds of nanometers. However, such large MZMs would be sensitive to hybridization with zero modes in other vortices, as the typical intervortex distance falls below $100$~nm for fields exceeding $\sim 70$~mT. How vortex-vortex interactions transform the MZMs studied here for an isolated vortex is an interesting and numerically challenging question.

We have argued that poisoning by ordinary Caroli--de Gennes--Matricon states likely constitutes the most challenging obstacle in isolating the vortex MZMs. The gap to the first of these states is only a fraction of $\Delta_\mathrm{F}$. For the Bi$_2$Se$_3$/cuprate system, the optimistic estimate is $E_0\sim0.3$~K, three orders of magnitude below the cuprate critical temperature. This casts doubt on the practical advantage of this system, relative to those based on conventional superconductors with much smaller---but isotropic---order parameters.

Nevertheless, from a fundamental perspective these nontrivial MZMs can be interesting. They are, in fact, special cases of topological edge modes in the presence of supercurrents \cite{Donís-Vela.2021, Pathak.2021} and are therefore compelling targets for experimental detection. In addition to the standard  scanning-tunneling microscopy (STM), microwave impedance microscopy (MIM) has been shown to be able to detect topological edge states \cite{Wang.202341}. The angular momentum of the MZM might be detected using the quantum twisting microscope (QTM) \cite{Inbar.2023}, which is both local and also allows for interference between graphene plane waves and the MZM wavefunction.

MZMs have been proposed as the basis for topological quantum computation, using braiding as a way to perform quantum gates \cite{Yazdani.2023, Aasen.2025}. In most proposals, there is only one flavor of MZM, leading to specific braiding operators. It would be interesting to study braiding groups for systems with different flavors of MZMs---such as MZMs with different angular momenta---to understand how these flavors influence non-Abelian statistics.

While we have calculated the angular momentum of the MZMs in the $d+id+\phantom{}$Dirac model, we expect MZM angular momentum to be a generic feature that extends to other microscopic models. In fact, beyond angular momentum, one might construct more types of flavors \cite{Sedlmayr.2015}, especially in systems with multiple orbitals (as was proposed in oxide nanochannels where these flavors originate from the electronic orbital character \cite{Maiellaro.2023}), different types of order-parameter symmetries, as well as nontrivial sources of Berry curvature or chirality in the normal state Hamiltonian. Exploring these possibilities would open up new paths to topological vortex-core state engineering, beyond the original $p+ip$ wave proposals.

\section*{Acknowledgements}

We thank Roberta Citro, Titus Neupert, Mark Fischer, Carlo Beenakker, Fabian Hassler and Steve Johnston for fruitful discussions. G.V.\ acknowledges the Swiss National Science Foundation (SNSF) via Swiss Postdoctoral Fellowship ``Xtra-ordinary vortices'', Grant No.\ TMPFP2 224637. L.R.\ acknowledges the Swiss National Science Foundation (SNSF) via Starting Grant TMSGI2 211296.

\section*{Data availability}

The data that support the findings of this article are openly available \cite{data-zenodo}.

\appendix

\section{\boldmath Continuum limit of the $d\pm id$ order parameter}
\label{app:continuum limit}

We derive here the continuum limit for the $d_{x^2-y^2}\pm id_{xy}$ order parameter, starting from the lattice model and following the approach of Ref.~\cite{Vafek.2001}. In general, the lattice version of the BdG Hamiltonian in real space is:
\begin{equation*}
	H_\mathrm{latt} = \sum_{\vec{r} \vec{r}' s s'}
	t^{s s'}_{\vec{r}\vec{r}'} c^\dagger_{\vec{r} s} c\pdagger_{\vec{r}' s'}
	+ \sum_{\vec{r}\vec{r}' s s'} \left( \Delta^{s s'}_{\vec{r}\vec{r}'}
	c^\dagger_{\vec{r} s} c^\dagger_{\vec{r}' s'} + \mathrm{h.c.} \right),
\end{equation*}
where $t^{s s'}_{\vec{r}\vec{r}'}$ is the hopping matrix between  sites $\vec{r},\vec{r}'$ and spins $s, s'$ and $\Delta^{s s'}_{\vec{r}\vec{r}'}$ is the superconducting order parameter, whose symmetry is still unspecified. Expressing the eigenstates in terms of the $(u, v)^\intercal$ spinors, the pairing energy of a state can be written as (omitting spin indices for brevity)
\begin{equation}
	E_\Delta=\sum_{ \vec{r}\vec{r'} } \Delta_{\vec{r}\vec{r}'}
	\left( u^*_\vec{r} v^{}_{\vec{r}'} + u^*_{\vec{r}'} v^{}_{\vec{r}} \right) + \mathrm{c.c.}
	\label{eq:generalDeltablock}
\end{equation}
Our goal is to find the correct operator $\hat \Delta$ in the continuum, such that this energy becomes
\begin{equation}
	E_\Delta=\int d\vec{r}\, u^*(\vec{r}) \hat{\Delta} v(\vec{r}) + \mathrm{c.c.},
	\label{eq:deltaoperator}
\end{equation}
with $ \hat{\Delta} $ acting on $v(\vec{r})$ only. Here and in the following, $\vec{r}$ as an index, like in $u_\vec{r}$, is meant as a discrete lattice vector, while $\vec{r}$ in parentheses, like in $u(\vec{r})$, is meant as a continuous variable.

In the lattice formulation of the $d\pm id$ pairing, the order parameter $\Delta_{\vec{r}\vec{r}'} \equiv |\Delta_{\vec{r}\vec{r}'}| e^{i \phi_{\vec{r}\vec{r}'}}$ lives on the nearest- and next-nearest neighbor links with opposite signs along perpendicular directions. Assuming a constant magnitude $|\Delta_{\vec{r}\vec{r}'}|=|\Delta|$, we can then write in the chiral limit
\begin{equation}
    \Delta_{ \vec{r}\vec{r}'}=\frac{1}{4}\begin{cases}
        -|\Delta|e^{i\phi_{ \vec{r}\vec{r'} } } &\text{ if } \vec{r}'=\vec{r}\pm \hat{\vec{x}}\\
        +|\Delta|e^{i\phi_{ \vec{r}\vec{r'} } } &\text{ if } \vec{r}'=\vec{r}\pm \hat{\vec{y}}\\
        -i\frac{n_\Delta}{4}|\Delta|e^{i\phi_{ \vec{r}\vec{r'} } }
        	&\text{ if } \vec{r}'=\vec{r}\pm(\hat{\vec{x}}+\hat{\vec{y}})\\
        +i\frac{n_\Delta}{4}|\Delta|e^{i\phi_{ \vec{r}\vec{r'} } }
        	&\text{ if } \vec{r}'=\vec{r}\pm(\hat{\vec{x}}-\hat{\vec{y}})\\
        0 &\text{ otherwise},
    \end{cases}
\end{equation}
with $|\Delta|$ playing the role of $\Delta_d=2\Delta'_d$ in Eq.~\eqref{Eq:LatticeHBdGMomentumSpace} and $\Delta$ in Eq.~\eqref{Eq:OrderParameterContinuum}. Upon going to the continuum limit, the phase $\phi_{\vec{r}\vec{r}'}$ that lives on the bonds can be approximated by the average of phases associated with the sites $\vec{r}$ and $\vec{r}'$, according to
\begin{equation}
	e^{i \phi_{\vec{r}\vec{r}'}} \rightarrow e^{i (\phi_\vec{r} + \phi_{\vec{r}'})/2}.
\end{equation}
Redefining the order parameter on the sites as $\Delta_\vec{r}=|\Delta| e^{i\phi_\vec{r}}$, Eq.~\eqref{eq:generalDeltablock} can be explicitly rewritten as
\begin{multline}
	E_\Delta=\frac{1}{4}\sum_{\vec{r},\pm} \Big[
	- \Delta_\vec{r} e^{i(\phi_{\vec{r}\pm \hat{\vec{x}}} - \phi_\vec{r} )/2}
	\left( u^*_\vec{r} v^{}_{\vec{r}\pm \hat{\vec{x}}} + u^*_{\vec{r}\pm \hat{\vec{x}}} v^{}_{\vec{r}} \right) \\
	+ \Delta_\vec{r} e^{i(\phi_{\vec{r}\pm \hat{\vec{y}}} - \phi_\vec{r} )/2}
	\left( u^*_\vec{r} v^{}_{\vec{r}\pm \hat{\vec{y}}} + u^*_{\vec{r}\pm \hat{\vec{y}}} v^{}_{\vec{r}} \right) \\
	- i\frac{n_\Delta}{4}\Delta_\vec{r} e^{i(\phi_{\vec{r}\pm(\hat{\vec{x}}+\hat{\vec{y}})} - \phi_\vec{r} )/2}
	\left( u^*_\vec{r} v^{}_{\vec{r}\pm(\hat{\vec{x}}+\hat{\vec{y}})}
	+ u^*_{\vec{r}\pm(\hat{\vec{x}}+\hat{\vec{y}})} v^{}_{\vec{r}} \right) \\
	+ i\frac{n_\Delta}{4}\Delta_\vec{r} e^{i(\phi_{\vec{r}\pm(\hat{\vec{x}}-\hat{\vec{y}})} - \phi_\vec{r} )/2}
	\left( u^*_\vec{r} v^{}_{\vec{r}\pm(\hat{\vec{x}}-\hat{\vec{y}})}
	+ u^*_{\vec{r}\pm(\hat{\vec{x}}-\hat{\vec{y}})} v^{}_{\vec{r}} \right) \\
	+ \mathrm{c.c.}\Big].
\end{multline}
After expanding up to second order in the lattice parameter $a$, we can perform the continuum limit $\sum_\vec{r}\rightarrow a^{-2}\int d\vec{r}$, $u_\vec{r}\rightarrow a u(\vec{r})$, $v_\vec{r}\rightarrow a v(\vec{r})$, $\Delta_\vec{r}\rightarrow \Delta(\vec{r})$, $\phi_\vec{r}\rightarrow \phi(\vec{r})$. With this, it is tedious but straightforward to derive the following expression for the pairing energy:
\begin{multline*}
	E_\Delta=-\frac{a^2}{4} \int d\vec{r}\, \bigg\{ \\ 
	 u^*(\vec{r}) \Delta(\vec{r})
	\left[\partial_{n_\Delta/2} + \frac{i}{2}\big( \partial_{n_\Delta/2} \phi(\vec{r}) \big) \right]^2 v(\vec{r}) \\
	+ v(\vec{r}) \Delta(\vec{r})
	\left[\partial_{n_\Delta/2} + \frac{i}{2} \big(\partial_{n_\Delta/2} \phi(\vec{r})\big) \right]^2 u^*(\vec{r})
	+ \mathrm{c.c.}\bigg\},
\end{multline*}
where $\partial_\pm=\partial_x\pm i\partial_y$. We need one last step to arrive at the form in Eq.~\eqref{eq:deltaoperator}. This can be obtained using partial integration to convert the derivatives acting on $u^*(\vec{r})$ into derivatives on $v$ and $\Delta$. Remember, throughout, that the magnitude of $\Delta(\vec{r})$ is considered to be constant. The operator $\hat{\Delta}$ can then be written as
\begin{equation}
	\hat{\Delta} = 
	\frac{a^2}{2} \Delta(\vec{r})
		\left[ \frac{1}{4} (\partial_{\pm} \phi)^2 - \frac{i}{2} (\partial_{\pm}^2 \phi) 
		- i (\partial_{\pm} \phi) \partial_{\pm} 
		- \partial_{\pm}^2 \right],
	\label{Eq:DeltaHat}
\end{equation}
with $\pm$ standing for $n_\Delta/2$.

Finally, let us derive the expression of this operator in the case of a vortex, where the order-parameter  phase expressed in polar coordinates is $\phi(r,\vartheta)=n_V\vartheta$ and thus $\Delta(\vec{r})=|\Delta|e^{in_V\vartheta}$. Using $\partial_\pm = e^{\pm i \vartheta} (\partial_r \pm \frac{i}{r} \partial_\vartheta)$ and replacing again $\pm$ by $n_\Delta/2$ to make contact with the main text, we arrive at 
\begin{align}
    \nonumber
    \hat\Delta &= -|\Delta| e^{i(n_\Delta+n_V)\vartheta}\, \hat{\mathcal{D}} \\
    \nonumber
    \hat{\mathcal{D}} &= \frac{a^2}{2} \Bigg[\frac{(n_V+2n_\Delta)n_V}{4r^2} -\frac{(2n_V + n_\Delta)n_\Delta}{4r}
    \left( \partial_r + \frac{in_\Delta}{2r} \partial_\vartheta \right) \\
    &\quad +\left( \partial_r + \frac{in_\Delta}{2r}\partial_\vartheta \right)^2 \Bigg].
\end{align}
When acting on the wavefunction components of the form $u_{\up/\down}(r)e^{in_{\up/\down}\vartheta}$, the operator $\hat{\mathcal{D}}$ does not change their angular-momentum quantum number $n_{\up/\down}$.

\section{Calculation of the localization length}
\label{app:xi}

\begin{figure}[b]
	\includegraphics[width=\columnwidth]{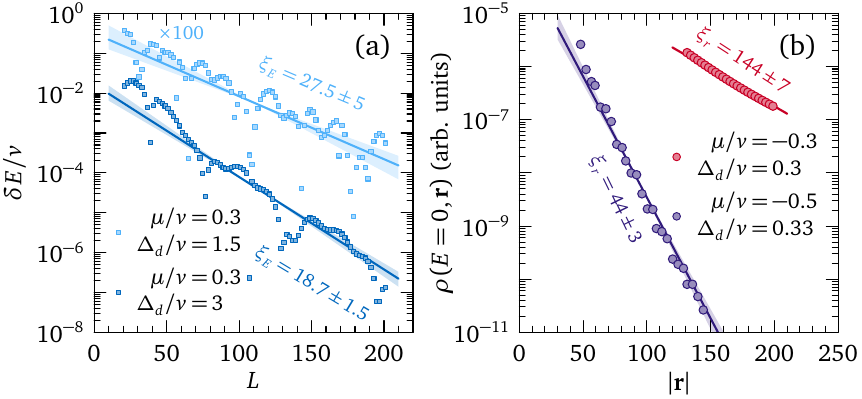}
	\caption{(a) Lattice-size dependence of the MZM eigenvalue obtained by ED. (b) Spatial dependence of the zero-energy LDOS calculated with the KPM. In each panel, the lines show the average value of the fitted $\xi_E$ ($\xi_r$) and the shaded ranges indicate the standard deviation.
	}
	\label{fig:xi}
\end{figure}

To study the localization length of the MZM, we use the parameters $(n_X, n_\Delta, n_V)=(+1, \pm 2, \pm 1)$, $m=0.5v$, $\Delta'_d=\Delta_d/2$, and we vary $\mu$ and $\Delta_d$ to scan the possible values provided by Eq.~\eqref{eq:loclength}. We have checked that equivalent results are obtained with all possible values of the winding numbers. We first perform ED as a function of $L$. By storing the Hamiltonian in sparse form and computing only the required eigenvalues, we can reach lattice sizes as large as $L=201$ with desktop-level resources. Figure~\ref{fig:xi}(a) shows how the MZM eigenvalue approaches zero with increasing system size. The overall behavior is exponential, with oscillations of approximate wavelength $2\pi/k_\mathrm{F}$. We fit the expression $e^{-L/\xi_E}$ to these data, using various ranges from $[51, 201]$ to $[119, 201]$, which delivers a distribution of values for $\xi_E$. The average and standard deviation of this distribution yield our estimate and error bar for $\xi_E$, as displayed in Fig.~\ref{fig:localization}. To probe larger values of the localization length, we perform calculations of $\rho(0,\vec{r})$ with a lattice of size $L=1001$ using the KPM and we fit the expression $e^{-2r/\xi_r}$ to these data. We compute the LDOS at 25 lattice points distributed along the diagonal of the lattice, in an interval that interpolates linearly between $[\xi_d,3\xi_d]$ if $\xi_d\leqslant50$ and $[\xi_d-25,\xi_d+25]$ if $\xi_d=200$, where $\xi_d$ is the value returned by Eq.~\eqref{eq:loclength}. Figure~\ref{fig:xi}(b) shows typical data sets. Again, we use increasing ranges for fitting, starting with the three farthest points, and we thus obtain an average value and a standard deviation for $\xi_r$.


\begin{thebibliography}{83}%
\makeatletter
\providecommand \@ifxundefined [1]{%
 \@ifx{#1\undefined}
}%
\providecommand \@ifnum [1]{%
 \ifnum #1\expandafter \@firstoftwo
 \else \expandafter \@secondoftwo
 \fi
}%
\providecommand \@ifx [1]{%
 \ifx #1\expandafter \@firstoftwo
 \else \expandafter \@secondoftwo
 \fi
}%
\providecommand \natexlab [1]{#1}%
\providecommand \enquote  [1]{``#1''}%
\providecommand \bibnamefont  [1]{#1}%
\providecommand \bibfnamefont [1]{#1}%
\providecommand \citenamefont [1]{#1}%
\providecommand \href@noop [0]{\@secondoftwo}%
\providecommand \href [0]{\begingroup \@sanitize@url \@href}%
\providecommand \@href[1]{\@@startlink{#1}\@@href}%
\providecommand \@@href[1]{\endgroup#1\@@endlink}%
\providecommand \@sanitize@url [0]{\catcode `\\12\catcode `\$12\catcode
  `\&12\catcode `\#12\catcode `\^12\catcode `\_12\catcode `\%12\relax}%
\providecommand \@@startlink[1]{}%
\providecommand \@@endlink[0]{}%
\providecommand \url  [0]{\begingroup\@sanitize@url \@url }%
\providecommand \@url [1]{\endgroup\@href {#1}{\urlprefix }}%
\providecommand \urlprefix  [0]{URL }%
\providecommand \Eprint [0]{\href }%
\providecommand \doibase [0]{https://doi.org/}%
\providecommand \selectlanguage [0]{\@gobble}%
\providecommand \bibinfo  [0]{\@secondoftwo}%
\providecommand \bibfield  [0]{\@secondoftwo}%
\providecommand \translation [1]{[#1]}%
\providecommand \BibitemOpen [0]{}%
\providecommand \bibitemStop [0]{}%
\providecommand \bibitemNoStop [0]{.\EOS\space}%
\providecommand \EOS [0]{\spacefactor3000\relax}%
\providecommand \BibitemShut  [1]{\csname bibitem#1\endcsname}%
\let\auto@bib@innerbib\@empty
\bibitem [{\citenamefont {Yazdani}\ \emph {et~al.}(2023)\citenamefont
  {Yazdani}, \citenamefont {von Oppen}, \citenamefont {Halperin},\ and\
  \citenamefont {Yacoby}}]{Yazdani.2023}%
  \BibitemOpen
  \bibfield  {author} {\bibinfo {author} {\bibfnamefont {A.}~\bibnamefont
  {Yazdani}}, \bibinfo {author} {\bibfnamefont {F.}~\bibnamefont {von Oppen}},
  \bibinfo {author} {\bibfnamefont {B.~I.}\ \bibnamefont {Halperin}},\ and\
  \bibinfo {author} {\bibfnamefont {A.}~\bibnamefont {Yacoby}},\ }\bibfield
  {title} {\bibinfo {title} {Hunting for {Majoranas}},\ }\href
  {https://doi.org/10.1126/science.ade0850} {\bibfield  {journal} {\bibinfo
  {journal} {Science}\ }\textbf {\bibinfo {volume} {380}},\ \bibinfo {pages}
  {eade0850} (\bibinfo {year} {2023})}\BibitemShut {NoStop}%
\bibitem [{\citenamefont {Aasen}\ \emph {et~al.}(2025)\citenamefont {Aasen}
  \emph {et~al.}}]{Aasen.2025}%
  \BibitemOpen
  \bibfield  {author} {\bibinfo {author} {\bibfnamefont {D.}~\bibnamefont
  {Aasen}} \emph {et~al.},\ }\bibfield  {title} {\bibinfo {title} {Blueprint
  for fault-tolerant quantum computation with topological qubit arrays},\
  }\href {https://doi.org/10.1103/qx36-4rv1} {\bibfield  {journal} {\bibinfo
  {journal} {Phys. Rev. Res.}\ }\textbf {\bibinfo {volume} {7}},\ \bibinfo
  {pages} {041002} (\bibinfo {year} {2025})}\BibitemShut {NoStop}%
\bibitem [{\citenamefont {Kitaev}(2001)}]{Kitaev.2001}%
  \BibitemOpen
  \bibfield  {author} {\bibinfo {author} {\bibfnamefont {A.}~\bibnamefont
  {Kitaev}},\ }\bibfield  {title} {\bibinfo {title} {Unpaired {Majorana}
  fermions in quantum wires},\ }\href
  {https://doi.org/10.1070/1063-7869/44/10s/s29} {\bibfield  {journal}
  {\bibinfo  {journal} {Usp. Fiz. Nauk}\ }\textbf {\bibinfo {volume} {17}},\
  \bibinfo {pages} {131} (\bibinfo {year} {2001})}\BibitemShut {NoStop}%
\bibitem [{\citenamefont {Fu}\ and\ \citenamefont {Kane}(2008)}]{Fu.2008}%
  \BibitemOpen
  \bibfield  {author} {\bibinfo {author} {\bibfnamefont {L.}~\bibnamefont
  {Fu}}\ and\ \bibinfo {author} {\bibfnamefont {C.~L.}\ \bibnamefont {Kane}},\
  }\bibfield  {title} {\bibinfo {title} {Superconducting proximity effect and
  {Majorana} fermions at the surface of a topological insulator},\ }\href
  {https://doi.org/10.1103/physrevlett.100.096407} {\bibfield  {journal}
  {\bibinfo  {journal} {Phys. Rev. Lett.}\ }\textbf {\bibinfo {volume} {100}},\
  \bibinfo {pages} {096407} (\bibinfo {year} {2008})}\BibitemShut {NoStop}%
\bibitem [{\citenamefont {Pathak}\ \emph {et~al.}(2021)\citenamefont {Pathak},
  \citenamefont {Plugge},\ and\ \citenamefont {Franz}}]{Pathak.2021}%
  \BibitemOpen
  \bibfield  {author} {\bibinfo {author} {\bibfnamefont {V.}~\bibnamefont
  {Pathak}}, \bibinfo {author} {\bibfnamefont {S.}~\bibnamefont {Plugge}},\
  and\ \bibinfo {author} {\bibfnamefont {M.}~\bibnamefont {Franz}},\ }\bibfield
   {title} {\bibinfo {title} {{Majorana} bound states in vortex lattices on
  iron-based superconductors},\ }\href
  {https://doi.org/10.1016/j.aop.2021.168431} {\bibfield  {journal} {\bibinfo
  {journal} {Ann. Phys.}\ }\textbf {\bibinfo {volume} {435}},\ \bibinfo {pages}
  {168431} (\bibinfo {year} {2021})}\BibitemShut {NoStop}%
\bibitem [{\citenamefont {Zhang}\ \emph {et~al.}(2018)\citenamefont {Zhang},
  \citenamefont {Yaji}, \citenamefont {Hashimoto}, \citenamefont {Ota},
  \citenamefont {Kondo}, \citenamefont {Okazaki}, \citenamefont {Wang},
  \citenamefont {Wen}, \citenamefont {Gu}, \citenamefont {Ding},\ and\
  \citenamefont {Shin}}]{PengZhang.2018}%
  \BibitemOpen
  \bibfield  {author} {\bibinfo {author} {\bibfnamefont {P.}~\bibnamefont
  {Zhang}}, \bibinfo {author} {\bibfnamefont {K.}~\bibnamefont {Yaji}},
  \bibinfo {author} {\bibfnamefont {T.}~\bibnamefont {Hashimoto}}, \bibinfo
  {author} {\bibfnamefont {Y.}~\bibnamefont {Ota}}, \bibinfo {author}
  {\bibfnamefont {T.}~\bibnamefont {Kondo}}, \bibinfo {author} {\bibfnamefont
  {K.}~\bibnamefont {Okazaki}}, \bibinfo {author} {\bibfnamefont
  {Z.}~\bibnamefont {Wang}}, \bibinfo {author} {\bibfnamefont {J.}~\bibnamefont
  {Wen}}, \bibinfo {author} {\bibfnamefont {G.~D.}\ \bibnamefont {Gu}},
  \bibinfo {author} {\bibfnamefont {H.}~\bibnamefont {Ding}},\ and\ \bibinfo
  {author} {\bibfnamefont {S.}~\bibnamefont {Shin}},\ }\bibfield  {title}
  {\bibinfo {title} {Observation of topological superconductivity on the
  surface of an iron-based superconductor},\ }\href
  {https://doi.org/10.1126/science.aan4596} {\bibfield  {journal} {\bibinfo
  {journal} {Science}\ }\textbf {\bibinfo {volume} {360}},\ \bibinfo {pages}
  {182} (\bibinfo {year} {2018})}\BibitemShut {NoStop}%
\bibitem [{\citenamefont {Wang}\ \emph {et~al.}(2018)\citenamefont {Wang},
  \citenamefont {Kong}, \citenamefont {Fan}, \citenamefont {Chen},
  \citenamefont {Zhu}, \citenamefont {Liu}, \citenamefont {Cao}, \citenamefont
  {Sun}, \citenamefont {Du}, \citenamefont {Schneeloch}, \citenamefont {Zhong},
  \citenamefont {Gu}, \citenamefont {Fu}, \citenamefont {Ding},\ and\
  \citenamefont {Gao}}]{DongfeiWang.2018}%
  \BibitemOpen
  \bibfield  {author} {\bibinfo {author} {\bibfnamefont {D.}~\bibnamefont
  {Wang}}, \bibinfo {author} {\bibfnamefont {L.}~\bibnamefont {Kong}}, \bibinfo
  {author} {\bibfnamefont {P.}~\bibnamefont {Fan}}, \bibinfo {author}
  {\bibfnamefont {H.}~\bibnamefont {Chen}}, \bibinfo {author} {\bibfnamefont
  {S.}~\bibnamefont {Zhu}}, \bibinfo {author} {\bibfnamefont {W.}~\bibnamefont
  {Liu}}, \bibinfo {author} {\bibfnamefont {L.}~\bibnamefont {Cao}}, \bibinfo
  {author} {\bibfnamefont {Y.}~\bibnamefont {Sun}}, \bibinfo {author}
  {\bibfnamefont {S.}~\bibnamefont {Du}}, \bibinfo {author} {\bibfnamefont
  {J.}~\bibnamefont {Schneeloch}}, \bibinfo {author} {\bibfnamefont
  {R.}~\bibnamefont {Zhong}}, \bibinfo {author} {\bibfnamefont
  {G.}~\bibnamefont {Gu}}, \bibinfo {author} {\bibfnamefont {L.}~\bibnamefont
  {Fu}}, \bibinfo {author} {\bibfnamefont {H.}~\bibnamefont {Ding}},\ and\
  \bibinfo {author} {\bibfnamefont {H.-J.}\ \bibnamefont {Gao}},\ }\bibfield
  {title} {\bibinfo {title} {Evidence for {Majorana} bound states in an
  iron-based superconductor},\ }\href {https://doi.org/10.1126/science.aao1797}
  {\bibfield  {journal} {\bibinfo  {journal} {Science}\ }\textbf {\bibinfo
  {volume} {362}},\ \bibinfo {pages} {333} (\bibinfo {year}
  {2018})}\BibitemShut {NoStop}%
\bibitem [{\citenamefont {Kong}\ \emph {et~al.}(2019)\citenamefont {Kong},
  \citenamefont {Zhu}, \citenamefont {Papaj}, \citenamefont {Chen},
  \citenamefont {Cao}, \citenamefont {Isobe}, \citenamefont {Xing},
  \citenamefont {Liu}, \citenamefont {Wang}, \citenamefont {Fan}, \citenamefont
  {Sun}, \citenamefont {Du}, \citenamefont {Schneeloch}, \citenamefont {Zhong},
  \citenamefont {Gu}, \citenamefont {Fu}, \citenamefont {Gao},\ and\
  \citenamefont {Ding}}]{Kong.2019}%
  \BibitemOpen
  \bibfield  {author} {\bibinfo {author} {\bibfnamefont {L.}~\bibnamefont
  {Kong}}, \bibinfo {author} {\bibfnamefont {S.}~\bibnamefont {Zhu}}, \bibinfo
  {author} {\bibfnamefont {M.}~\bibnamefont {Papaj}}, \bibinfo {author}
  {\bibfnamefont {H.}~\bibnamefont {Chen}}, \bibinfo {author} {\bibfnamefont
  {L.}~\bibnamefont {Cao}}, \bibinfo {author} {\bibfnamefont {H.}~\bibnamefont
  {Isobe}}, \bibinfo {author} {\bibfnamefont {Y.}~\bibnamefont {Xing}},
  \bibinfo {author} {\bibfnamefont {W.}~\bibnamefont {Liu}}, \bibinfo {author}
  {\bibfnamefont {D.}~\bibnamefont {Wang}}, \bibinfo {author} {\bibfnamefont
  {P.}~\bibnamefont {Fan}}, \bibinfo {author} {\bibfnamefont {Y.}~\bibnamefont
  {Sun}}, \bibinfo {author} {\bibfnamefont {S.}~\bibnamefont {Du}}, \bibinfo
  {author} {\bibfnamefont {J.}~\bibnamefont {Schneeloch}}, \bibinfo {author}
  {\bibfnamefont {R.}~\bibnamefont {Zhong}}, \bibinfo {author} {\bibfnamefont
  {G.}~\bibnamefont {Gu}}, \bibinfo {author} {\bibfnamefont {L.}~\bibnamefont
  {Fu}}, \bibinfo {author} {\bibfnamefont {H.-J.}\ \bibnamefont {Gao}},\ and\
  \bibinfo {author} {\bibfnamefont {H.}~\bibnamefont {Ding}},\ }\bibfield
  {title} {\bibinfo {title} {Half-integer level shift of vortex bound states in
  an iron-based superconductor},\ }\href
  {https://doi.org/10.1038/s41567-019-0630-5} {\bibfield  {journal} {\bibinfo
  {journal} {Nat. Phys.}\ }\textbf {\bibinfo {volume} {15}},\ \bibinfo {pages}
  {1181} (\bibinfo {year} {2019})}\BibitemShut {NoStop}%
\bibitem [{\citenamefont {Xu}\ \emph {et~al.}(2015)\citenamefont {Xu},
  \citenamefont {Wang}, \citenamefont {Liu}, \citenamefont {Ge}, \citenamefont
  {Yang}, \citenamefont {Liu}, \citenamefont {Xu}, \citenamefont {Guan},
  \citenamefont {Gao}, \citenamefont {Qian}, \citenamefont {Liu}, \citenamefont
  {Wang}, \citenamefont {Zhang}, \citenamefont {Xue},\ and\ \citenamefont
  {Jia}}]{Xu.2015}%
  \BibitemOpen
  \bibfield  {author} {\bibinfo {author} {\bibfnamefont {J.-P.}\ \bibnamefont
  {Xu}}, \bibinfo {author} {\bibfnamefont {M.-X.}\ \bibnamefont {Wang}},
  \bibinfo {author} {\bibfnamefont {Z.~L.}\ \bibnamefont {Liu}}, \bibinfo
  {author} {\bibfnamefont {J.-F.}\ \bibnamefont {Ge}}, \bibinfo {author}
  {\bibfnamefont {X.}~\bibnamefont {Yang}}, \bibinfo {author} {\bibfnamefont
  {C.}~\bibnamefont {Liu}}, \bibinfo {author} {\bibfnamefont {Z.~A.}\
  \bibnamefont {Xu}}, \bibinfo {author} {\bibfnamefont {D.}~\bibnamefont
  {Guan}}, \bibinfo {author} {\bibfnamefont {C.~L.}\ \bibnamefont {Gao}},
  \bibinfo {author} {\bibfnamefont {D.}~\bibnamefont {Qian}}, \bibinfo {author}
  {\bibfnamefont {Y.}~\bibnamefont {Liu}}, \bibinfo {author} {\bibfnamefont
  {Q.-H.}\ \bibnamefont {Wang}}, \bibinfo {author} {\bibfnamefont {F.-C.}\
  \bibnamefont {Zhang}}, \bibinfo {author} {\bibfnamefont {Q.-K.}\ \bibnamefont
  {Xue}},\ and\ \bibinfo {author} {\bibfnamefont {J.-F.}\ \bibnamefont {Jia}},\
  }\bibfield  {title} {\bibinfo {title} {Experimental detection of a {Majorana}
  mode in the core of a magnetic vortex inside a topological
  insulator-superconductor {Bi$_2$Te$_3$/NbSe$_2$} heterostructure},\ }\href
  {https://doi.org/10.1103/PhysRevLett.114.017001} {\bibfield  {journal}
  {\bibinfo  {journal} {Phys. Rev. Lett.}\ }\textbf {\bibinfo {volume} {114}},\
  \bibinfo {pages} {017001} (\bibinfo {year} {2015})}\BibitemShut {NoStop}%
\bibitem [{\citenamefont {Sun}\ \emph {et~al.}(2016)\citenamefont {Sun},
  \citenamefont {Zhang}, \citenamefont {Hu}, \citenamefont {Li}, \citenamefont
  {Wang}, \citenamefont {Ma}, \citenamefont {Xu}, \citenamefont {Gao},
  \citenamefont {Guan}, \citenamefont {Li}, \citenamefont {Liu}, \citenamefont
  {Qian}, \citenamefont {Zhou}, \citenamefont {Fu}, \citenamefont {Li},
  \citenamefont {Zhang},\ and\ \citenamefont {Jia}}]{Sun.2016}%
  \BibitemOpen
  \bibfield  {author} {\bibinfo {author} {\bibfnamefont {H.-H.}\ \bibnamefont
  {Sun}}, \bibinfo {author} {\bibfnamefont {K.-W.}\ \bibnamefont {Zhang}},
  \bibinfo {author} {\bibfnamefont {L.-H.}\ \bibnamefont {Hu}}, \bibinfo
  {author} {\bibfnamefont {C.}~\bibnamefont {Li}}, \bibinfo {author}
  {\bibfnamefont {G.-Y.}\ \bibnamefont {Wang}}, \bibinfo {author}
  {\bibfnamefont {H.-Y.}\ \bibnamefont {Ma}}, \bibinfo {author} {\bibfnamefont
  {Z.-A.}\ \bibnamefont {Xu}}, \bibinfo {author} {\bibfnamefont {C.-L.}\
  \bibnamefont {Gao}}, \bibinfo {author} {\bibfnamefont {D.-D.}\ \bibnamefont
  {Guan}}, \bibinfo {author} {\bibfnamefont {Y.-Y.}\ \bibnamefont {Li}},
  \bibinfo {author} {\bibfnamefont {C.}~\bibnamefont {Liu}}, \bibinfo {author}
  {\bibfnamefont {D.}~\bibnamefont {Qian}}, \bibinfo {author} {\bibfnamefont
  {Y.}~\bibnamefont {Zhou}}, \bibinfo {author} {\bibfnamefont {L.}~\bibnamefont
  {Fu}}, \bibinfo {author} {\bibfnamefont {S.-C.}\ \bibnamefont {Li}}, \bibinfo
  {author} {\bibfnamefont {F.-C.}\ \bibnamefont {Zhang}},\ and\ \bibinfo
  {author} {\bibfnamefont {J.-F.}\ \bibnamefont {Jia}},\ }\bibfield  {title}
  {\bibinfo {title} {{Majorana} zero mode detected with spin selective
  {Andreev} reflection in the vortex of a topological superconductor},\ }\href
  {https://doi.org/10.1103/PhysRevLett.116.257003} {\bibfield  {journal}
  {\bibinfo  {journal} {Phys. Rev. Lett.}\ }\textbf {\bibinfo {volume} {116}},\
  \bibinfo {pages} {257003} (\bibinfo {year} {2016})}\BibitemShut {NoStop}%
\bibitem [{\citenamefont {Liu}\ \emph {et~al.}(2024)\citenamefont {Liu},
  \citenamefont {Wan}, \citenamefont {Yang}, \citenamefont {Zhao},
  \citenamefont {Xie}, \citenamefont {Zheng}, \citenamefont {Yi}, \citenamefont
  {Guan}, \citenamefont {Wang}, \citenamefont {Zheng}, \citenamefont {Liu},
  \citenamefont {Fu}, \citenamefont {Liu}, \citenamefont {Li},\ and\
  \citenamefont {Jia}}]{Liu.2024}%
  \BibitemOpen
  \bibfield  {author} {\bibinfo {author} {\bibfnamefont {T.}~\bibnamefont
  {Liu}}, \bibinfo {author} {\bibfnamefont {C.~Y.}\ \bibnamefont {Wan}},
  \bibinfo {author} {\bibfnamefont {H.}~\bibnamefont {Yang}}, \bibinfo {author}
  {\bibfnamefont {Y.}~\bibnamefont {Zhao}}, \bibinfo {author} {\bibfnamefont
  {B.}~\bibnamefont {Xie}}, \bibinfo {author} {\bibfnamefont {W.}~\bibnamefont
  {Zheng}}, \bibinfo {author} {\bibfnamefont {Z.}~\bibnamefont {Yi}}, \bibinfo
  {author} {\bibfnamefont {D.}~\bibnamefont {Guan}}, \bibinfo {author}
  {\bibfnamefont {S.}~\bibnamefont {Wang}}, \bibinfo {author} {\bibfnamefont
  {H.}~\bibnamefont {Zheng}}, \bibinfo {author} {\bibfnamefont
  {C.}~\bibnamefont {Liu}}, \bibinfo {author} {\bibfnamefont {L.}~\bibnamefont
  {Fu}}, \bibinfo {author} {\bibfnamefont {J.}~\bibnamefont {Liu}}, \bibinfo
  {author} {\bibfnamefont {Y.}~\bibnamefont {Li}},\ and\ \bibinfo {author}
  {\bibfnamefont {J.}~\bibnamefont {Jia}},\ }\bibfield  {title} {\bibinfo
  {title} {Signatures of hybridization of multiple {Majorana} zero modes in a
  vortex},\ }\href {https://doi.org/10.1038/s41586-024-07857-4} {\bibfield
  {journal} {\bibinfo  {journal} {Nature}\ }\textbf {\bibinfo {volume} {633}},\
  \bibinfo {pages} {71} (\bibinfo {year} {2024})}\BibitemShut {NoStop}%
\bibitem [{\citenamefont {Tsuei}\ and\ \citenamefont
  {Kirtley}(2000)}]{Tsuei.2000}%
  \BibitemOpen
  \bibfield  {author} {\bibinfo {author} {\bibfnamefont {C.~C.}\ \bibnamefont
  {Tsuei}}\ and\ \bibinfo {author} {\bibfnamefont {J.~R.}\ \bibnamefont
  {Kirtley}},\ }\bibfield  {title} {\bibinfo {title} {Pairing symmetry in
  cuprate superconductors},\ }\href {https://doi.org/10.1103/RevModPhys.72.969}
  {\bibfield  {journal} {\bibinfo  {journal} {Rev. Mod. Phys.}\ }\textbf
  {\bibinfo {volume} {72}},\ \bibinfo {pages} {969} (\bibinfo {year}
  {2000})}\BibitemShut {NoStop}%
\bibitem [{\citenamefont {Yan}\ \emph {et~al.}(2018)\citenamefont {Yan},
  \citenamefont {Song},\ and\ \citenamefont {Wang}}]{Zhongbo.2018}%
  \BibitemOpen
  \bibfield  {author} {\bibinfo {author} {\bibfnamefont {Z.}~\bibnamefont
  {Yan}}, \bibinfo {author} {\bibfnamefont {F.}~\bibnamefont {Song}},\ and\
  \bibinfo {author} {\bibfnamefont {Z.}~\bibnamefont {Wang}},\ }\bibfield
  {title} {\bibinfo {title} {{Majorana} corner modes in a high-temperature
  platform},\ }\href {https://doi.org/10.1103/PhysRevLett.121.096803}
  {\bibfield  {journal} {\bibinfo  {journal} {Phys. Rev. Lett.}\ }\textbf
  {\bibinfo {volume} {121}},\ \bibinfo {pages} {096803} (\bibinfo {year}
  {2018})}\BibitemShut {NoStop}%
\bibitem [{\citenamefont {Liu}\ \emph {et~al.}(2018)\citenamefont {Liu},
  \citenamefont {He},\ and\ \citenamefont {Nori}}]{TaoLiu.2018}%
  \BibitemOpen
  \bibfield  {author} {\bibinfo {author} {\bibfnamefont {T.}~\bibnamefont
  {Liu}}, \bibinfo {author} {\bibfnamefont {J.~J.}\ \bibnamefont {He}},\ and\
  \bibinfo {author} {\bibfnamefont {F.}~\bibnamefont {Nori}},\ }\bibfield
  {title} {\bibinfo {title} {{Majorana} corner states in a two-dimensional
  magnetic topological insulator on a high-temperature superconductor},\ }\href
  {https://doi.org/10.1103/PhysRevB.98.245413} {\bibfield  {journal} {\bibinfo
  {journal} {Phys. Rev. B}\ }\textbf {\bibinfo {volume} {98}},\ \bibinfo
  {pages} {245413} (\bibinfo {year} {2018})}\BibitemShut {NoStop}%
\bibitem [{\citenamefont {Wang}\ \emph {et~al.}(2013)\citenamefont {Wang},
  \citenamefont {Ding}, \citenamefont {Fedorov}, \citenamefont {Yao},
  \citenamefont {Li}, \citenamefont {Lv}, \citenamefont {Zhao}, \citenamefont
  {Zhang}, \citenamefont {Xu}, \citenamefont {Schneeloch}, \citenamefont
  {Zhong}, \citenamefont {Ji}, \citenamefont {Wang}, \citenamefont {He},
  \citenamefont {Ma}, \citenamefont {Gu}, \citenamefont {Yao}, \citenamefont
  {Xue}, \citenamefont {Chen},\ and\ \citenamefont {Zhou}}]{Wang.2013}%
  \BibitemOpen
  \bibfield  {author} {\bibinfo {author} {\bibfnamefont {E.}~\bibnamefont
  {Wang}}, \bibinfo {author} {\bibfnamefont {H.}~\bibnamefont {Ding}}, \bibinfo
  {author} {\bibfnamefont {A.~V.}\ \bibnamefont {Fedorov}}, \bibinfo {author}
  {\bibfnamefont {W.}~\bibnamefont {Yao}}, \bibinfo {author} {\bibfnamefont
  {Z.}~\bibnamefont {Li}}, \bibinfo {author} {\bibfnamefont {Y.-F.}\
  \bibnamefont {Lv}}, \bibinfo {author} {\bibfnamefont {K.}~\bibnamefont
  {Zhao}}, \bibinfo {author} {\bibfnamefont {L.-G.}\ \bibnamefont {Zhang}},
  \bibinfo {author} {\bibfnamefont {Z.}~\bibnamefont {Xu}}, \bibinfo {author}
  {\bibfnamefont {J.}~\bibnamefont {Schneeloch}}, \bibinfo {author}
  {\bibfnamefont {R.}~\bibnamefont {Zhong}}, \bibinfo {author} {\bibfnamefont
  {S.-H.}\ \bibnamefont {Ji}}, \bibinfo {author} {\bibfnamefont
  {L.}~\bibnamefont {Wang}}, \bibinfo {author} {\bibfnamefont {K.}~\bibnamefont
  {He}}, \bibinfo {author} {\bibfnamefont {X.}~\bibnamefont {Ma}}, \bibinfo
  {author} {\bibfnamefont {G.}~\bibnamefont {Gu}}, \bibinfo {author}
  {\bibfnamefont {H.}~\bibnamefont {Yao}}, \bibinfo {author} {\bibfnamefont
  {Q.-K.}\ \bibnamefont {Xue}}, \bibinfo {author} {\bibfnamefont
  {X.}~\bibnamefont {Chen}},\ and\ \bibinfo {author} {\bibfnamefont
  {S.}~\bibnamefont {Zhou}},\ }\bibfield  {title} {\bibinfo {title} {Fully
  gapped topological surface states in {Bi$_2$Se$_3$} films induced by a
  $d$-wave high-temperature superconductor},\ }\href
  {https://doi.org/10.1038/nphys2744} {\bibfield  {journal} {\bibinfo
  {journal} {Nat. Phys.}\ }\textbf {\bibinfo {volume} {9}},\ \bibinfo {pages}
  {621} (\bibinfo {year} {2013})}\BibitemShut {NoStop}%
\bibitem [{\citenamefont {Xu}\ \emph {et~al.}(2014)\citenamefont {Xu},
  \citenamefont {Liu}, \citenamefont {Richardella}, \citenamefont {Belopolski},
  \citenamefont {Alidoust}, \citenamefont {Neupane}, \citenamefont {Bian},
  \citenamefont {Samarth},\ and\ \citenamefont {Hasan}}]{Su-YangXu.2014}%
  \BibitemOpen
  \bibfield  {author} {\bibinfo {author} {\bibfnamefont {S.-Y.}\ \bibnamefont
  {Xu}}, \bibinfo {author} {\bibfnamefont {C.}~\bibnamefont {Liu}}, \bibinfo
  {author} {\bibfnamefont {A.}~\bibnamefont {Richardella}}, \bibinfo {author}
  {\bibfnamefont {I.}~\bibnamefont {Belopolski}}, \bibinfo {author}
  {\bibfnamefont {N.}~\bibnamefont {Alidoust}}, \bibinfo {author}
  {\bibfnamefont {M.}~\bibnamefont {Neupane}}, \bibinfo {author} {\bibfnamefont
  {G.}~\bibnamefont {Bian}}, \bibinfo {author} {\bibfnamefont {N.}~\bibnamefont
  {Samarth}},\ and\ \bibinfo {author} {\bibfnamefont {M.~Z.}\ \bibnamefont
  {Hasan}},\ }\bibfield  {title} {\bibinfo {title} {{Fermi}-level electronic
  structure of a topological-insulator/cuprate-superconductor based
  heterostructure in the superconducting proximity effect regime},\ }\href
  {https://doi.org/10.1103/PhysRevB.90.085128} {\bibfield  {journal} {\bibinfo
  {journal} {Phys. Rev. B}\ }\textbf {\bibinfo {volume} {90}},\ \bibinfo
  {pages} {085128} (\bibinfo {year} {2014})}\BibitemShut {NoStop}%
\bibitem [{\citenamefont {Li}\ \emph {et~al.}(2015)\citenamefont {Li},
  \citenamefont {Chan},\ and\ \citenamefont {Yao}}]{Zi-XiangLi.2015}%
  \BibitemOpen
  \bibfield  {author} {\bibinfo {author} {\bibfnamefont {Z.-X.}\ \bibnamefont
  {Li}}, \bibinfo {author} {\bibfnamefont {C.}~\bibnamefont {Chan}},\ and\
  \bibinfo {author} {\bibfnamefont {H.}~\bibnamefont {Yao}},\ }\bibfield
  {title} {\bibinfo {title} {Realizing {Majorana} zero modes by proximity
  effect between topological insulators and $d$-wave high-temperature
  superconductors},\ }\href {https://doi.org/10.1103/PhysRevB.91.235143}
  {\bibfield  {journal} {\bibinfo  {journal} {Phys. Rev. B}\ }\textbf {\bibinfo
  {volume} {91}},\ \bibinfo {pages} {235143} (\bibinfo {year}
  {2015})}\BibitemShut {NoStop}%
\bibitem [{\citenamefont {Zareapour}\ \emph {et~al.}(2012)\citenamefont
  {Zareapour}, \citenamefont {Hayat}, \citenamefont {Zhao}, \citenamefont
  {Kreshchuk}, \citenamefont {Jain}, \citenamefont {Kwok}, \citenamefont {Lee},
  \citenamefont {Cheong}, \citenamefont {Xu}, \citenamefont {Yang},
  \citenamefont {Gu}, \citenamefont {Jia}, \citenamefont {Cava},\ and\
  \citenamefont {Burch}}]{Zareapour.2012}%
  \BibitemOpen
  \bibfield  {author} {\bibinfo {author} {\bibfnamefont {P.}~\bibnamefont
  {Zareapour}}, \bibinfo {author} {\bibfnamefont {A.}~\bibnamefont {Hayat}},
  \bibinfo {author} {\bibfnamefont {S.~Y.~F.}\ \bibnamefont {Zhao}}, \bibinfo
  {author} {\bibfnamefont {M.}~\bibnamefont {Kreshchuk}}, \bibinfo {author}
  {\bibfnamefont {A.}~\bibnamefont {Jain}}, \bibinfo {author} {\bibfnamefont
  {D.~C.}\ \bibnamefont {Kwok}}, \bibinfo {author} {\bibfnamefont
  {N.}~\bibnamefont {Lee}}, \bibinfo {author} {\bibfnamefont {S.-W.}\
  \bibnamefont {Cheong}}, \bibinfo {author} {\bibfnamefont {Z.}~\bibnamefont
  {Xu}}, \bibinfo {author} {\bibfnamefont {A.}~\bibnamefont {Yang}}, \bibinfo
  {author} {\bibfnamefont {G.}~\bibnamefont {Gu}}, \bibinfo {author}
  {\bibfnamefont {S.}~\bibnamefont {Jia}}, \bibinfo {author} {\bibfnamefont
  {R.~J.}\ \bibnamefont {Cava}},\ and\ \bibinfo {author} {\bibfnamefont
  {K.~S.}\ \bibnamefont {Burch}},\ }\bibfield  {title} {\bibinfo {title}
  {Proximity-induced high-temperature superconductivity in the topological
  insulators {Bi$_2$Se$_3$} and {Bi$_2$Te$_3$}},\ }\href
  {https://doi.org/10.1038/ncomms2042} {\bibfield  {journal} {\bibinfo
  {journal} {Nat. Commun.}\ }\textbf {\bibinfo {volume} {3}},\ \bibinfo {pages}
  {1056} (\bibinfo {year} {2012})}\BibitemShut {NoStop}%
\bibitem [{\citenamefont {Mercado}\ \emph {et~al.}(2022)\citenamefont
  {Mercado}, \citenamefont {Sahoo},\ and\ \citenamefont
  {Franz}}]{Mercado.2022}%
  \BibitemOpen
  \bibfield  {author} {\bibinfo {author} {\bibfnamefont {A.}~\bibnamefont
  {Mercado}}, \bibinfo {author} {\bibfnamefont {S.}~\bibnamefont {Sahoo}},\
  and\ \bibinfo {author} {\bibfnamefont {M.}~\bibnamefont {Franz}},\ }\bibfield
   {title} {\bibinfo {title} {High-temperature {Majorana} zero modes},\ }\href
  {https://doi.org/10.1103/physrevlett.128.137002} {\bibfield  {journal}
  {\bibinfo  {journal} {Phys. Rev. Lett.}\ }\textbf {\bibinfo {volume} {128}},\
  \bibinfo {pages} {137002} (\bibinfo {year} {2022})}\BibitemShut {NoStop}%
\bibitem [{\citenamefont {Can}\ \emph {et~al.}(2021)\citenamefont {Can},
  \citenamefont {Tummuru}, \citenamefont {Day}, \citenamefont {Elfimov},
  \citenamefont {Damascelli},\ and\ \citenamefont {Franz}}]{Can.2021}%
  \BibitemOpen
  \bibfield  {author} {\bibinfo {author} {\bibfnamefont {O.}~\bibnamefont
  {Can}}, \bibinfo {author} {\bibfnamefont {T.}~\bibnamefont {Tummuru}},
  \bibinfo {author} {\bibfnamefont {R.~P.}\ \bibnamefont {Day}}, \bibinfo
  {author} {\bibfnamefont {I.}~\bibnamefont {Elfimov}}, \bibinfo {author}
  {\bibfnamefont {A.}~\bibnamefont {Damascelli}},\ and\ \bibinfo {author}
  {\bibfnamefont {M.}~\bibnamefont {Franz}},\ }\bibfield  {title} {\bibinfo
  {title} {High-temperature topological superconductivity in twisted
  double-layer copper oxides},\ }\href
  {https://doi.org/10.1038/s41567-020-01142-7} {\bibfield  {journal} {\bibinfo
  {journal} {Nat. Phys.}\ }\textbf {\bibinfo {volume} {17}},\ \bibinfo {pages}
  {519 } (\bibinfo {year} {2021})}\BibitemShut {NoStop}%
\bibitem [{\citenamefont {Volkov}\ \emph {et~al.}(2025)\citenamefont {Volkov},
  \citenamefont {Zhao}, \citenamefont {Poccia}, \citenamefont {Cui},
  \citenamefont {Kim},\ and\ \citenamefont {Pixley}}]{Volkov.2025}%
  \BibitemOpen
  \bibfield  {author} {\bibinfo {author} {\bibfnamefont {P.~A.}\ \bibnamefont
  {Volkov}}, \bibinfo {author} {\bibfnamefont {S.~Y.~F.}\ \bibnamefont {Zhao}},
  \bibinfo {author} {\bibfnamefont {N.}~\bibnamefont {Poccia}}, \bibinfo
  {author} {\bibfnamefont {X.}~\bibnamefont {Cui}}, \bibinfo {author}
  {\bibfnamefont {P.}~\bibnamefont {Kim}},\ and\ \bibinfo {author}
  {\bibfnamefont {J.~H.}\ \bibnamefont {Pixley}},\ }\bibfield  {title}
  {\bibinfo {title} {{Josephson} effects in twisted nodal superconductors},\
  }\href {https://doi.org/10.1103/PhysRevB.111.014514} {\bibfield  {journal}
  {\bibinfo  {journal} {Phys. Rev. B}\ }\textbf {\bibinfo {volume} {111}},\
  \bibinfo {pages} {014514} (\bibinfo {year} {2025})}\BibitemShut {NoStop}%
\bibitem [{\citenamefont {Zhao}\ \emph {et~al.}(2023)\citenamefont {Zhao},
  \citenamefont {Cui}, \citenamefont {Volkov}, \citenamefont {Yoo},
  \citenamefont {Lee}, \citenamefont {Gardener}, \citenamefont {Akey},
  \citenamefont {Engelke}, \citenamefont {Ronen}, \citenamefont {Zhong},
  \citenamefont {Gu}, \citenamefont {Plugge}, \citenamefont {Tummuru},
  \citenamefont {Kim}, \citenamefont {Franz}, \citenamefont {Pixley},
  \citenamefont {Poccia},\ and\ \citenamefont {Kim}}]{Zhao.2023}%
  \BibitemOpen
  \bibfield  {author} {\bibinfo {author} {\bibfnamefont {S.~Y.~F.}\
  \bibnamefont {Zhao}}, \bibinfo {author} {\bibfnamefont {X.}~\bibnamefont
  {Cui}}, \bibinfo {author} {\bibfnamefont {P.~A.}\ \bibnamefont {Volkov}},
  \bibinfo {author} {\bibfnamefont {H.}~\bibnamefont {Yoo}}, \bibinfo {author}
  {\bibfnamefont {S.}~\bibnamefont {Lee}}, \bibinfo {author} {\bibfnamefont
  {J.~A.}\ \bibnamefont {Gardener}}, \bibinfo {author} {\bibfnamefont {A.~J.}\
  \bibnamefont {Akey}}, \bibinfo {author} {\bibfnamefont {R.}~\bibnamefont
  {Engelke}}, \bibinfo {author} {\bibfnamefont {Y.}~\bibnamefont {Ronen}},
  \bibinfo {author} {\bibfnamefont {R.}~\bibnamefont {Zhong}}, \bibinfo
  {author} {\bibfnamefont {G.}~\bibnamefont {Gu}}, \bibinfo {author}
  {\bibfnamefont {S.}~\bibnamefont {Plugge}}, \bibinfo {author} {\bibfnamefont
  {T.}~\bibnamefont {Tummuru}}, \bibinfo {author} {\bibfnamefont
  {M.}~\bibnamefont {Kim}}, \bibinfo {author} {\bibfnamefont {M.}~\bibnamefont
  {Franz}}, \bibinfo {author} {\bibfnamefont {J.~H.}\ \bibnamefont {Pixley}},
  \bibinfo {author} {\bibfnamefont {N.}~\bibnamefont {Poccia}},\ and\ \bibinfo
  {author} {\bibfnamefont {P.}~\bibnamefont {Kim}},\ }\bibfield  {title}
  {\bibinfo {title} {Time-reversal symmetry breaking superconductivity between
  twisted cuprate superconductors},\ }\href
  {https://doi.org/10.1126/science.abl8371} {\bibfield  {journal} {\bibinfo
  {journal} {Science}\ }\textbf {\bibinfo {volume} {382}},\ \bibinfo {pages}
  {1422} (\bibinfo {year} {2023})}\BibitemShut {NoStop}%
\bibitem [{\citenamefont {Zhu}\ \emph {et~al.}(2021)\citenamefont {Zhu},
  \citenamefont {Liao}, \citenamefont {Zhang}, \citenamefont {Xie},
  \citenamefont {Meng}, \citenamefont {Liu}, \citenamefont {Bai}, \citenamefont
  {Ji}, \citenamefont {Zhang}, \citenamefont {Jiang}, \citenamefont {Zhong},
  \citenamefont {Schneeloch}, \citenamefont {Gu}, \citenamefont {Gu},
  \citenamefont {Ma}, \citenamefont {Zhang},\ and\ \citenamefont
  {Xue}}]{Zhu.2021}%
  \BibitemOpen
  \bibfield  {author} {\bibinfo {author} {\bibfnamefont {Y.}~\bibnamefont
  {Zhu}}, \bibinfo {author} {\bibfnamefont {M.}~\bibnamefont {Liao}}, \bibinfo
  {author} {\bibfnamefont {Q.}~\bibnamefont {Zhang}}, \bibinfo {author}
  {\bibfnamefont {H.-Y.}\ \bibnamefont {Xie}}, \bibinfo {author} {\bibfnamefont
  {F.}~\bibnamefont {Meng}}, \bibinfo {author} {\bibfnamefont {Y.}~\bibnamefont
  {Liu}}, \bibinfo {author} {\bibfnamefont {Z.}~\bibnamefont {Bai}}, \bibinfo
  {author} {\bibfnamefont {S.}~\bibnamefont {Ji}}, \bibinfo {author}
  {\bibfnamefont {J.}~\bibnamefont {Zhang}}, \bibinfo {author} {\bibfnamefont
  {K.}~\bibnamefont {Jiang}}, \bibinfo {author} {\bibfnamefont
  {R.}~\bibnamefont {Zhong}}, \bibinfo {author} {\bibfnamefont
  {J.}~\bibnamefont {Schneeloch}}, \bibinfo {author} {\bibfnamefont
  {G.}~\bibnamefont {Gu}}, \bibinfo {author} {\bibfnamefont {L.}~\bibnamefont
  {Gu}}, \bibinfo {author} {\bibfnamefont {X.}~\bibnamefont {Ma}}, \bibinfo
  {author} {\bibfnamefont {D.}~\bibnamefont {Zhang}},\ and\ \bibinfo {author}
  {\bibfnamefont {Q.-K.}\ \bibnamefont {Xue}},\ }\bibfield  {title} {\bibinfo
  {title} {Presence of $s$-wave pairing in {Josephson} junctions made of
  twisted ultrathin {Bi$_2$Sr$_2$CaCu$_2$O$_{8+x}$} flakes},\ }\href
  {https://doi.org/10.1103/physrevx.11.031011} {\bibfield  {journal} {\bibinfo
  {journal} {Phys. Rev. X}\ }\textbf {\bibinfo {volume} {11}},\ \bibinfo
  {pages} {031011} (\bibinfo {year} {2021})}\BibitemShut {NoStop}%
\bibitem [{\citenamefont {Volkov}\ \emph {et~al.}(2023)\citenamefont {Volkov},
  \citenamefont {Wilson}, \citenamefont {Lucht},\ and\ \citenamefont
  {Pixley}}]{Volkov.2023}%
  \BibitemOpen
  \bibfield  {author} {\bibinfo {author} {\bibfnamefont {P.~A.}\ \bibnamefont
  {Volkov}}, \bibinfo {author} {\bibfnamefont {J.~H.}\ \bibnamefont {Wilson}},
  \bibinfo {author} {\bibfnamefont {K.~P.}\ \bibnamefont {Lucht}},\ and\
  \bibinfo {author} {\bibfnamefont {J.~H.}\ \bibnamefont {Pixley}},\ }\bibfield
   {title} {\bibinfo {title} {Magic angles and correlations in twisted nodal
  superconductors},\ }\href {https://doi.org/10.1103/PhysRevB.107.174506}
  {\bibfield  {journal} {\bibinfo  {journal} {Phys. Rev. B}\ }\textbf {\bibinfo
  {volume} {107}},\ \bibinfo {pages} {174506} (\bibinfo {year}
  {2023})}\BibitemShut {NoStop}%
\bibitem [{\citenamefont {Martini}\ \emph {et~al.}(2023)\citenamefont
  {Martini}, \citenamefont {Lee}, \citenamefont {Confalone}, \citenamefont
  {Shokri}, \citenamefont {Saggau}, \citenamefont {Wolf}, \citenamefont {Gu},
  \citenamefont {Watanabe}, \citenamefont {Taniguchi}, \citenamefont
  {Montemurro}, \citenamefont {Vinokur}, \citenamefont {Nielsch},\ and\
  \citenamefont {Poccia}}]{Martini.2023}%
  \BibitemOpen
  \bibfield  {author} {\bibinfo {author} {\bibfnamefont {M.}~\bibnamefont
  {Martini}}, \bibinfo {author} {\bibfnamefont {Y.}~\bibnamefont {Lee}},
  \bibinfo {author} {\bibfnamefont {T.}~\bibnamefont {Confalone}}, \bibinfo
  {author} {\bibfnamefont {S.}~\bibnamefont {Shokri}}, \bibinfo {author}
  {\bibfnamefont {C.~N.}\ \bibnamefont {Saggau}}, \bibinfo {author}
  {\bibfnamefont {D.}~\bibnamefont {Wolf}}, \bibinfo {author} {\bibfnamefont
  {G.}~\bibnamefont {Gu}}, \bibinfo {author} {\bibfnamefont {K.}~\bibnamefont
  {Watanabe}}, \bibinfo {author} {\bibfnamefont {T.}~\bibnamefont {Taniguchi}},
  \bibinfo {author} {\bibfnamefont {D.}~\bibnamefont {Montemurro}}, \bibinfo
  {author} {\bibfnamefont {V.~M.}\ \bibnamefont {Vinokur}}, \bibinfo {author}
  {\bibfnamefont {K.}~\bibnamefont {Nielsch}},\ and\ \bibinfo {author}
  {\bibfnamefont {N.}~\bibnamefont {Poccia}},\ }\bibfield  {title} {\bibinfo
  {title} {Twisted cuprate van der {Waals} heterostructures with controlled
  {Josephson} coupling},\ }\href {https://doi.org/10.1016/j.mattod.2023.06.007}
  {\bibfield  {journal} {\bibinfo  {journal} {Mater. Today}\ }\textbf {\bibinfo
  {volume} {67}},\ \bibinfo {pages} {106} (\bibinfo {year} {2023})}\BibitemShut
  {NoStop}%
\bibitem [{\citenamefont {Wu}\ \emph {et~al.}(2025)\citenamefont {Wu},
  \citenamefont {Hao}, \citenamefont {Chen}, \citenamefont {Cai}, \citenamefont
  {Wu}, \citenamefont {Chen}, \citenamefont {Wang}, \citenamefont {Ming},
  \citenamefont {Johnston}, \citenamefont {Zhang},\ and\ \citenamefont
  {Weitering}}]{Wu.2025}%
  \BibitemOpen
  \bibfield  {author} {\bibinfo {author} {\bibfnamefont {X.}~\bibnamefont
  {Wu}}, \bibinfo {author} {\bibfnamefont {X.}~\bibnamefont {Hao}}, \bibinfo
  {author} {\bibfnamefont {Z.}~\bibnamefont {Chen}}, \bibinfo {author}
  {\bibfnamefont {Y.}~\bibnamefont {Cai}}, \bibinfo {author} {\bibfnamefont
  {M.}~\bibnamefont {Wu}}, \bibinfo {author} {\bibfnamefont {C.}~\bibnamefont
  {Chen}}, \bibinfo {author} {\bibfnamefont {K.}~\bibnamefont {Wang}}, \bibinfo
  {author} {\bibfnamefont {F.}~\bibnamefont {Ming}}, \bibinfo {author}
  {\bibfnamefont {S.}~\bibnamefont {Johnston}}, \bibinfo {author}
  {\bibfnamefont {R.-X.}\ \bibnamefont {Zhang}},\ and\ \bibinfo {author}
  {\bibfnamefont {H.~H.}\ \bibnamefont {Weitering}},\ }\bibfield  {title}
  {\bibinfo {title} {Microscopic fingerprint of chiral superconductivity},\
  }\href {https://doi.org/10.48550/arxiv.2507.18693} {\bibfield  {journal}
  {\bibinfo  {journal} {arXiv:2507.18693}\ } (\bibinfo {year}
  {2025})}\BibitemShut {NoStop}%
\bibitem [{\citenamefont {Ming}\ \emph {et~al.}(2017)\citenamefont {Ming},
  \citenamefont {Johnston}, \citenamefont {Mulugeta}, \citenamefont {Smith},
  \citenamefont {Vilmercati}, \citenamefont {Lee}, \citenamefont {Maier},
  \citenamefont {Snijders},\ and\ \citenamefont {Weitering}}]{Ming.2017}%
  \BibitemOpen
  \bibfield  {author} {\bibinfo {author} {\bibfnamefont {F.}~\bibnamefont
  {Ming}}, \bibinfo {author} {\bibfnamefont {S.}~\bibnamefont {Johnston}},
  \bibinfo {author} {\bibfnamefont {D.}~\bibnamefont {Mulugeta}}, \bibinfo
  {author} {\bibfnamefont {T.~S.}\ \bibnamefont {Smith}}, \bibinfo {author}
  {\bibfnamefont {P.}~\bibnamefont {Vilmercati}}, \bibinfo {author}
  {\bibfnamefont {G.}~\bibnamefont {Lee}}, \bibinfo {author} {\bibfnamefont
  {T.~A.}\ \bibnamefont {Maier}}, \bibinfo {author} {\bibfnamefont {P.~C.}\
  \bibnamefont {Snijders}},\ and\ \bibinfo {author} {\bibfnamefont {H.~H.}\
  \bibnamefont {Weitering}},\ }\bibfield  {title} {\bibinfo {title}
  {Realization of a hole-doped {Mott} insulator on a triangular silicon
  lattice},\ }\href {https://doi.org/10.1103/PhysRevLett.119.266802} {\bibfield
   {journal} {\bibinfo  {journal} {Phys. Rev. Lett.}\ }\textbf {\bibinfo
  {volume} {119}},\ \bibinfo {pages} {266802} (\bibinfo {year}
  {2017})}\BibitemShut {NoStop}%
\bibitem [{\citenamefont {Ming}\ \emph {et~al.}(2023)\citenamefont {Ming},
  \citenamefont {Wu}, \citenamefont {Chen}, \citenamefont {Wang}, \citenamefont
  {Mai}, \citenamefont {Maier}, \citenamefont {Strockoz}, \citenamefont
  {Venderbos}, \citenamefont {Gonz\'{a}lez}, \citenamefont {Ortega},
  \citenamefont {Johnston},\ and\ \citenamefont {Weitering}}]{Ming.2023}%
  \BibitemOpen
  \bibfield  {author} {\bibinfo {author} {\bibfnamefont {F.}~\bibnamefont
  {Ming}}, \bibinfo {author} {\bibfnamefont {X.}~\bibnamefont {Wu}}, \bibinfo
  {author} {\bibfnamefont {C.}~\bibnamefont {Chen}}, \bibinfo {author}
  {\bibfnamefont {K.~D.}\ \bibnamefont {Wang}}, \bibinfo {author}
  {\bibfnamefont {P.}~\bibnamefont {Mai}}, \bibinfo {author} {\bibfnamefont
  {T.~A.}\ \bibnamefont {Maier}}, \bibinfo {author} {\bibfnamefont
  {J.}~\bibnamefont {Strockoz}}, \bibinfo {author} {\bibfnamefont {J.~W.~F.}\
  \bibnamefont {Venderbos}}, \bibinfo {author} {\bibfnamefont {C.}~\bibnamefont
  {Gonz\'{a}lez}}, \bibinfo {author} {\bibfnamefont {J.}~\bibnamefont
  {Ortega}}, \bibinfo {author} {\bibfnamefont {S.}~\bibnamefont {Johnston}},\
  and\ \bibinfo {author} {\bibfnamefont {H.~H.}\ \bibnamefont {Weitering}},\
  }\bibfield  {title} {\bibinfo {title} {Evidence for chiral superconductivity
  on a silicon surface},\ }\href {https://doi.org/10.1038/s41567-022-01889-1}
  {\bibfield  {journal} {\bibinfo  {journal} {Nat. Phys.}\ }\textbf {\bibinfo
  {volume} {19}},\ \bibinfo {pages} {500} (\bibinfo {year} {2023})}\BibitemShut
  {NoStop}%
\bibitem [{\citenamefont {Kennes}\ \emph {et~al.}(2018)\citenamefont {Kennes},
  \citenamefont {Lischner},\ and\ \citenamefont {Karrasch}}]{Kennes.2018}%
  \BibitemOpen
  \bibfield  {author} {\bibinfo {author} {\bibfnamefont {D.~M.}\ \bibnamefont
  {Kennes}}, \bibinfo {author} {\bibfnamefont {J.}~\bibnamefont {Lischner}},\
  and\ \bibinfo {author} {\bibfnamefont {C.}~\bibnamefont {Karrasch}},\
  }\bibfield  {title} {\bibinfo {title} {Strong correlations and $d+id$
  superconductivity in twisted bilayer graphene},\ }\href
  {https://doi.org/10.1103/physrevb.98.241407} {\bibfield  {journal} {\bibinfo
  {journal} {Phys. Rev. B}\ }\textbf {\bibinfo {volume} {98}},\ \bibinfo
  {pages} {241407} (\bibinfo {year} {2018})}\BibitemShut {NoStop}%
\bibitem [{\citenamefont {Törmä}\ \emph {et~al.}(2022)\citenamefont
  {Törmä}, \citenamefont {Peotta},\ and\ \citenamefont
  {Bernevig}}]{Törmä.2022}%
  \BibitemOpen
  \bibfield  {author} {\bibinfo {author} {\bibfnamefont {P.}~\bibnamefont
  {Törmä}}, \bibinfo {author} {\bibfnamefont {S.}~\bibnamefont {Peotta}},\
  and\ \bibinfo {author} {\bibfnamefont {B.~A.}\ \bibnamefont {Bernevig}},\
  }\bibfield  {title} {\bibinfo {title} {Superconductivity, superfluidity and
  quantum geometry in twisted multilayer systems},\ }\href
  {https://doi.org/10.1038/s42254-022-00466-y} {\bibfield  {journal} {\bibinfo
  {journal} {Nat. Rev. Phys.}\ }\textbf {\bibinfo {volume} {4}},\ \bibinfo
  {pages} {528} (\bibinfo {year} {2022})}\BibitemShut {NoStop}%
\bibitem [{\citenamefont {Pantaleón}\ \emph {et~al.}(2023)\citenamefont
  {Pantaleón}, \citenamefont {Jimeno-Pozo}, \citenamefont {Sainz-Cruz},
  \citenamefont {Phong}, \citenamefont {Cea},\ and\ \citenamefont
  {Guinea}}]{Pantaleón.2023}%
  \BibitemOpen
  \bibfield  {author} {\bibinfo {author} {\bibfnamefont {P.~A.}\ \bibnamefont
  {Pantaleón}}, \bibinfo {author} {\bibfnamefont {A.}~\bibnamefont
  {Jimeno-Pozo}}, \bibinfo {author} {\bibfnamefont {H.}~\bibnamefont
  {Sainz-Cruz}}, \bibinfo {author} {\bibfnamefont {V.~T.}\ \bibnamefont
  {Phong}}, \bibinfo {author} {\bibfnamefont {T.}~\bibnamefont {Cea}},\ and\
  \bibinfo {author} {\bibfnamefont {F.}~\bibnamefont {Guinea}},\ }\bibfield
  {title} {\bibinfo {title} {Superconductivity and correlated phases in
  non-twisted bilayer and trilayer graphene},\ }\href
  {https://doi.org/10.1038/s42254-023-00575-2} {\bibfield  {journal} {\bibinfo
  {journal} {Nat. Rev. Phys.}\ }\textbf {\bibinfo {volume} {5}},\ \bibinfo
  {pages} {304} (\bibinfo {year} {2023})}\BibitemShut {NoStop}%
\bibitem [{\citenamefont {Akbar}\ \emph {et~al.}(2024)\citenamefont {Akbar},
  \citenamefont {Biborski}, \citenamefont {Rademaker},\ and\ \citenamefont
  {Zegrodnik}}]{Akbar.2024}%
  \BibitemOpen
  \bibfield  {author} {\bibinfo {author} {\bibfnamefont {W.}~\bibnamefont
  {Akbar}}, \bibinfo {author} {\bibfnamefont {A.}~\bibnamefont {Biborski}},
  \bibinfo {author} {\bibfnamefont {L.}~\bibnamefont {Rademaker}},\ and\
  \bibinfo {author} {\bibfnamefont {M.}~\bibnamefont {Zegrodnik}},\ }\bibfield
  {title} {\bibinfo {title} {Topological superconductivity with mixed
  singlet-triplet pairing in moir{\'{e}} transition metal dichalcogenide
  bilayers},\ }\href {https://doi.org/10.1103/physrevb.110.064516} {\bibfield
  {journal} {\bibinfo  {journal} {Phys. Rev. B}\ }\textbf {\bibinfo {volume}
  {110}},\ \bibinfo {pages} {064516} (\bibinfo {year} {2024})}\BibitemShut
  {NoStop}%
\bibitem [{\citenamefont {Xia}\ \emph {et~al.}(2025)\citenamefont {Xia},
  \citenamefont {Han}, \citenamefont {Watanabe}, \citenamefont {Taniguchi},
  \citenamefont {Shan},\ and\ \citenamefont {Mak}}]{Xia.2025}%
  \BibitemOpen
  \bibfield  {author} {\bibinfo {author} {\bibfnamefont {Y.}~\bibnamefont
  {Xia}}, \bibinfo {author} {\bibfnamefont {Z.}~\bibnamefont {Han}}, \bibinfo
  {author} {\bibfnamefont {K.}~\bibnamefont {Watanabe}}, \bibinfo {author}
  {\bibfnamefont {T.}~\bibnamefont {Taniguchi}}, \bibinfo {author}
  {\bibfnamefont {J.}~\bibnamefont {Shan}},\ and\ \bibinfo {author}
  {\bibfnamefont {K.~F.}\ \bibnamefont {Mak}},\ }\bibfield  {title} {\bibinfo
  {title} {Superconductivity in twisted bilayer {WSe$_2$}},\ }\href
  {https://doi.org/10.1038/s41586-024-08116-2} {\bibfield  {journal} {\bibinfo
  {journal} {Nature}\ }\textbf {\bibinfo {volume} {637}},\ \bibinfo {pages}
  {833} (\bibinfo {year} {2025})}\BibitemShut {NoStop}%
\bibitem [{\citenamefont {Guo}\ \emph {et~al.}(2025)\citenamefont {Guo},
  \citenamefont {Pack}, \citenamefont {Swann}, \citenamefont {Holtzman},
  \citenamefont {Cothrine}, \citenamefont {Watanabe}, \citenamefont
  {Taniguchi}, \citenamefont {Mandrus}, \citenamefont {Barmak}, \citenamefont
  {Hone}, \citenamefont {Millis}, \citenamefont {Pasupathy},\ and\
  \citenamefont {Dean}}]{Guo.2025}%
  \BibitemOpen
  \bibfield  {author} {\bibinfo {author} {\bibfnamefont {Y.}~\bibnamefont
  {Guo}}, \bibinfo {author} {\bibfnamefont {J.}~\bibnamefont {Pack}}, \bibinfo
  {author} {\bibfnamefont {J.}~\bibnamefont {Swann}}, \bibinfo {author}
  {\bibfnamefont {L.}~\bibnamefont {Holtzman}}, \bibinfo {author}
  {\bibfnamefont {M.}~\bibnamefont {Cothrine}}, \bibinfo {author}
  {\bibfnamefont {K.}~\bibnamefont {Watanabe}}, \bibinfo {author}
  {\bibfnamefont {T.}~\bibnamefont {Taniguchi}}, \bibinfo {author}
  {\bibfnamefont {D.~G.}\ \bibnamefont {Mandrus}}, \bibinfo {author}
  {\bibfnamefont {K.}~\bibnamefont {Barmak}}, \bibinfo {author} {\bibfnamefont
  {J.}~\bibnamefont {Hone}}, \bibinfo {author} {\bibfnamefont {A.~J.}\
  \bibnamefont {Millis}}, \bibinfo {author} {\bibfnamefont {A.}~\bibnamefont
  {Pasupathy}},\ and\ \bibinfo {author} {\bibfnamefont {C.~R.}\ \bibnamefont
  {Dean}},\ }\bibfield  {title} {\bibinfo {title} {Superconductivity in
  5.0\textdegree{} twisted bilayer {WSe$_2$}},\ }\href
  {https://doi.org/10.1038/s41586-024-08381-1} {\bibfield  {journal} {\bibinfo
  {journal} {Nature}\ }\textbf {\bibinfo {volume} {637}},\ \bibinfo {pages}
  {839} (\bibinfo {year} {2025})}\BibitemShut {NoStop}%
\bibitem [{\citenamefont {R\o{}ising}\ \emph {et~al.}(2019)\citenamefont
  {R\o{}ising}, \citenamefont {Scaffidi}, \citenamefont {Flicker},
  \citenamefont {Lange},\ and\ \citenamefont {Simon}}]{Roising.2019}%
  \BibitemOpen
  \bibfield  {author} {\bibinfo {author} {\bibfnamefont {H.~S.}\ \bibnamefont
  {R\o{}ising}}, \bibinfo {author} {\bibfnamefont {T.}~\bibnamefont
  {Scaffidi}}, \bibinfo {author} {\bibfnamefont {F.}~\bibnamefont {Flicker}},
  \bibinfo {author} {\bibfnamefont {G.~F.}\ \bibnamefont {Lange}},\ and\
  \bibinfo {author} {\bibfnamefont {S.~H.}\ \bibnamefont {Simon}},\ }\bibfield
  {title} {\bibinfo {title} {Superconducting order of
  $\mathrm{Sr}_{2}\mathrm{RuO}_{4}$ from a three-dimensional microscopic
  model},\ }\href {https://doi.org/10.1103/PhysRevResearch.1.033108} {\bibfield
   {journal} {\bibinfo  {journal} {Phys. Rev. Res.}\ }\textbf {\bibinfo
  {volume} {1}},\ \bibinfo {pages} {033108} (\bibinfo {year}
  {2019})}\BibitemShut {NoStop}%
\bibitem [{\citenamefont {Grinenko}\ \emph {et~al.}(2021)\citenamefont
  {Grinenko}, \citenamefont {Das}, \citenamefont {Gupta}, \citenamefont
  {Zinkl}, \citenamefont {Kikugawa}, \citenamefont {Maeno}, \citenamefont
  {Hicks}, \citenamefont {Klauss}, \citenamefont {Sigrist},\ and\ \citenamefont
  {Khasanov}}]{Grinenko.2021}%
  \BibitemOpen
  \bibfield  {author} {\bibinfo {author} {\bibfnamefont {V.}~\bibnamefont
  {Grinenko}}, \bibinfo {author} {\bibfnamefont {D.}~\bibnamefont {Das}},
  \bibinfo {author} {\bibfnamefont {R.}~\bibnamefont {Gupta}}, \bibinfo
  {author} {\bibfnamefont {B.}~\bibnamefont {Zinkl}}, \bibinfo {author}
  {\bibfnamefont {N.}~\bibnamefont {Kikugawa}}, \bibinfo {author}
  {\bibfnamefont {Y.}~\bibnamefont {Maeno}}, \bibinfo {author} {\bibfnamefont
  {C.~W.}\ \bibnamefont {Hicks}}, \bibinfo {author} {\bibfnamefont {H.-H.}\
  \bibnamefont {Klauss}}, \bibinfo {author} {\bibfnamefont {M.}~\bibnamefont
  {Sigrist}},\ and\ \bibinfo {author} {\bibfnamefont {R.}~\bibnamefont
  {Khasanov}},\ }\bibfield  {title} {\bibinfo {title} {Unsplit superconducting
  and time reversal symmetry breaking transitions in
  $\mathrm{Sr}_{2}\mathrm{RuO}_{4}$ under hydrostatic pressure and disorder},\
  }\href {https://doi.org/10.1038/s41467-021-24176-8} {\bibfield  {journal}
  {\bibinfo  {journal} {Nat. Commun.}\ }\textbf {\bibinfo {volume} {12}},\
  \bibinfo {pages} {3920} (\bibinfo {year} {2021})}\BibitemShut {NoStop}%
\bibitem [{\citenamefont {Caroli}\ \emph {et~al.}(1964)\citenamefont {Caroli},
  \citenamefont {de~Gennes},\ and\ \citenamefont {Matricon}}]{Caroli.1964}%
  \BibitemOpen
  \bibfield  {author} {\bibinfo {author} {\bibfnamefont {C.}~\bibnamefont
  {Caroli}}, \bibinfo {author} {\bibfnamefont {P.~G.}\ \bibnamefont
  {de~Gennes}},\ and\ \bibinfo {author} {\bibfnamefont {J.}~\bibnamefont
  {Matricon}},\ }\bibfield  {title} {\bibinfo {title} {Bound fermion states on
  a vortex line in a type {II} superconductor},\ }\href
  {https://doi.org/10.1016/0031-9163(64)90375-0} {\bibfield  {journal}
  {\bibinfo  {journal} {Phys. Lett.}\ }\textbf {\bibinfo {volume} {9}},\
  \bibinfo {pages} {307} (\bibinfo {year} {1964})}\BibitemShut {NoStop}%
\bibitem [{\citenamefont {Ohashi}\ \emph {et~al.}(2021)\citenamefont {Ohashi},
  \citenamefont {Kobayashi},\ and\ \citenamefont {Tanaka}}]{Ohashi.2021}%
  \BibitemOpen
  \bibfield  {author} {\bibinfo {author} {\bibfnamefont {R.}~\bibnamefont
  {Ohashi}}, \bibinfo {author} {\bibfnamefont {S.}~\bibnamefont {Kobayashi}},\
  and\ \bibinfo {author} {\bibfnamefont {Y.}~\bibnamefont {Tanaka}},\
  }\bibfield  {title} {\bibinfo {title} {Possible topological phases in quantum
  anomalous {Hall} insulator/unconventional superconductor hybrid systems},\
  }\href {https://doi.org/10.1103/PhysRevB.104.134518} {\bibfield  {journal}
  {\bibinfo  {journal} {Phys. Rev. B}\ }\textbf {\bibinfo {volume} {104}},\
  \bibinfo {pages} {134518} (\bibinfo {year} {2021})}\BibitemShut {NoStop}%
\bibitem [{\citenamefont {Hasan}\ and\ \citenamefont
  {Kane}(2010)}]{Hasan.2010}%
  \BibitemOpen
  \bibfield  {author} {\bibinfo {author} {\bibfnamefont {M.~Z.}\ \bibnamefont
  {Hasan}}\ and\ \bibinfo {author} {\bibfnamefont {C.~L.}\ \bibnamefont
  {Kane}},\ }\bibfield  {title} {\bibinfo {title} {Colloquium: {T}opological
  insulators},\ }\href {https://doi.org/10.1103/revmodphys.82.3045} {\bibfield
  {journal} {\bibinfo  {journal} {Rev. Mod. Phys.}\ }\textbf {\bibinfo {volume}
  {82}},\ \bibinfo {pages} {3045 } (\bibinfo {year} {2010})}\BibitemShut
  {NoStop}%
\bibitem [{\citenamefont {Teo}\ and\ \citenamefont
  {Kane}(2010)}]{Teo.Kane.2008}%
  \BibitemOpen
  \bibfield  {author} {\bibinfo {author} {\bibfnamefont {J.~C.~Y.}\
  \bibnamefont {Teo}}\ and\ \bibinfo {author} {\bibfnamefont {C.~L.}\
  \bibnamefont {Kane}},\ }\bibfield  {title} {\bibinfo {title} {Topological
  defects and gapless modes in insulators and superconductors},\ }\href
  {https://doi.org/10.1103/PhysRevB.82.115120} {\bibfield  {journal} {\bibinfo
  {journal} {Phys. Rev. B}\ }\textbf {\bibinfo {volume} {82}},\ \bibinfo
  {pages} {115120} (\bibinfo {year} {2010})}\BibitemShut {NoStop}%
\bibitem [{Note1()}]{Note1}%
  \BibitemOpen
  \bibinfo {note} {Note that this transformation should have a smooth gauge;
  otherwise new sources of Berry curvature are introduced in the normal part of
  the BdG Hamiltonian. In that case, the winding of the order parameter would
  no longer be equal to the Chern number.}\BibitemShut {Stop}%
\bibitem [{\citenamefont {Fukui}\ \emph {et~al.}(2005)\citenamefont {Fukui},
  \citenamefont {Hatsugai},\ and\ \citenamefont {Suzuki}}]{Fukui.2005}%
  \BibitemOpen
  \bibfield  {author} {\bibinfo {author} {\bibfnamefont {T.}~\bibnamefont
  {Fukui}}, \bibinfo {author} {\bibfnamefont {Y.}~\bibnamefont {Hatsugai}},\
  and\ \bibinfo {author} {\bibfnamefont {H.}~\bibnamefont {Suzuki}},\
  }\bibfield  {title} {\bibinfo {title} {{Chern} numbers in discretized
  {Brillouin} zone: Efficient method of computing (spin) {Hall} conductances},\
  }\href {https://doi.org/10.1143/JPSJ.74.1674} {\bibfield  {journal} {\bibinfo
   {journal} {J. Phys. Soc. Jpn.}\ }\textbf {\bibinfo {volume} {74}},\ \bibinfo
  {pages} {1674} (\bibinfo {year} {2005})}\BibitemShut {NoStop}%
\bibitem [{\citenamefont {Read}\ and\ \citenamefont {Green}(2000)}]{Read.2000}%
  \BibitemOpen
  \bibfield  {author} {\bibinfo {author} {\bibfnamefont {N.}~\bibnamefont
  {Read}}\ and\ \bibinfo {author} {\bibfnamefont {D.}~\bibnamefont {Green}},\
  }\bibfield  {title} {\bibinfo {title} {Paired states of fermions in two
  dimensions with breaking of parity and time-reversal symmetries and the
  fractional quantum {Hall} effect},\ }\href
  {https://doi.org/10.1103/PhysRevB.61.10267} {\bibfield  {journal} {\bibinfo
  {journal} {Phys. Rev. B}\ }\textbf {\bibinfo {volume} {61}},\ \bibinfo
  {pages} {10267} (\bibinfo {year} {2000})}\BibitemShut {NoStop}%
\bibitem [{\citenamefont {Fukui}(2010)}]{Fukui.2010}%
  \BibitemOpen
  \bibfield  {author} {\bibinfo {author} {\bibfnamefont {T.}~\bibnamefont
  {Fukui}},\ }\bibfield  {title} {\bibinfo {title} {Majorana zero modes bound
  to a vortex line in a topological superconductor},\ }\href
  {https://doi.org/10.1103/PhysRevB.81.214516} {\bibfield  {journal} {\bibinfo
  {journal} {Phys. Rev. B}\ }\textbf {\bibinfo {volume} {81}},\ \bibinfo
  {pages} {214516} (\bibinfo {year} {2010})}\BibitemShut {NoStop}%
\bibitem [{\citenamefont {Chamon}\ \emph {et~al.}(2010)\citenamefont {Chamon},
  \citenamefont {Jackiw}, \citenamefont {Nishida}, \citenamefont {Pi},\ and\
  \citenamefont {Santos}}]{Chamon.2010}%
  \BibitemOpen
  \bibfield  {author} {\bibinfo {author} {\bibfnamefont {C.}~\bibnamefont
  {Chamon}}, \bibinfo {author} {\bibfnamefont {R.}~\bibnamefont {Jackiw}},
  \bibinfo {author} {\bibfnamefont {Y.}~\bibnamefont {Nishida}}, \bibinfo
  {author} {\bibfnamefont {S.-Y.}\ \bibnamefont {Pi}},\ and\ \bibinfo {author}
  {\bibfnamefont {L.}~\bibnamefont {Santos}},\ }\bibfield  {title} {\bibinfo
  {title} {Quantizing {Majorana} fermions in a superconductor},\ }\href
  {https://doi.org/10.1103/PhysRevB.81.224515} {\bibfield  {journal} {\bibinfo
  {journal} {Phys. Rev. B}\ }\textbf {\bibinfo {volume} {81}},\ \bibinfo
  {pages} {224515} (\bibinfo {year} {2010})}\BibitemShut {NoStop}%
\bibitem [{\citenamefont {Bernevig}\ and\ \citenamefont
  {Hughes}(2013)}]{Bernevig-Hughes.2013}%
  \BibitemOpen
  \bibfield  {author} {\bibinfo {author} {\bibfnamefont {B.~A.}\ \bibnamefont
  {Bernevig}}\ and\ \bibinfo {author} {\bibfnamefont {T.~L.}\ \bibnamefont
  {Hughes}},\ }\href
  {https://press.princeton.edu/books/hardcover/9780691151755/topological-insulators-and-topological-superconductors}
  {\emph {\bibinfo {title} {Topological insulators and topological
  superconductors}}}\ (\bibinfo  {publisher} {Princeton University Press},\
  \bibinfo {address} {Princeton},\ \bibinfo {year} {2013})\BibitemShut
  {NoStop}%
\bibitem [{\citenamefont {Jackiw}\ and\ \citenamefont
  {Rebbi}(1976)}]{Jackiw.1976}%
  \BibitemOpen
  \bibfield  {author} {\bibinfo {author} {\bibfnamefont {R.}~\bibnamefont
  {Jackiw}}\ and\ \bibinfo {author} {\bibfnamefont {C.}~\bibnamefont {Rebbi}},\
  }\bibfield  {title} {\bibinfo {title} {Solitons with fermion number $1/2$},\
  }\href {https://doi.org/10.1103/PhysRevD.13.3398} {\bibfield  {journal}
  {\bibinfo  {journal} {Phys. Rev. D}\ }\textbf {\bibinfo {volume} {13}},\
  \bibinfo {pages} {3398} (\bibinfo {year} {1976})}\BibitemShut {NoStop}%
\bibitem [{\citenamefont {Jackiw}\ and\ \citenamefont
  {Rossi}(1981)}]{Jackiw.1981}%
  \BibitemOpen
  \bibfield  {author} {\bibinfo {author} {\bibfnamefont {R.}~\bibnamefont
  {Jackiw}}\ and\ \bibinfo {author} {\bibfnamefont {P.}~\bibnamefont {Rossi}},\
  }\bibfield  {title} {\bibinfo {title} {Zero modes of the vortex-fermion
  system},\ }\href {https://doi.org/10.1016/0550-3213(81)90044-4} {\bibfield
  {journal} {\bibinfo  {journal} {Nucl. Phys. B}\ }\textbf {\bibinfo {volume}
  {190}},\ \bibinfo {pages} {681} (\bibinfo {year} {1981})}\BibitemShut
  {NoStop}%
\bibitem [{\citenamefont {Volovik}(1999)}]{Volovik.1999}%
  \BibitemOpen
  \bibfield  {author} {\bibinfo {author} {\bibfnamefont {G.~E.}\ \bibnamefont
  {Volovik}},\ }\bibfield  {title} {\bibinfo {title} {Fermion zero modes on
  vortices in chiral superconductors},\ }\href
  {https://doi.org/10.1134/1.568223} {\bibfield  {journal} {\bibinfo  {journal}
  {JETP Lett.}\ }\textbf {\bibinfo {volume} {70}},\ \bibinfo {pages} {609}
  (\bibinfo {year} {1999})}\BibitemShut {NoStop}%
\bibitem [{\citenamefont {Zlotnikov}(2023)}]{Zlotnikov.2023}%
  \BibitemOpen
  \bibfield  {author} {\bibinfo {author} {\bibfnamefont {A.~O.}\ \bibnamefont
  {Zlotnikov}},\ }\bibfield  {title} {\bibinfo {title} {Majorana vortex modes
  in spin-singlet chiral superconductors with noncollinear spin ordering: Local
  density of states study},\ }\href
  {https://doi.org/10.1103/PhysRevB.107.144513} {\bibfield  {journal} {\bibinfo
   {journal} {Phys. Rev. B}\ }\textbf {\bibinfo {volume} {107}},\ \bibinfo
  {pages} {144513} (\bibinfo {year} {2023})}\BibitemShut {NoStop}%
\bibitem [{\citenamefont {Chiu}\ \emph {et~al.}(2016)\citenamefont {Chiu},
  \citenamefont {Teo}, \citenamefont {Schnyder},\ and\ \citenamefont
  {Ryu}}]{Chiu.2016}%
  \BibitemOpen
  \bibfield  {author} {\bibinfo {author} {\bibfnamefont {C.-K.}\ \bibnamefont
  {Chiu}}, \bibinfo {author} {\bibfnamefont {J.~C.~Y.}\ \bibnamefont {Teo}},
  \bibinfo {author} {\bibfnamefont {A.~P.}\ \bibnamefont {Schnyder}},\ and\
  \bibinfo {author} {\bibfnamefont {S.}~\bibnamefont {Ryu}},\ }\bibfield
  {title} {\bibinfo {title} {Classification of topological quantum matter with
  symmetries},\ }\href {https://doi.org/10.1103/RevModPhys.88.035005}
  {\bibfield  {journal} {\bibinfo  {journal} {Rev. Mod. Phys.}\ }\textbf
  {\bibinfo {volume} {88}},\ \bibinfo {pages} {035005} (\bibinfo {year}
  {2016})}\BibitemShut {NoStop}%
\bibitem [{\citenamefont {Pavarini}\ and\ \citenamefont
  {Koch}(2020)}]{Pavarini:884084}%
  \BibitemOpen
  \bibinfo {editor} {\bibfnamefont {E.}~\bibnamefont {Pavarini}}\ and\ \bibinfo
  {editor} {\bibfnamefont {E.}~\bibnamefont {Koch}},\ eds.,\ \href
  {https://juser.fz-juelich.de/record/884084} {\emph {\bibinfo {title}
  {{T}opology, {E}ntanglement, and {S}trong {C}orrelations}}},\ \bibinfo
  {series} {Schriften des Forschungszentrums Jülich. Reihe modeling and
  simulation}, Vol.~\bibinfo {volume} {10},\ \bibinfo {organization} {Autumn
  School on Correlated Electrons, Jülich (Germany), 21 Sep 2020 - 25 Sep
  2020}\ (\bibinfo  {publisher} {Forschungszentrum Jülich GmbH
  Zentralbibliothek, Verlag},\ \bibinfo {address} {Jülich},\ \bibinfo {year}
  {2020})\BibitemShut {NoStop}%
\bibitem [{\citenamefont {Ptok}\ \emph {et~al.}(2020)\citenamefont {Ptok},
  \citenamefont {Alspaugh}, \citenamefont {G\l{}odzik}, \citenamefont
  {Kobia\l{}ka}, \citenamefont {Ole\ifmmode~\acute{s}\else \'{s}\fi{}},
  \citenamefont {Simon},\ and\ \citenamefont {Piekarz}}]{Ptok.2020}%
  \BibitemOpen
  \bibfield  {author} {\bibinfo {author} {\bibfnamefont {A.}~\bibnamefont
  {Ptok}}, \bibinfo {author} {\bibfnamefont {D.~J.}\ \bibnamefont {Alspaugh}},
  \bibinfo {author} {\bibfnamefont {S.}~\bibnamefont {G\l{}odzik}}, \bibinfo
  {author} {\bibfnamefont {A.}~\bibnamefont {Kobia\l{}ka}}, \bibinfo {author}
  {\bibfnamefont {A.~M.}\ \bibnamefont {Ole\ifmmode~\acute{s}\else
  \'{s}\fi{}}}, \bibinfo {author} {\bibfnamefont {P.}~\bibnamefont {Simon}},\
  and\ \bibinfo {author} {\bibfnamefont {P.}~\bibnamefont {Piekarz}},\
  }\bibfield  {title} {\bibinfo {title} {Probing the chirality of
  one-dimensional {M}ajorana edge states around a two-dimensional nanoflake in
  a superconductor},\ }\href {https://doi.org/10.1103/PhysRevB.102.245405}
  {\bibfield  {journal} {\bibinfo  {journal} {Phys. Rev. B}\ }\textbf {\bibinfo
  {volume} {102}},\ \bibinfo {pages} {245405} (\bibinfo {year}
  {2020})}\BibitemShut {NoStop}%
\bibitem [{\citenamefont {Gygi}\ and\ \citenamefont
  {Schl{\"u}ter}(1990)}]{Gygi.1990}%
  \BibitemOpen
  \bibfield  {author} {\bibinfo {author} {\bibfnamefont {F.}~\bibnamefont
  {Gygi}}\ and\ \bibinfo {author} {\bibfnamefont {M.}~\bibnamefont
  {Schl{\"u}ter}},\ }\bibfield  {title} {\bibinfo {title} {Angular band
  structure of a vortex line in a type-{II} superconductor},\ }\href
  {https://doi.org/10.1103/PhysRevLett.65.1820} {\bibfield  {journal} {\bibinfo
   {journal} {Phys. Rev. Lett.}\ }\textbf {\bibinfo {volume} {65}},\ \bibinfo
  {pages} {1820} (\bibinfo {year} {1990})}\BibitemShut {NoStop}%
\bibitem [{\citenamefont {Hayashi}\ \emph {et~al.}(1998)\citenamefont
  {Hayashi}, \citenamefont {Isoshima}, \citenamefont {Ichioka},\ and\
  \citenamefont {Machida}}]{Hayashi.1998}%
  \BibitemOpen
  \bibfield  {author} {\bibinfo {author} {\bibfnamefont {N.}~\bibnamefont
  {Hayashi}}, \bibinfo {author} {\bibfnamefont {T.}~\bibnamefont {Isoshima}},
  \bibinfo {author} {\bibfnamefont {M.}~\bibnamefont {Ichioka}},\ and\ \bibinfo
  {author} {\bibfnamefont {K.}~\bibnamefont {Machida}},\ }\bibfield  {title}
  {\bibinfo {title} {Low-lying quasiparticle excitations around a vortex core
  in quantum limit},\ }\href {https://doi.org/10.1103/PhysRevLett.80.2921}
  {\bibfield  {journal} {\bibinfo  {journal} {Phys. Rev. Lett.}\ }\textbf
  {\bibinfo {volume} {80}},\ \bibinfo {pages} {2921} (\bibinfo {year}
  {1998})}\BibitemShut {NoStop}%
\bibitem [{\citenamefont {Franz}\ and\ \citenamefont
  {Te{\v{s}}anovi{\'{c}}}(1998)}]{Franz.1998}%
  \BibitemOpen
  \bibfield  {author} {\bibinfo {author} {\bibfnamefont {M.}~\bibnamefont
  {Franz}}\ and\ \bibinfo {author} {\bibfnamefont {Z.}~\bibnamefont
  {Te{\v{s}}anovi{\'{c}}}},\ }\bibfield  {title} {\bibinfo {title}
  {Self-consistent electronic structure of a $d_{x^2-y^2}$ and a
  $d_{x^2-y^2}+id_{xy}$ vortex},\ }\href
  {https://doi.org/10.1103/PhysRevLett.80.4763} {\bibfield  {journal} {\bibinfo
   {journal} {Phys. Rev. Lett.}\ }\textbf {\bibinfo {volume} {80}},\ \bibinfo
  {pages} {4763} (\bibinfo {year} {1998})}\BibitemShut {NoStop}%
\bibitem [{\citenamefont {Weisse}\ \emph {et~al.}(2006)\citenamefont {Weisse},
  \citenamefont {Wellein}, \citenamefont {Alvermann},\ and\ \citenamefont
  {Fehske}}]{Weisse.2006}%
  \BibitemOpen
  \bibfield  {author} {\bibinfo {author} {\bibfnamefont {A.}~\bibnamefont
  {Weisse}}, \bibinfo {author} {\bibfnamefont {G.}~\bibnamefont {Wellein}},
  \bibinfo {author} {\bibfnamefont {A.}~\bibnamefont {Alvermann}},\ and\
  \bibinfo {author} {\bibfnamefont {H.}~\bibnamefont {Fehske}},\ }\bibfield
  {title} {\bibinfo {title} {The kernel polynomial method},\ }\href
  {https://doi.org/10.1103/RevModPhys.78.275} {\bibfield  {journal} {\bibinfo
  {journal} {Rev. Mod. Phys.}\ }\textbf {\bibinfo {volume} {78}},\ \bibinfo
  {pages} {275} (\bibinfo {year} {2006})}\BibitemShut {NoStop}%
\bibitem [{\citenamefont {Covaci}\ \emph {et~al.}(2010)\citenamefont {Covaci},
  \citenamefont {Peeters},\ and\ \citenamefont {Berciu}}]{Covaci.2010}%
  \BibitemOpen
  \bibfield  {author} {\bibinfo {author} {\bibfnamefont {L.}~\bibnamefont
  {Covaci}}, \bibinfo {author} {\bibfnamefont {F.~M.}\ \bibnamefont
  {Peeters}},\ and\ \bibinfo {author} {\bibfnamefont {M.}~\bibnamefont
  {Berciu}},\ }\bibfield  {title} {\bibinfo {title} {Efficient numerical
  approach to inhomogeneous superconductivity: The {Chebyshev-Bogoliubov–de
  Gennes} method},\ }\href {https://doi.org/10.1103/physrevlett.105.167006}
  {\bibfield  {journal} {\bibinfo  {journal} {Phys. Rev. Lett.}\ }\textbf
  {\bibinfo {volume} {105}},\ \bibinfo {pages} {167006} (\bibinfo {year}
  {2010})}\BibitemShut {NoStop}%
\bibitem [{\citenamefont {Beenakker}\ \emph {et~al.}(2023)\citenamefont
  {Beenakker}, \citenamefont {Don{\'\i}s~Vela}, \citenamefont {Lemut},
  \citenamefont {Pacholski},\ and\ \citenamefont
  {Tworzyd{\l}o}}]{Beenakker.2023}%
  \BibitemOpen
  \bibfield  {author} {\bibinfo {author} {\bibfnamefont {C.~W.}\ \bibnamefont
  {Beenakker}}, \bibinfo {author} {\bibfnamefont {A.}~\bibnamefont
  {Don{\'\i}s~Vela}}, \bibinfo {author} {\bibfnamefont {G.}~\bibnamefont
  {Lemut}}, \bibinfo {author} {\bibfnamefont {M.}~\bibnamefont {Pacholski}},\
  and\ \bibinfo {author} {\bibfnamefont {J.}~\bibnamefont {Tworzyd{\l}o}},\
  }\bibfield  {title} {\bibinfo {title} {Tangent fermions: {Dirac} or
  {Majorana} fermions on a lattice without fermion doubling},\ }\href
  {https://doi.org/10.1002/andp.202300081} {\bibfield  {journal} {\bibinfo
  {journal} {Ann. Phys.}\ }\textbf {\bibinfo {volume} {535}},\ \bibinfo {pages}
  {2300081} (\bibinfo {year} {2023})}\BibitemShut {NoStop}%
\bibitem [{\citenamefont {Berthod}(2016)}]{Berthod.2016}%
  \BibitemOpen
  \bibfield  {author} {\bibinfo {author} {\bibfnamefont {C.}~\bibnamefont
  {Berthod}},\ }\bibfield  {title} {\bibinfo {title} {Vortex spectroscopy in
  the vortex glass: {A} real-space numerical approach},\ }\href
  {https://doi.org/10.1103/physrevb.94.184510} {\bibfield  {journal} {\bibinfo
  {journal} {Phys. Rev. B}\ }\textbf {\bibinfo {volume} {94}},\ \bibinfo
  {pages} {184510} (\bibinfo {year} {2016})}\BibitemShut {NoStop}%
\bibitem [{\citenamefont {Berthod}(2018)}]{Berthod.2018}%
  \BibitemOpen
  \bibfield  {author} {\bibinfo {author} {\bibfnamefont {C.}~\bibnamefont
  {Berthod}},\ }\bibfield  {title} {\bibinfo {title} {Signatures of nodeless
  multiband superconductivity and particle-hole crossover in the vortex cores
  of {FeTe$_{0.55}$Se$_{0.45}$}},\ }\href
  {https://doi.org/10.1103/PhysRevB.98.144519} {\bibfield  {journal} {\bibinfo
  {journal} {Phys. Rev. B}\ }\textbf {\bibinfo {volume} {98}},\ \bibinfo
  {pages} {144519} (\bibinfo {year} {2018})}\BibitemShut {NoStop}%
\bibitem [{Note2()}]{Note2}%
  \BibitemOpen
  \bibinfo {note} {On a square lattice, the winding number conditions are all
  modulo 4, so the precise condition is $2n_R~\protect \mathrm {mod}~4 =
  0$.}\BibitemShut {Stop}%
\bibitem [{\citenamefont {Fang}\ \emph {et~al.}(2014)\citenamefont {Fang},
  \citenamefont {Gilbert},\ and\ \citenamefont {Bernevig}}]{Fang.2014}%
  \BibitemOpen
  \bibfield  {author} {\bibinfo {author} {\bibfnamefont {C.}~\bibnamefont
  {Fang}}, \bibinfo {author} {\bibfnamefont {M.~J.}\ \bibnamefont {Gilbert}},\
  and\ \bibinfo {author} {\bibfnamefont {B.~A.}\ \bibnamefont {Bernevig}},\
  }\bibfield  {title} {\bibinfo {title} {New class of topological
  superconductors protected by magnetic group symmetries},\ }\href
  {https://doi.org/http://dx.doi.org/10.1103/PhysRevLett.112.106401} {\bibfield
   {journal} {\bibinfo  {journal} {Phys. Rev. Lett.}\ }\textbf {\bibinfo
  {volume} {112}},\ \bibinfo {pages} {106401} (\bibinfo {year}
  {2014})}\BibitemShut {NoStop}%
\bibitem [{\citenamefont {Liu}\ \emph {et~al.}(2014)\citenamefont {Liu},
  \citenamefont {He},\ and\ \citenamefont {Law}}]{Liu.2014}%
  \BibitemOpen
  \bibfield  {author} {\bibinfo {author} {\bibfnamefont {X.-J.}\ \bibnamefont
  {Liu}}, \bibinfo {author} {\bibfnamefont {J.~J.}\ \bibnamefont {He}},\ and\
  \bibinfo {author} {\bibfnamefont {K.~T.}\ \bibnamefont {Law}},\ }\bibfield
  {title} {\bibinfo {title} {Demonstrating lattice symmetry protection in
  topological crystalline superconductors},\ }\href
  {https://doi.org/10.1103/PhysRevB.90.235141} {\bibfield  {journal} {\bibinfo
  {journal} {Phys. Rev. B}\ }\textbf {\bibinfo {volume} {90}},\ \bibinfo
  {pages} {235141} (\bibinfo {year} {2014})}\BibitemShut {NoStop}%
\bibitem [{\citenamefont {K\"{o}nig}\ and\ \citenamefont
  {Coleman}(2019)}]{Konig.2019}%
  \BibitemOpen
  \bibfield  {author} {\bibinfo {author} {\bibfnamefont {E.~J.}\ \bibnamefont
  {K\"{o}nig}}\ and\ \bibinfo {author} {\bibfnamefont {P.}~\bibnamefont
  {Coleman}},\ }\bibfield  {title} {\bibinfo {title}
  {Crystalline-symmetry-protected helical {Majorana} modes in the iron
  pnictides},\ }\href {https://doi.org/10.1103/PhysRevLett.122.207001}
  {\bibfield  {journal} {\bibinfo  {journal} {Phys. Rev. Lett}\ }\textbf
  {\bibinfo {volume} {122}},\ \bibinfo {pages} {207001} (\bibinfo {year}
  {2019})}\BibitemShut {NoStop}%
\bibitem [{\citenamefont {Qin}\ \emph {et~al.}(2019)\citenamefont {Qin},
  \citenamefont {Hu}, \citenamefont {Le}, \citenamefont {Zeng}, \citenamefont
  {Zhang}, \citenamefont {Fang},\ and\ \citenamefont {Hu}}]{Qin.2019}%
  \BibitemOpen
  \bibfield  {author} {\bibinfo {author} {\bibfnamefont {S.}~\bibnamefont
  {Qin}}, \bibinfo {author} {\bibfnamefont {L.}~\bibnamefont {Hu}}, \bibinfo
  {author} {\bibfnamefont {C.}~\bibnamefont {Le}}, \bibinfo {author}
  {\bibfnamefont {J.}~\bibnamefont {Zeng}}, \bibinfo {author} {\bibfnamefont
  {F.-c.}\ \bibnamefont {Zhang}}, \bibinfo {author} {\bibfnamefont
  {C.}~\bibnamefont {Fang}},\ and\ \bibinfo {author} {\bibfnamefont
  {J.}~\bibnamefont {Hu}},\ }\bibfield  {title} {\bibinfo {title} {Quasi-{1D}
  topological nodal vortex line phase in doped superconducting {3D} {Dirac}
  semimetals},\ }\href {https://doi.org/10.1103/PhysRevLett.123.027003}
  {\bibfield  {journal} {\bibinfo  {journal} {Phys. Rev. Lett}\ }\textbf
  {\bibinfo {volume} {123}},\ \bibinfo {pages} {027003} (\bibinfo {year}
  {2019})}\BibitemShut {NoStop}%
\bibitem [{\citenamefont {Kobayashi}\ and\ \citenamefont
  {Furusaki}(2020)}]{Kobayashi.2020}%
  \BibitemOpen
  \bibfield  {author} {\bibinfo {author} {\bibfnamefont {S.}~\bibnamefont
  {Kobayashi}}\ and\ \bibinfo {author} {\bibfnamefont {A.}~\bibnamefont
  {Furusaki}},\ }\bibfield  {title} {\bibinfo {title} {Double {Majorana} vortex
  zero modes in superconducting topological crystalline insulators with surface
  rotation anomaly},\ }\href {https://doi.org/10.1103/PhysRevB.102.180505}
  {\bibfield  {journal} {\bibinfo  {journal} {Phys. Rev. B}\ }\textbf {\bibinfo
  {volume} {102}},\ \bibinfo {pages} {180505(R)} (\bibinfo {year}
  {2020})}\BibitemShut {NoStop}%
\bibitem [{\citenamefont {Hu}\ \emph {et~al.}(2022)\citenamefont {Hu},
  \citenamefont {Wu}, \citenamefont {Liu},\ and\ \citenamefont
  {Zhang}}]{Hu.2022}%
  \BibitemOpen
  \bibfield  {author} {\bibinfo {author} {\bibfnamefont {L.-H.}\ \bibnamefont
  {Hu}}, \bibinfo {author} {\bibfnamefont {X.}~\bibnamefont {Wu}}, \bibinfo
  {author} {\bibfnamefont {C.-X.}\ \bibnamefont {Liu}},\ and\ \bibinfo {author}
  {\bibfnamefont {R.-X.}\ \bibnamefont {Zhang}},\ }\bibfield  {title} {\bibinfo
  {title} {Competing vortex topologies in iron-based superconductors},\ }\href
  {https://doi.org/10.1103/PhysRevLett.129.277001} {\bibfield  {journal}
  {\bibinfo  {journal} {Phys. Rev. Lett.}\ }\textbf {\bibinfo {volume} {129}},\
  \bibinfo {pages} {277001} (\bibinfo {year} {2022})}\BibitemShut {NoStop}%
\bibitem [{\citenamefont {Kobayashi}\ \emph {et~al.}(2023)\citenamefont
  {Kobayashi}, \citenamefont {Sumita}, \citenamefont {Hirayama},\ and\
  \citenamefont {Furusaki}}]{Kobayashi.2023}%
  \BibitemOpen
  \bibfield  {author} {\bibinfo {author} {\bibfnamefont {S.}~\bibnamefont
  {Kobayashi}}, \bibinfo {author} {\bibfnamefont {S.}~\bibnamefont {Sumita}},
  \bibinfo {author} {\bibfnamefont {M.}~\bibnamefont {Hirayama}},\ and\
  \bibinfo {author} {\bibfnamefont {A.}~\bibnamefont {Furusaki}},\ }\bibfield
  {title} {\bibinfo {title} {Crystal symmetry protected gapless vortex line
  phases in superconducting {Dirac} semimetals},\ }\href
  {https://doi.org/10.1103/PhysRevB.107.214518} {\bibfield  {journal} {\bibinfo
   {journal} {Phys. Rev. B}\ }\textbf {\bibinfo {volume} {107}},\ \bibinfo
  {pages} {214518} (\bibinfo {year} {2023})}\BibitemShut {NoStop}%
\bibitem [{\citenamefont {Kraus}\ \emph {et~al.}(2009)\citenamefont {Kraus},
  \citenamefont {Auerbach}, \citenamefont {Fertig},\ and\ \citenamefont
  {Simon}}]{Kraus.2009}%
  \BibitemOpen
  \bibfield  {author} {\bibinfo {author} {\bibfnamefont {Y.~E.}\ \bibnamefont
  {Kraus}}, \bibinfo {author} {\bibfnamefont {A.}~\bibnamefont {Auerbach}},
  \bibinfo {author} {\bibfnamefont {H.~A.}\ \bibnamefont {Fertig}},\ and\
  \bibinfo {author} {\bibfnamefont {S.~H.}\ \bibnamefont {Simon}},\ }\bibfield
  {title} {\bibinfo {title} {Majorana fermions of a two-dimensional $p_x+ip_y$
  superconductor},\ }\href {https://doi.org/10.1103/PhysRevB.79.134515}
  {\bibfield  {journal} {\bibinfo  {journal} {Phys. Rev. B}\ }\textbf {\bibinfo
  {volume} {79}},\ \bibinfo {pages} {134515} (\bibinfo {year}
  {2009})}\BibitemShut {NoStop}%
\bibitem [{\citenamefont {Lee}\ and\ \citenamefont
  {Schnyder}(2016)}]{Lee.2016}%
  \BibitemOpen
  \bibfield  {author} {\bibinfo {author} {\bibfnamefont {D.}~\bibnamefont
  {Lee}}\ and\ \bibinfo {author} {\bibfnamefont {A.~P.}\ \bibnamefont
  {Schnyder}},\ }\bibfield  {title} {\bibinfo {title} {Structure of
  vortex-bound states in spin-singlet chiral superconductors},\ }\href
  {https://doi.org/10.1103/PhysRevB.93.064522} {\bibfield  {journal} {\bibinfo
  {journal} {Phys. Rev. B}\ }\textbf {\bibinfo {volume} {93}},\ \bibinfo
  {pages} {064522} (\bibinfo {year} {2016})}\BibitemShut {NoStop}%
\bibitem [{\citenamefont {Hanaguri}\ \emph {et~al.}(2012)\citenamefont
  {Hanaguri}, \citenamefont {Kitagawa}, \citenamefont {Matsubayashi},
  \citenamefont {Mazaki}, \citenamefont {Uwatoko},\ and\ \citenamefont
  {Takagi}}]{Hanaguri.2012}%
  \BibitemOpen
  \bibfield  {author} {\bibinfo {author} {\bibfnamefont {T.}~\bibnamefont
  {Hanaguri}}, \bibinfo {author} {\bibfnamefont {K.}~\bibnamefont {Kitagawa}},
  \bibinfo {author} {\bibfnamefont {K.}~\bibnamefont {Matsubayashi}}, \bibinfo
  {author} {\bibfnamefont {Y.}~\bibnamefont {Mazaki}}, \bibinfo {author}
  {\bibfnamefont {Y.}~\bibnamefont {Uwatoko}},\ and\ \bibinfo {author}
  {\bibfnamefont {H.}~\bibnamefont {Takagi}},\ }\bibfield  {title} {\bibinfo
  {title} {Scanning tunneling microscopy/spectroscopy of vortices in
  {LiFeAs}},\ }\href {https://doi.org/10.1103/PhysRevB.85.214505} {\bibfield
  {journal} {\bibinfo  {journal} {Phys. Rev. B}\ }\textbf {\bibinfo {volume}
  {85}},\ \bibinfo {pages} {214505} (\bibinfo {year} {2012})}\BibitemShut
  {NoStop}%
\bibitem [{\citenamefont {Liu}\ \emph {et~al.}(2020)\citenamefont {Liu},
  \citenamefont {Cao}, \citenamefont {Zhu}, \citenamefont {Kong}, \citenamefont
  {Wang}, \citenamefont {Papaj}, \citenamefont {Zhang}, \citenamefont {Liu},
  \citenamefont {Chen}, \citenamefont {Li}, \citenamefont {Yang}, \citenamefont
  {Kondo}, \citenamefont {Du}, \citenamefont {Cao}, \citenamefont {Shin},
  \citenamefont {Fu}, \citenamefont {Yin}, \citenamefont {Gao},\ and\
  \citenamefont {Ding}}]{Liu.2020}%
  \BibitemOpen
  \bibfield  {author} {\bibinfo {author} {\bibfnamefont {W.}~\bibnamefont
  {Liu}}, \bibinfo {author} {\bibfnamefont {L.}~\bibnamefont {Cao}}, \bibinfo
  {author} {\bibfnamefont {S.}~\bibnamefont {Zhu}}, \bibinfo {author}
  {\bibfnamefont {L.}~\bibnamefont {Kong}}, \bibinfo {author} {\bibfnamefont
  {G.}~\bibnamefont {Wang}}, \bibinfo {author} {\bibfnamefont {M.}~\bibnamefont
  {Papaj}}, \bibinfo {author} {\bibfnamefont {P.}~\bibnamefont {Zhang}},
  \bibinfo {author} {\bibfnamefont {Y.-B.}\ \bibnamefont {Liu}}, \bibinfo
  {author} {\bibfnamefont {H.}~\bibnamefont {Chen}}, \bibinfo {author}
  {\bibfnamefont {G.}~\bibnamefont {Li}}, \bibinfo {author} {\bibfnamefont
  {F.}~\bibnamefont {Yang}}, \bibinfo {author} {\bibfnamefont {T.}~\bibnamefont
  {Kondo}}, \bibinfo {author} {\bibfnamefont {S.}~\bibnamefont {Du}}, \bibinfo
  {author} {\bibfnamefont {G.-H.}\ \bibnamefont {Cao}}, \bibinfo {author}
  {\bibfnamefont {S.}~\bibnamefont {Shin}}, \bibinfo {author} {\bibfnamefont
  {L.}~\bibnamefont {Fu}}, \bibinfo {author} {\bibfnamefont {Z.}~\bibnamefont
  {Yin}}, \bibinfo {author} {\bibfnamefont {H.-J.}\ \bibnamefont {Gao}},\ and\
  \bibinfo {author} {\bibfnamefont {H.}~\bibnamefont {Ding}},\ }\bibfield
  {title} {\bibinfo {title} {A new {Majorana} platform in an {Fe-As} bilayer
  superconductor},\ }\href {https://doi.org/10.1038/s41467-020-19487-1}
  {\bibfield  {journal} {\bibinfo  {journal} {Nat. Commun.}\ }\textbf {\bibinfo
  {volume} {11}},\ \bibinfo {pages} {5688} (\bibinfo {year}
  {2020})}\BibitemShut {NoStop}%
\bibitem [{\citenamefont {Chen}\ \emph {et~al.}(2021)\citenamefont {Chen},
  \citenamefont {Duan}, \citenamefont {Fan}, \citenamefont {Hong},
  \citenamefont {Chen}, \citenamefont {Yang}, \citenamefont {Li}, \citenamefont
  {Luo},\ and\ \citenamefont {Wen}}]{Chen.2021}%
  \BibitemOpen
  \bibfield  {author} {\bibinfo {author} {\bibfnamefont {X.}~\bibnamefont
  {Chen}}, \bibinfo {author} {\bibfnamefont {W.}~\bibnamefont {Duan}}, \bibinfo
  {author} {\bibfnamefont {X.}~\bibnamefont {Fan}}, \bibinfo {author}
  {\bibfnamefont {W.}~\bibnamefont {Hong}}, \bibinfo {author} {\bibfnamefont
  {K.}~\bibnamefont {Chen}}, \bibinfo {author} {\bibfnamefont {H.}~\bibnamefont
  {Yang}}, \bibinfo {author} {\bibfnamefont {S.}~\bibnamefont {Li}}, \bibinfo
  {author} {\bibfnamefont {H.}~\bibnamefont {Luo}},\ and\ \bibinfo {author}
  {\bibfnamefont {H.-H.}\ \bibnamefont {Wen}},\ }\bibfield  {title} {\bibinfo
  {title} {Friedel oscillations of vortex bound states under extreme quantum
  limit in {KCa$_2$Fe$_4$As$_4$F$_2$}},\ }\href
  {https://doi.org/10.1103/PhysRevLett.126.257002} {\bibfield  {journal}
  {\bibinfo  {journal} {Phys. Rev. Lett.}\ }\textbf {\bibinfo {volume} {126}},\
  \bibinfo {pages} {257002} (\bibinfo {year} {2021})}\BibitemShut {NoStop}%
\bibitem [{\citenamefont {Wiesendanger}(2009)}]{Wiesendanger.2009}%
  \BibitemOpen
  \bibfield  {author} {\bibinfo {author} {\bibfnamefont {R.}~\bibnamefont
  {Wiesendanger}},\ }\bibfield  {title} {\bibinfo {title} {Spin mapping at the
  nanoscale and atomic scale},\ }\href
  {https://doi.org/10.1103/RevModPhys.81.1495} {\bibfield  {journal} {\bibinfo
  {journal} {Rev. Mod. Phys.}\ }\textbf {\bibinfo {volume} {81}},\ \bibinfo
  {pages} {1495} (\bibinfo {year} {2009})}\BibitemShut {NoStop}%
\bibitem [{\citenamefont {Zhang}\ \emph {et~al.}(2009)\citenamefont {Zhang},
  \citenamefont {Liu}, \citenamefont {Qi}, \citenamefont {Dai}, \citenamefont
  {Fang},\ and\ \citenamefont {Zhang}}]{Zhang.2009}%
  \BibitemOpen
  \bibfield  {author} {\bibinfo {author} {\bibfnamefont {H.}~\bibnamefont
  {Zhang}}, \bibinfo {author} {\bibfnamefont {C.-X.}\ \bibnamefont {Liu}},
  \bibinfo {author} {\bibfnamefont {X.-L.}\ \bibnamefont {Qi}}, \bibinfo
  {author} {\bibfnamefont {X.}~\bibnamefont {Dai}}, \bibinfo {author}
  {\bibfnamefont {Z.}~\bibnamefont {Fang}},\ and\ \bibinfo {author}
  {\bibfnamefont {S.-C.}\ \bibnamefont {Zhang}},\ }\bibfield  {title} {\bibinfo
  {title} {Topological insulators in {Bi$_2$Se$_3$}, {Bi$_2$Te$_3$} and
  {Sb$_2$Te$_3$} with a single {Dirac} cone on the surface},\ }\href
  {https://doi.org/10.1038/nphys1270} {\bibfield  {journal} {\bibinfo
  {journal} {Nat. Phys.}\ }\textbf {\bibinfo {volume} {5}},\ \bibinfo {pages}
  {438} (\bibinfo {year} {2009})}\BibitemShut {NoStop}%
\bibitem [{\citenamefont {Donís~Vela}\ \emph {et~al.}(2021)\citenamefont
  {Donís~Vela}, \citenamefont {Lemut}, \citenamefont {Pacholski},\ and\
  \citenamefont {Beenakker}}]{Donís-Vela.2021}%
  \BibitemOpen
  \bibfield  {author} {\bibinfo {author} {\bibfnamefont {A.}~\bibnamefont
  {Donís~Vela}}, \bibinfo {author} {\bibfnamefont {G.}~\bibnamefont {Lemut}},
  \bibinfo {author} {\bibfnamefont {M.~J.}\ \bibnamefont {Pacholski}},\ and\
  \bibinfo {author} {\bibfnamefont {C.~W.~J.}\ \bibnamefont {Beenakker}},\
  }\bibfield  {title} {\bibinfo {title} {Chirality inversion of {Majorana} edge
  modes in a {Fu–Kane} heterostructure},\ }\href
  {https://doi.org/10.1088/1367-2630/ac265f} {\bibfield  {journal} {\bibinfo
  {journal} {New J. Phys.}\ }\textbf {\bibinfo {volume} {23}},\ \bibinfo
  {pages} {103006} (\bibinfo {year} {2021})}\BibitemShut {NoStop}%
\bibitem [{\citenamefont {Wang}\ \emph {et~al.}(2023)\citenamefont {Wang},
  \citenamefont {Wu}, \citenamefont {Mogi}, \citenamefont {Kawamura},
  \citenamefont {Tokura}, \citenamefont {Shen}, \citenamefont {You},\ and\
  \citenamefont {Allen}}]{Wang.202341}%
  \BibitemOpen
  \bibfield  {author} {\bibinfo {author} {\bibfnamefont {T.}~\bibnamefont
  {Wang}}, \bibinfo {author} {\bibfnamefont {C.}~\bibnamefont {Wu}}, \bibinfo
  {author} {\bibfnamefont {M.}~\bibnamefont {Mogi}}, \bibinfo {author}
  {\bibfnamefont {M.}~\bibnamefont {Kawamura}}, \bibinfo {author}
  {\bibfnamefont {Y.}~\bibnamefont {Tokura}}, \bibinfo {author} {\bibfnamefont
  {Z.-X.}\ \bibnamefont {Shen}}, \bibinfo {author} {\bibfnamefont
  {Y.}~\bibnamefont {You}},\ and\ \bibinfo {author} {\bibfnamefont {M.~T.}\
  \bibnamefont {Allen}},\ }\bibfield  {title} {\bibinfo {title} {Probing the
  edge states of {Chern} insulators using microwave impedance microscopy},\
  }\href {https://doi.org/10.1103/physrevb.108.235432} {\bibfield  {journal}
  {\bibinfo  {journal} {Phys. Rev. B}\ }\textbf {\bibinfo {volume} {108}},\
  \bibinfo {pages} {235432} (\bibinfo {year} {2023})}\BibitemShut {NoStop}%
\bibitem [{\citenamefont {Inbar}\ \emph {et~al.}(2023)\citenamefont {Inbar},
  \citenamefont {Birkbeck}, \citenamefont {Xiao}, \citenamefont {Taniguchi},
  \citenamefont {Watanabe}, \citenamefont {Yan}, \citenamefont {Oreg},
  \citenamefont {Stern}, \citenamefont {Berg},\ and\ \citenamefont
  {Ilani}}]{Inbar.2023}%
  \BibitemOpen
  \bibfield  {author} {\bibinfo {author} {\bibfnamefont {A.}~\bibnamefont
  {Inbar}}, \bibinfo {author} {\bibfnamefont {J.}~\bibnamefont {Birkbeck}},
  \bibinfo {author} {\bibfnamefont {J.}~\bibnamefont {Xiao}}, \bibinfo {author}
  {\bibfnamefont {T.}~\bibnamefont {Taniguchi}}, \bibinfo {author}
  {\bibfnamefont {K.}~\bibnamefont {Watanabe}}, \bibinfo {author}
  {\bibfnamefont {B.}~\bibnamefont {Yan}}, \bibinfo {author} {\bibfnamefont
  {Y.}~\bibnamefont {Oreg}}, \bibinfo {author} {\bibfnamefont {A.}~\bibnamefont
  {Stern}}, \bibinfo {author} {\bibfnamefont {E.}~\bibnamefont {Berg}},\ and\
  \bibinfo {author} {\bibfnamefont {S.}~\bibnamefont {Ilani}},\ }\bibfield
  {title} {\bibinfo {title} {The quantum twisting microscope},\ }\href
  {https://doi.org/10.1038/s41586-022-05685-y} {\bibfield  {journal} {\bibinfo
  {journal} {Nature}\ }\textbf {\bibinfo {volume} {614}},\ \bibinfo {pages}
  {682} (\bibinfo {year} {2023})}\BibitemShut {NoStop}%
\bibitem [{\citenamefont {Sedlmayr}\ \emph {et~al.}(2015)\citenamefont
  {Sedlmayr}, \citenamefont {Guigou}, \citenamefont {Simon},\ and\
  \citenamefont {Bena}}]{Sedlmayr.2015}%
  \BibitemOpen
  \bibfield  {author} {\bibinfo {author} {\bibfnamefont {N.}~\bibnamefont
  {Sedlmayr}}, \bibinfo {author} {\bibfnamefont {M.}~\bibnamefont {Guigou}},
  \bibinfo {author} {\bibfnamefont {P.}~\bibnamefont {Simon}},\ and\ \bibinfo
  {author} {\bibfnamefont {C.}~\bibnamefont {Bena}},\ }\bibfield  {title}
  {\bibinfo {title} {Majoranas with and without a ‘character’:
  hybridization, braiding and chiral {Majorana} number},\ }\href
  {https://doi.org/10.1088/0953-8984/27/45/455601} {\bibfield  {journal}
  {\bibinfo  {journal} {J. Phys.: Condens. Matter}\ }\textbf {\bibinfo {volume}
  {27}},\ \bibinfo {pages} {455601} (\bibinfo {year} {2015})}\BibitemShut
  {NoStop}%
\bibitem [{\citenamefont {Maiellaro}\ \emph {et~al.}(2023)\citenamefont
  {Maiellaro}, \citenamefont {Settino}, \citenamefont {Guarcello},
  \citenamefont {Romeo},\ and\ \citenamefont {Citro}}]{Maiellaro.2023}%
  \BibitemOpen
  \bibfield  {author} {\bibinfo {author} {\bibfnamefont {A.}~\bibnamefont
  {Maiellaro}}, \bibinfo {author} {\bibfnamefont {J.}~\bibnamefont {Settino}},
  \bibinfo {author} {\bibfnamefont {C.}~\bibnamefont {Guarcello}}, \bibinfo
  {author} {\bibfnamefont {F.}~\bibnamefont {Romeo}},\ and\ \bibinfo {author}
  {\bibfnamefont {R.}~\bibnamefont {Citro}},\ }\bibfield  {title} {\bibinfo
  {title} {Hallmarks of orbital-flavored {M}ajorana states in {J}osephson
  junctions based on oxide nanochannels},\ }\href
  {https://doi.org/10.1103/PhysRevB.107.L201405} {\bibfield  {journal}
  {\bibinfo  {journal} {Phys. Rev. B}\ }\textbf {\bibinfo {volume} {107}},\
  \bibinfo {pages} {L201405} (\bibinfo {year} {2023})}\BibitemShut {NoStop}%
\bibitem [{\citenamefont {Venditti}\ \emph {et~al.}()\citenamefont {Venditti},
  \citenamefont {Berthod},\ and\ \citenamefont {Rademaker}}]{data-zenodo}%
  \BibitemOpen
  \bibfield  {author} {\bibinfo {author} {\bibfnamefont {G.}~\bibnamefont
  {Venditti}}, \bibinfo {author} {\bibfnamefont {C.}~\bibnamefont {Berthod}},\
  and\ \bibinfo {author} {\bibfnamefont {L.}~\bibnamefont {Rademaker}},\
  }\bibfield  {title} {\bibinfo {title} {Open data to ``{A}ngular momentum of
  vortex-core {M}ajorana zero modes'' [{D}ata set]},\ }\bibfield  {journal}
  {\bibinfo  {journal} {Zenodo (2025),}\ }\href
  {https://doi.org/https://doi.org/10.5281/zenodo.17868235}
  {https://doi.org/10.5281/zenodo.17868235}\BibitemShut {NoStop}%
\bibitem [{\citenamefont {Vafek}\ \emph {et~al.}(2001)\citenamefont {Vafek},
  \citenamefont {Melikyan}, \citenamefont {Franz},\ and\ \citenamefont
  {Tešanović}}]{Vafek.2001}%
  \BibitemOpen
  \bibfield  {author} {\bibinfo {author} {\bibfnamefont {O.}~\bibnamefont
  {Vafek}}, \bibinfo {author} {\bibfnamefont {A.}~\bibnamefont {Melikyan}},
  \bibinfo {author} {\bibfnamefont {M.}~\bibnamefont {Franz}},\ and\ \bibinfo
  {author} {\bibfnamefont {Z.}~\bibnamefont {Tešanović}},\ }\bibfield
  {title} {\bibinfo {title} {Quasiparticles and vortices in unconventional
  superconductors},\ }\href {https://doi.org/10.1103/physrevb.63.134509}
  {\bibfield  {journal} {\bibinfo  {journal} {Phys. Rev. B}\ }\textbf {\bibinfo
  {volume} {63}},\ \bibinfo {pages} {134509} (\bibinfo {year}
  {2001})}\BibitemShut {NoStop}%
\end{thebibliography}
\end{document}